\documentclass[longauth]{aa}
\usepackage{graphicx}
\usepackage{txfonts}
\usepackage{booktabs}
\usepackage{float}
\usepackage{pdflscape}
\usepackage{caption}
\usepackage{comment}
\usepackage{placeins}
\usepackage{stfloats}

\usepackage{hyperref}
\begin{document} 
\defcitealias{2023A&A...676A..76B}{B23}

\title{Ultra High-Redshift or Closer-by, Dust-Obscured Galaxies? Deciphering the Nature of Faint, Previously Missed F200W-Dropouts in CEERS}

\titlerunning{Ultra High-Redshift or Dust? Faint, Previously Missed Dropouts in CEERS}

\author{G. Gandolfi$^{1,2}$\thanks{Corresponding author, \email{giovanni.gandolfi@unipd.it}}
\and G. Rodighiero$^{1,2}$
\and L. Bisigello$^{2}$
\and A. Grazian$^{2}$
\and S. L. Finkelstein$^{3}$
\and M. Dickinson$^{4}$
\and M. Castellano$^{5}$
\and E. Merlin$^{5}$
\and A. Calabr\`{o}$^{5}$
\and C. Papovich$^{6,7}$
\and A. Bianchetti$^{1,2}$
\and E. Ba\~{n}ados$^{8}$
\and P. Benotto$^{1,2}$
\and M. Catone$^{1}$
\and F. Buitrago$^{9,10}$
\and E. Daddi$^{11}$
\and G. Girardi$^{1,2}$
\and M. Giulietti$^{3}$
\and M. Hirschmann$^{12,13}$
\and B. W. Holwerda$^{14}$
\and P. Arrabal Haro$^{15}$\thanks{NASA Postdoctoral Fellow}
\and A. Lapi$^{16,3}$
\and R. A. Lucas$^{17}$
\and Y. Lyu$^{11}$
\and M. Massardi$^{18,16}$
\and F. Pacucci$^{19,20}$
\and P. G. P\'{e}rez-Gonz\'{a}lez$^{21}$
\and T. Ronconi$^{16}$
\and M. Tarrasse$^{11}$
\and S. Wilkins$^{22}$
\and B. Vulcani$^{2}$
\and L. Y. A. Yung$^{17}$
\and J. A. Zavala$^{23,24}$
\and B. Backhaus$^{25}$
\and M. Bagley$^{3}$
\and V. Buat$^{26}$
\and D. Burgarella$^{26}$
\and J. Kartaltepe$^{27}$
\and Y. Khusanova$^{8}$
\and A. Kirkpatrick$^{28}$
\and D. Kocevski$^{29}$
\and A. M. Koekemoer$^{17}$
\and E. Lambrides$^{15}$
\and N. Pirzkal$^{30}$
\and G. Yang$^{31,32}$
}

\institute{
$^{1}$ Dipartimento di Fisica e Astronomia "G. Galilei", Universit\`a di Padova, Vicolo dell'Osservatorio 3, 35131 Padova, Italy \\
$^{2}$ INAF, Osservatorio Astronomico di Padova, Vicolo dell'Osservatorio 5, 35122 Padova, Italy \\
$^{3}$ Department of Astronomy, The University of Texas at Austin, Austin, TX, USA \\
$^{4}$ NSF's National Optical-Infrared Astronomy Research Laboratory, 950 N. Cherry Ave., Tucson, AZ 85719, USA \\
$^{5}$ INAF, Osservatorio Astronomico di Roma, Via Frascati 33, 00078 Monteporzio Catone, Roma, Italy \\
$^{6}$ Department of Physics and Astronomy, Texas A\&M University, College Station, TX, 77843-4242 USA \\
$^{7}$ George P. and Cynthia Woods Mitchell Institute for Fundamental Physics and Astronomy, Texas A\&M University, College Station, TX, 77843-4242 USA \\
$^{8}$ Max-Planck-Institut f\"ur Astronomie, K\"onigstuhl 17, D-69117, Heidelberg, Germany \\
$^{9}$ Departamento de F\'isica Te\'orica, At\'omica y \"Optica, Universidad de Valladolid, 47011 Valladolid, Spain \\
$^{10}$ Instituto de Astrof\'isica e Ci\^encias do Espa\c{c}o, Universidade de Lisboa, OAL, Tapada da Ajuda, PT1349-018 Lisbon, Portugal \\
$^{11}$ Universit\'e Paris-Saclay, Universit\'e Paris Cit\'e, CEA, CNRS, AIM, 91191, Gif-sur-Yvette, France \\
$^{12}$ Institute for Physics, Laboratory for Galaxy Evolution and Spectral modelling, \'Ecole Polytechnique F\'ed\'erale de Lausanne, Observatoire de Sauverny, Chemin P\'egasi 51, 1290 Versoix, Switzerland \\
$^{13}$ INAF, Osservatorio Astronomico di Trieste, Via Tiepolo 11, 34131 Trieste, Italy \\
$^{14}$ Department of Physics \& Astronomy, University of Louisville, Natural Science Building 102, Louisville, KY 40292, USA \\
$^{15}$ Astrophysics Science Division, NASA Goddard Space Flight Center, 8800 Greenbelt Rd, Greenbelt, MD 20771, USA \\
$^{16}$ SISSA, Via Bonomea 265, 34136 Trieste, Italy \\
$^{17}$ Space Telescope Science Institute, 3700 San Martin Drive, Baltimore, MD 21218, USA \\
$^{18}$ IRA-INAF / Italian ARC, via Gobetti 101, 40129 Bologna \\
$^{19}$ Center for Astrophysics $\vert$ Harvard \& Smithsonian, 60 Garden St, Cambridge, MA 02138, USA \\
$^{20}$ Black Hole Initiative, Harvard University, 20 Garden St, Cambridge, MA 02138, USA \\
$^{21}$ Centro de Astrobiolog\'ia (CAB), CSIC-INTA, Ctra. de Ajalvir km 4, Torrej\'on de Ardoz, E-28850, Madrid, Spain \\
$^{22}$ Astronomy Centre, Department of Physics \& Astronomy, University of Sussex, Brighton, BN1 9RH, United Kingdom \\
$^{23}$ National Astronomical Observatory of Japan, 2-21-1 Osawa, Mitaka, Tokyo 181-8588, Japan \\
$^{24}$ Department of Astronomy, University of Massachusetts, 710 North Pleasant Street, Amherst MA 01003, USA \\
$^{25}$ Department of Physics, 196 Auditorium Road, Unit 3046, University of Connecticut, Storrs, CT 06269 \\
$^{26}$ Aix Marseille Universit\'e, CNRS, CNES, LAM, Marseille, France \\
$^{27}$ Laboratory for Multiwavelength Astrophysics, School of Physics and Astronomy, Rochester Institute of Technology, 84 Lomb Memorial Drive, Rochester, NY 14623, USA \\
$^{28}$ Department of Physics and Astronomy, University of Kansas, Lawrence, KS 66045, USA \\
$^{29}$ Department of Physics and Astronomy, Colby College, Waterville, ME 04901, USA \\
$^{30}$ ESA/AURA Space Telescope Science Institute, USA \\
$^{31}$ Kapteyn Astronomical Institute, University of Groningen, P.O. Box 800, 9700 AV Groningen, The Netherlands \\
$^{32}$ SRON Netherlands Institute for Space Research, Postbus 800, 9700 AV Groningen, The Netherlands
}

   \date{Received -; accepted -}

\authorrunning{Gandolfi et al.}

 \abstract {The James Webb Space Telescope (JWST) is revolutionizing our understanding of the Universe by unveiling faint, near-infrared dropouts previously beyond our reach, ranging from exceptionally dusty sources to galaxies up to redshift $z \sim 14$.} {In this paper, we identify F200W-dropout objects in the Cosmic Evolution Early Release Science (CEERS) survey which are absent from existing catalogs. Our selection method can effectively identify obscured low-mass ($\log \text{M}_* / \text{M}_\odot$$\leq 9$) objects at $z \leq 6$, massive dust-rich sources up to $z \sim 12$, and ultra-high-redshift ($z > 15$) candidates. Our goal is to uncover promising targets for further studies using deep mid-infrared imaging and/or spectroscopic follow-ups.} {We utilize two photometric catalogs optimized for detecting faint, red objects. Primarily relying on NIRCam photometry from the latest CEERS data release and supplementing with Mid-Infrared/(sub-)mm data when available, our analysis pipeline combines multiple SED-fitting codes, star formation histories, and the novel \texttt{CosMix} tool for astronomical stacking to maximize available photometric information.} {Our work highlights three $2<z<3$ dusty dwarf galaxies which have larger masses compared to the typical dusty dwarfs previously identified in CEERS. Additionally, we reveal five faint sources with significant probability of lying above $z>15$, with best-fit masses compatible with $\Lambda$CDM and a standard baryons-to-star conversion efficiency. We exploit these candidates to compute the $z\!\sim\!17$ UV luminosity function, finding estimates in good agreement with other similar studies. Their bi-modal redshift probability distributions suggest they could also be $z<1.5$ dwarf galaxies with extreme dust extinction. We also identify a strong line emitter galaxy at $z \sim 5$ mimicking the near-infrared emission of a $z \sim 13$ galaxy.} {Our sample holds promising candidates for future follow-ups. Confirming ultra high-redshift galaxies or lower-z dusty dwarfs will offer valuable insights into early galaxy formation, evolution with their central black holes and the nature of dark matter, and/or cosmic dust production mechanisms in low-mass galaxies, and will help us to understand degeneracies and contamination in high-z object searches.}

   \maketitle

\section{Introduction}
The James Webb Space Telescope (JWST; \citealt{2006SSRv..123..485G, 2023PASP..135f8001G}) has revolutionized our ability to detect faint galaxies at high redshifts, thanks to its unprecedented sensitivity and wavelength coverage extending beyond 2~$\mu$m. In particular, the identification of near-infrared (NIR) dropout sources --- objects that drop out of detection in the bluest available NIRCam filters --- has emerged as a powerful approach for uncovering previously hidden galaxy populations. These dropouts can trace a wide range of astrophysical phenomena across cosmic time.

NIR dropouts can signal the presence of high-redshift ($z>10$) galaxies, where the dropout is caused by the Lyman break redshifted into the NIR bands. These galaxies offer a unique window into the epoch of reionization and the early formation of stars, black holes, and large-scale structure (e.g., \citealt{2001PhR...349..125B, 2022ApJ...940L..14N, 2024arXiv240518485C}). Moreover, pre-launch simulations had predicted JWST's capability to detect galaxies even at ultra-high redshifts (UHR; $z>15$, e.g., \citealt{Cowley2017Predictions, Behroozi2020The, 2020MNRAS.496.4574Y}), and several candidates have been photometrically identified (e.g., \citealt{2023ApJ...952L...7A, 2023ApJS..265....5H, 2024arXiv240714973C, 2025arXiv250315594P, Castellano2025}), whose brightness has been compared with theoretical models and predictions (e.g., \citealt{2023MNRAS.522.3986F, 2025arXiv250318850M, 2025A&A...697A..65M}). However, no spectroscopic confirmations have yet been achieved \citep{2024ApJ...970...31R, 2024ApJ...969L..10P}, due to the extreme faintness of these objects and degeneracies in photometric redshift estimates.

At intermediate redshifts ($1 \lesssim z \lesssim 6$), NIR dropouts can unveil heavily dust-obscured galaxies, where the lack of detection in bluer bands arises from strong attenuation by cosmic dust. These include massive, ultra-violet (UV) and/or optically dark galaxies (e.g., \citealt{2016ApJ...816...84W, 2022ApJ...927..204E}) as well as the recently identified High-Extinction, Low-Mass (HELM) population \citep{2023A&A...676A..76B}. The latter are especially intriguing, as their combination of low stellar mass and high dust content challenges standard dust production models which associate significant dust enrichment with high stellar mass and strong star formation (e.g., \citealt{Brinchmann2004, Dayal2013, 2024arXiv241010954B}).

Moreover, other sources such as galaxies with strong line emission at intermediate redshift \citep{2023Natur.622..707A, 2023MNRAS.518.6011D} can reproduce the photometric properties expected for $z>15$ galaxies. A prominent example is CEERS-93316, a $z=4.9$ dusty star-forming galaxy whose emission-line spectrum mimicked the Lyman break of a $z \sim 16$ source \citep{2023ApJS..265....5H, 2023ApJ...946L..16P}.

Finally, even cold, Milky Way sub-stellar objects such as brown dwarfs \citep{2024MNRAS.529.1067H} characterized by strong molecular absorption features can appear as NIR-dropouts, mimicking the photometric colors of both dusty galaxies and UHR systems.

The overlapping photometric signatures of these diverse populations lead to substantial degeneracies in spectral energy distribution (SED) fitting and complicate the interpretation of NIR dropouts. To disentangle these overlapping populations, robust photometric analysis pipelines must account for a wide range of physical scenarios, including variation in dust attenuation, star formation histories, and possible contamination from stars or AGN. Identifying the most promising candidates for future follow-up requires not only broad and deep multi-wavelength photometry, but also careful modeling that evaluates and compares alternative redshift and physical parameter solutions.

In this study, we conduct a targeted search for faint, previously unpublished F200W-dropout sources in the CEERS survey \citep{2023ApJ...946L..13F, 2023ApJ...946L..12B, Yang2023, 2025arXiv250104085F}. Our goal is to identify and characterize candidates that may belong to one of the extreme populations described above: either dusty galaxies at low to intermediate redshift, or UHR systems at $z>15$. To this end, we apply a multi-pronged SED fitting analysis incorporating diverse star formation histories and dust laws, and we introduce \texttt{CosMix}, a novel tool designed to stack the photometric signal of similar sources to improve the robustness of the inferred physical parameters. Our framework explicitly accounts for alternative solutions, including intermediate-$z$ strong line emitters and potential AGN.

The paper is organized as follows: Section \ref{2|sec:ceers} provides an overview of the CEERS survey, describing the data exploited in this work; Section \ref{3|sec:photometry} presents the catalogs used in the analysis and provides relevant information on the extraction of the photometry; Section \ref{4|sec:sampleselection} details the selection technique used to obtain our sample of objects and a discussion of their photometric colors; Section \ref{5|sec:sedfitting} outlines the SED-fitting procedure adopted on our objects; Section \ref{6|sec:results} delivers the core results of our analysis showing the results of the adopted SED-fitting procedure, as well as computing the UV luminosity function at ultra high-redshifts; Section \ref{discussion} contains a discussion of our results and Section \ref{7|sec:conclusions} draws the relevant conclusions, giving prospects for future spectroscopic studies that could target our sources.

Throughout this work, we adopt the cosmological parameters yielded by the latest Planck collaboration release \citep{2020A&A...641A...6P} and a \cite{10.1046/j.1365-8711.2001.04022.x} initial mass function (IMF).

\section{CEERS data}
\label{2|sec:ceers}
This work uses imaging data obtained by JWST's NIRCam as part of the CEERS survey (P.I. S. Finkelstein; \citealt{2017jwst.prop.1345F, 2023ApJ...946L..13F, 2023ApJ...946L..12B, Yang2023, finkelstein2025cosmicevolutionearlyrelease}). CEERS covers $\sim$90 arcmin$^2$ of the Extended Groth Strip \citep[EGS;][]{2007ApJ...660L...1D} field with JWST imaging and spectroscopy through a 77.2 hours Director's Discretionary Early Release Science Program. CEERS NIRCam observations are available in three short-wavelength bands (F115W, F150W, F200W) and four long-wavelength bands (F277W, F356W, F410M, F444W). Supplementary coverage in the F090W band is provided by the Cycle 1 GO program ``The JWST-legacy narrow-band survey of H-alpha and [OIII] emitters in the epoch of reionization'' (P.I. E.o Ba\~nados). This program also includes observations in the F470N band, which will not be utilized in this work due to their low sensitivity. CEERS NIRCam pointings were previously targeted by Hubble Space Telescope (HST) observations both with the Advanced Camera for Surveys (ACS; covering F435W, F606W and F814W) and Wide Field Camera 3 (WFC3; covering F105W, F125W, F140W and F160W) as part of the Cosmic Assembly Near-infrared Deep Extragalactic Legacy Survey (CANDELS; \citealt{2011ApJS..197...35G,2011ApJS..197...36K}).

This work utilizes the latest release of CEERS NIRCam imaging data (v.1.0), processed with the JWST Calibration Pipeline (v.1.13.4; \citealt{bushouse2024jwst}) and a customized CEERS pipeline for enhanced $1/f$ noise subtraction and artifact removal, as detailed in \cite{2023ApJ...946L..12B}. Compared to earlier releases (v.0.51 or v.0.2), Data Release (DR) v.1.0 provides significant improvements in removing ``snowballs'' (cosmic-ray-induced visual artifacts, cleaned using James Davies' \texttt{snowblind} software\footnote{\url{https://github.com/mpi-astronomy/snowblind}}) and ``wisps'' (stray-light artifacts; see \citealt{2023ApJ...952...74T, 2023ApJS..269...16R}), along with enhanced handling of saturated and defective pixels. These refinements are crucial, as some artifacts can mimic compact, red galaxies, potentially overlapping with our dropout selection criterion, though visual inspection often suffices for exclusion. NIRCam images were pixel-aligned to the HST data \citep{2011ApJS..197...36K, 2023zndo...7892935R}, and all HST images were converted to a pixel scale of 0.03''. The Full Width at Half Maximum (FWHM), Point Spread Function (PSF) and 5$\sigma$ depth of the relevant CEERS NIRCam and HST observations are reported in Table~\ref{tab:depths}. With respect to previous versions, CEERS DR v.1.0 offers improved astrometric alignment across multi-band images and superior background subtraction.

\begin{figure*}
    \centering
    \includegraphics[width=\linewidth]{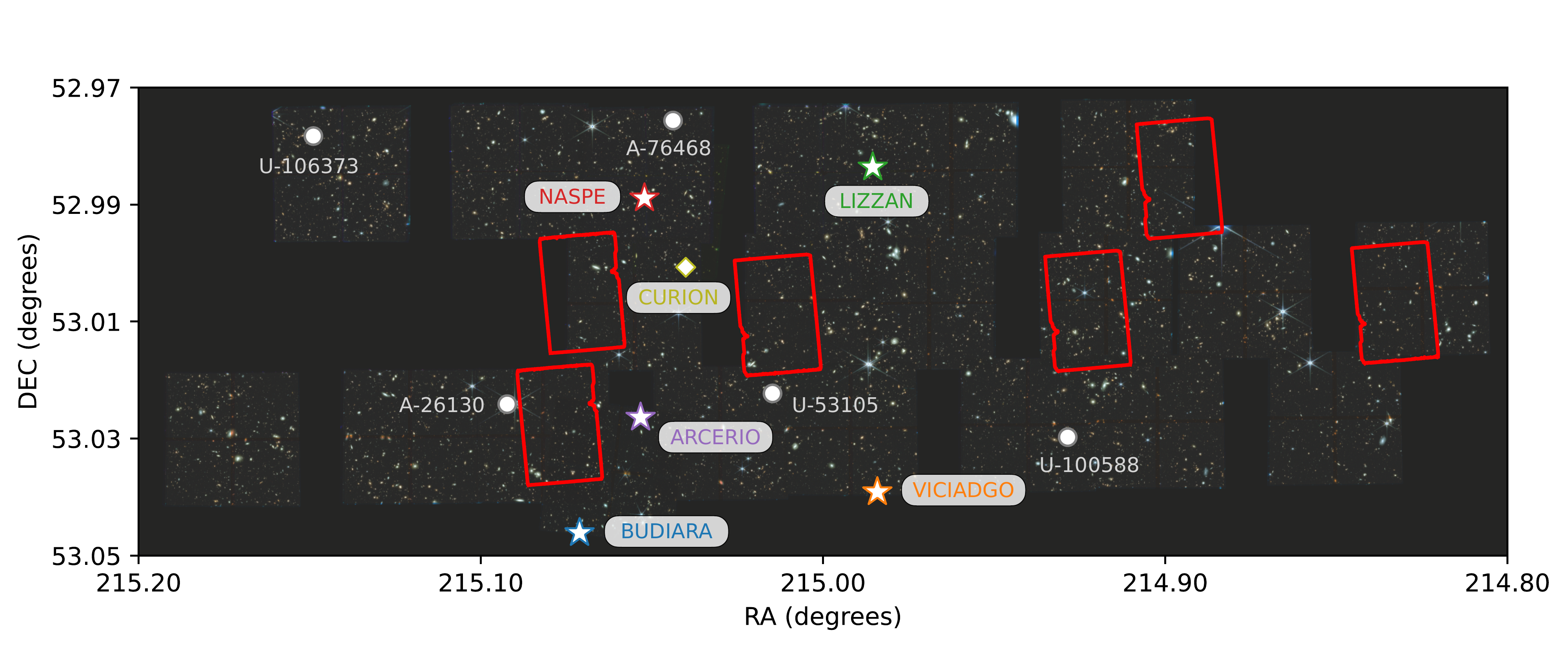}
    \caption{RGB mosaic of CEERS DR v.1.0 NIRCam data showing the position of our sample of dropouts, which is described in Section~\ref{4|sec:sampleselection}. Sources belonging to the Ultra High-Redshift (UHR) galaxy candidates sample are represented by colored star markers (see Section~\ref{uhrproperties}), while the strong line emitter CURION is depicted with a diamond marker (see Section~\ref{sec:curion}). All other F200W-dropouts are instead indicated by gray circles. The available CEERS DR 0.6 MIRI pointings overlapping with the EGS NIRCam-covered area are shown as red contours.}
    \label{fig:ceerscoverage}
\end{figure*}

CEERS JWST MIRI data are limited \citep{Yang2023}, with the available filters varying based on the pointing considered. In some cases, only two deep short-wavelength filters (F560W and F770W) are included, while in others, six broad-band filters spanning from 5.6 $\mu$m to 21 $\mu$m are available but with shallower depths. However, as we will discuss in Section~\ref{4|sec:sampleselection}, our sources are not covered by CEERS MIRI data (see also Figure~\ref{fig:ceerscoverage}).

\begin{table*}[t]
\centering
\caption{CEERS survey bands (plus F090W observations) and technical details.}
\begin{tabular}{cccccc}
\hline \hline Camera & Filter & $\lambda_\text{pivot} (\mu m)$ & FWHM & \begin{tabular}{c} 
PSF Enclosed \\
Flux $\left(\mathrm{d}=0.2^{\prime \prime}\right)$
\end{tabular} & \begin{tabular}{c} 
Point-Source Limiting \\
AB Magnitude $(5 \sigma)$
\end{tabular} \\
\hline
JWST/NIRCam SW & F090W & 0.9022 & $0.056^{\prime \prime}$ & 0.70 & 29.06 \\
JWST/NIRCam SW & F115W & 1.1543 & $0.066^{\prime \prime}$ & 0.80 & 29.16 \\
JWST/NIRCam SW & F150W & 1.6592 & $0.070^{\prime \prime}$ & 0.80 & 29.12 \\
JWST/NIRCam SW & F200W & 1.9886 & $0.077^{\prime \prime}$ & 0.76 & 29.32 \\
JWST/NIRCam LW & F277W & 2.7617 & $0.123^{\prime \prime}$ & 0.64 & 29.53 \\
JWST/NIRCam LW & F356W & 3.5684 & $0.142^{\prime \prime}$ & 0.58 & 29.40 \\
JWST/NIRCam LW & F410M & 4.0822 & $0.155^{\prime \prime}$ & 0.56 & 28.70 \\
JWST/NIRCam LW & F444W & 4.4043 & $0.161^{\prime \prime}$ & 0.52 & 29.03 \\
HST/ACS & F435W & 0.4329 & $0.112^{\prime \prime}$ & 0.66 & 28.72 \\
HST/ACS & F606W & 0.5922 & $0.118^{\prime \prime}$ & 0.70 & 28.77 \\
HST/ACS & F814W & 0.8046 & $0.124^{\prime \prime}$ & 0.63 & 28.49 \\
HST/WFC3 & F105W & 1.0550 & $0.235^{\prime \prime}$ & 0.35 & 27.55 \\
HST/WFC3 & F125W & 1.2486 & $0.244^{\prime \prime}$ & 0.33 & 27.67 \\
HST/WFC3 & F140W & 1.3923 & $0.247^{\prime \prime}$ & 0.32 & 26.99 \\
HST/WFC3 & F160W & 1.5370 & $0.254^{\prime \prime}$ & 0.30 & 27.68 \\
\hline
\end{tabular}
\label{tab:depths}
\tablefoot{The first and second columns list the available instruments and band name, while the third column $\lambda_\text{pivot}$ shows the pivotal wavelengths expressed in $\mu m$. The fourth column displays each filter's FWHM in units of arcseconds, whereas the fifth column contains the PSF enclosed flux in a 0.2'' diameter (d) aperture. Finally, we report in the last columns the $5\sigma$ depths associated with the CEERS catalog described in Section~\ref{3|sec:photometry}, calculated as medians of the 5$\sigma$ depths in each CEERS pointing for each band (see Table 1 of \citealt{2024ApJ...969L...2F} for a comprehensive overview). These values were estimated for a point source by utilizing the measured noise function in the field. This was done using an aperture with a 0.2'' diameter, with a correction applied to account for the total flux based on the enclosed flux within the PSF of each band.}
\end{table*}

\section{Photometry}\label{3|sec:photometry}

To ensure a comprehensive selection of F200W-dropout sources, we rely on two catalogs constructed from CEERS survey data. The first consists of an updated version of the v0.51 CEERS catalog, originally described in \cite{2024ApJ...969L...2F} (Steven Finkelstein, private communication). The second is the public CEERS ASTRODEEP-JWST catalog\footnote{\url{http://www.astrodeep.eu/astrodeep-jwst-catalogs/}} \citep{2024arXiv240900169M}. Using both catalogs is essential for our study, as each employs a different photometric extraction technique, meaning some faint high-redshift sources may be missed in one catalog but captured in the other (as discussed in Section \ref{4|sec:sampleselection}). Both catalogs are optimized for detecting objects (even very faint ones) in the long-wavelength NIRCam bands, a critical feature given our focus on identifying faint, very red objects. Appendix~\ref{A0|catalogs} provides an overview of the key steps involved in creating both catalogs, as well as the methods used for estimating fluxes and photometric uncertainties.

\section{Sample Selection}
\label{4|sec:sampleselection}
We aim to select F200W-dropout galaxies in both the updated CEERS and CEERS ASTRODEEP-JWST catalogs. This selection is specifically designed to include HELM galaxies as well as $z > 10$ galaxies without restricting the sample to only blue, relatively dust-free objects (see the discussion in Section~2.3 of \citetalias{2023A&A...676A..76B}), allowing the selection of UHR candidates with breaks falling between F200W and F277W (or even at longer wavelengths). We hence require our objects to have:

\begin{itemize}
    \item $\rm \text{S/N} \geq 3$ in NIRCam/F444W;
    \item $\rm \text{S/N} \leq 2$ in every NIRCam/HST filter equal or below 2 $\mu$m;
    \item $\rm \text{S/N} < 3$ in the coadded F090W+F115W+F150W+F200W image.
\end{itemize}

We enforced a S/N$<$2 for all HST and short NIRCam wavelength bands to ensure our selected objects are all F200W-dropouts, enforcing a non-detection in the coadded NIRCam short wavelength images to further ensure the dropout nature of our objects. We created the coadded NIRCam short wavelength by a weighted sum of the F090W, F115W, F150W and F200W mosaics pixel by pixel. The coadded image photometry was subsequently measured using the \texttt{photutils} Python package \citep{larry_bradley_2024_12585239}, employing circular apertures sized to match the average fiducial aperture for each source used in the updated CEERS catalog's and CEERS ASTRODEEP's JWST photometry. We applied a slightly more lenient threshold on the coadded images (S/N < 3), as the summing process includes some of the background noise. The S/N$\,\geq\,$3 requirement in F444W corresponds to a detection threshold of AB$\,\lesssim\,$29.6, while the S/N$\,\leq\,$2 non-detection criterion applied to all NIRCam/HST filters with $\lambda\leq2\,\mu$m implies approximate AB magnitude thresholds of F090W$\,\gtrsim\,$30.8, F115W$\,\gtrsim\,$30.9, F150W$\,\gtrsim\,$30.9, F200W$\,\gtrsim\,$31.1, F435W$\,\gtrsim\,$30.5, F606W$\,\gtrsim\,$30.5, F814W$\,\gtrsim\,$30.4, F105W$\,\gtrsim\,$29.5, F125W$\,\gtrsim\,$29.4, F140W$\,\gtrsim\,$29.4, and F160W$\,\gtrsim\,$29.4.

While our F200W-dropout selection shares similarities with the HELM galaxy selection of \citetalias{2023A&A...676A..76B}, a key difference lies in the underlying adopted catalogs and images in the analysis. The HELM galaxies of \citetalias{2023A&A...676A..76B} are identified in the v.0.51 CEERS Photometric Catalog \citep{2023ApJ...946L..13F}, which is based on a co-added F277W+F356W detection image. Our work, on the other hand, utilizes the updated CEERS and CEERS ASTRODEEP-JWST catalogs, which include sources extracted using the coadded F356W+F444W image as a detection image. This approach enables us to identify F444W-detected F200W-dropout sources that were not captured in previous CEERS catalogs. To focus on these novel objects, we exclude any sources appearing in the latest, publicly available release of the CEERS catalog (v0.51.4) within a 0.3'' search radius. This procedure grants that there are not overlap between our sample of sources and the updated CEERS HELM catalog from Bisigello et al. (2025; \textit{in prep.}), obtained via private communication, which is based on a co-added F277W+F356W detection image covering the total CEERS area.

Our selection criteria leave us with 636 (in the updated CEERS catalog) and 941 (in the CEERS ASTRODEEP-JWST catalog) sources. After a careful visual inspection to remove diffraction spikes, residual artifacts, noise peaks in the F444W band and objects with positions falling in the chip gap of all NIRCam short-wavelength observations, our final sample of F200W-dropouts includes 8 updated CEERS objects and 3 CEERS ASTRODEEP-JWST objects, totaling 11 objects. 5 of these 11 F200W-dropout objects lack coverage in either the F090W or F115W bands. We list the IDs, coordinates and names of our F200W-dropouts sample in Table~\ref{tab:samples} and their photometry in Table~\ref{tab:photometry}.

\subsection{Cross-match with other catalogs}
We further checked that our sample of sources has no overlap with the CEERS Little Red Dots (LRDs) sample found in \cite{2024arXiv240403576K} and \cite{2024arXiv240906772T}, nor the high-z sample presented in \cite{2024ApJ...969L...2F}\footnote{We also checked that none of our sources were contained in the object list removed from such high-z sample during visual inspection.}, nor the CEERS UHR sample presented in \cite{2023arXiv231115121Y} (highlighted via dropout selections; see their Section~5.1) and neither with the sample of HST-dark galaxies revealed by \cite{2023MNRAS.522..449B} in CEERS. We checked for potential spectroscopic counterparts of the sources in v0.5 of the CEERS NIRCam F356W Wide Field Spectroscopy (WFSS) catalog covering pointings 5, 7, 8 and 9 and CEERS DR 0.7 NIRSpec catalog, finding that our objects are not covered by spectroscopic observations (no objects were discarded during this step). Moreover, despite the few objects falling near the edges of the available CEERS MIRI pointings, no MIRI observations are available for our sources (with the only exception of ``CURION'', i.e., U-112842; as we will discuss in Section~\ref{sec:curion}). A few of our sources are also covered by (sub-)mm data. Source A-76468 falls near the low-sensitivity edge of a proprietary NOEMA 1.1 mm map (NOEMA program W20CK, PIs V. Buat and J. Zavala, obtained via private communication), and it is undetected at a 3 $\sigma$ level (with a 3$\sigma$ upper limit of 0.9 mJy). Sources VICIADGO (U-34120), LIZZAN (U-75985), NASPE (U-80918) and CURION (U-112842) are instead covered by 850 $\mu$m SCUBA-2 observations (described in \citealt{2017MNRAS.464.3369Z} and \citealt{2017MNRAS.465.1789G}), and all of these are undetected as well (with 3$\sigma$ upper limits of $\sim 3.4$ mJy/beam). We checked during our SED-fitting procedure that the inclusion of these (sub-)mm upper limits in the available photometry does not change our results in any way. To determine the physical properties of these objects we will therefore rely on broadband JWST/NIRCam, HST/ACS and HST/WFC3 photometry, with the aim of corroborating the present analysis with deep MIRI imaging, wider and deeper NOEMA observations and/or future spectroscopic data targeting our sample.

\begin{table}[t]
\centering
\caption{Our sample of F200W-dropouts.}
\setlength{\tabcolsep}{6pt}
\begin{tabular}{ccccc}
\hline \hline
ID & Name\footnotemark & RA & DEC\\
\hline
\vspace{2pt}
U-31863 & BUDIARA & 215.064009 & 52.882608\\
U-34120 & VICIADGO & 214.962236 & 52.827796\\
U-53105 & - &214.958983 & 52.867184\\
U-75985 & LIZZAN & 214.851223 & 52.886427\\
U-80918 & NASPE & 214.929089 & 52.928587\\
U-100588 & - & 214.887376 & 52.797809\\
U-106373 & - & 215.005197 & 53.008687\\
U-112842 & CURION & 214.940860 & 52.907705\\
A-22691 & ARCERIO & 215.006121 & 52.890428\\
A-26130 & - & 215.040845 & 52.920593\\
A-76468 & - & 214.893779 & 52.936404\\
\hline
\end{tabular}
\label{tab:samples}
\tablefoot{IDs starting with ``U-'' were selected in the updated CEERS catalog, whereas those starting with ``A-'' from the CEERS ASTRODEEP-JWST catalog.}
\end{table}
\footnotetext{Names were assigned only to objects in our UHR galaxy sample to distinguish them from other sources analyzed in this study. These names are derived from the ancient nomenclatures of old towns in the Tuscan-Aemilian Apennines in Italy.}

\begin{figure}
    \centering
    \includegraphics[width=\linewidth]{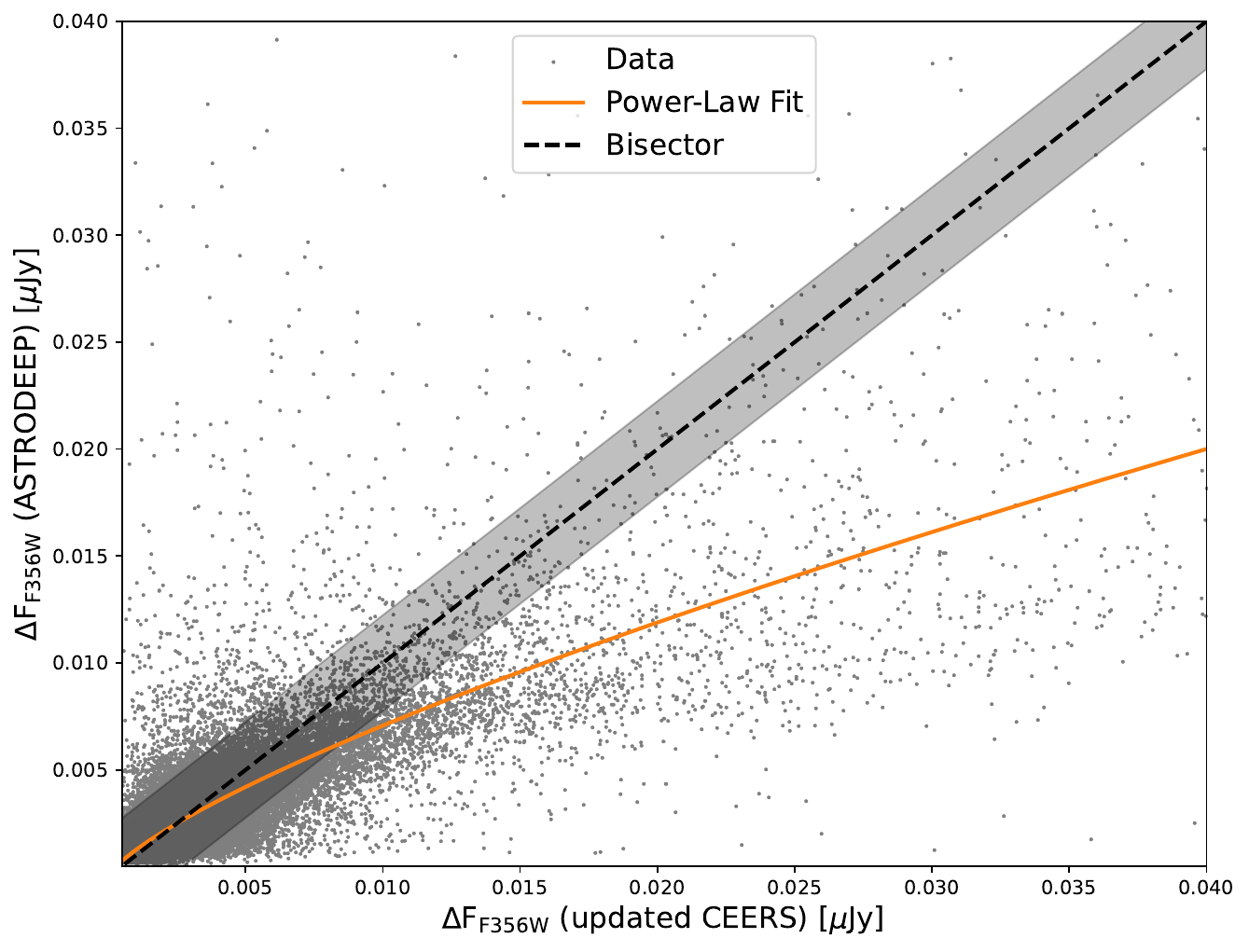}
    \caption{Updated CEERS catalog photometry errors vs CEERS ASTRODEEP-JWST photometry errors for the sources in common between the two catalogs (displayed as gray dots) for a sample band (F356W). The bisector of the plot is represented as a black dashed line surrounded by a 1$\sigma$ confidence interval (shaded black area) representing the typical dispersion of the data points around the bisector. A power-law fit is displayed to showcase how the updated CEERS catalog's errors tend to be higher with respect to CEERS ASTRODEEP-JWST ones.}
    \label{fig:errorscomparison}
\end{figure}

It is important to highlight that, since both the updated CEERS catalog and the CEERS ASTRODEEP-JWST catalog target the same field, a large overlap between them is expected, with the two catalogs having 63,034 objects in common (exploiting a search radius of 0.3''). As for our sample, we find that all three CEERS ASTRODEEP-JWST sources are also present in the updated CEERS catalog. This means that the updated CEERS catalog's photometry of these sources did not meet our selection criteria due to the different techniques exploited to compute photometry errors in the two catalogs (see Section~\ref{3|sec:photometry}), with the updated CEERS catalog's ones being higher on average. An example of this is illustrated in Figure~\ref{fig:errorscomparison}, which compares the updated CEERS catalog and CEERS ASTRODEEP-JWST photometric errors of common objects between the two catalogs in a reference band (F356W). The aim of this is to show how the updated CEERS catalog's photometric errors are systematically greater than CEERS ASTRODEEP-JWST ones, especially for higher error values. We will not delve into pinpointing the details behind these discrepancies in photometric uncertainty estimations, as it falls outside the scope of our paper.

Additionally, we confirmed that not all updated CEERS catalog sources in our sample are found in the CEERS ASTRODEEP-JWST catalog. This is the case for U-53105, which is absent from the CEERS ASTRODEEP-JWST catalog within a 0.5'' search radius from its updated CEERS catalog's position. This emphasizes the value of combining multiple catalogs with different source extraction methods to capture a broader array of sources.

\subsection{Comparison to red color-selected objects}
\begin{figure*}
    \centering
    \includegraphics[width=0.75\textwidth]{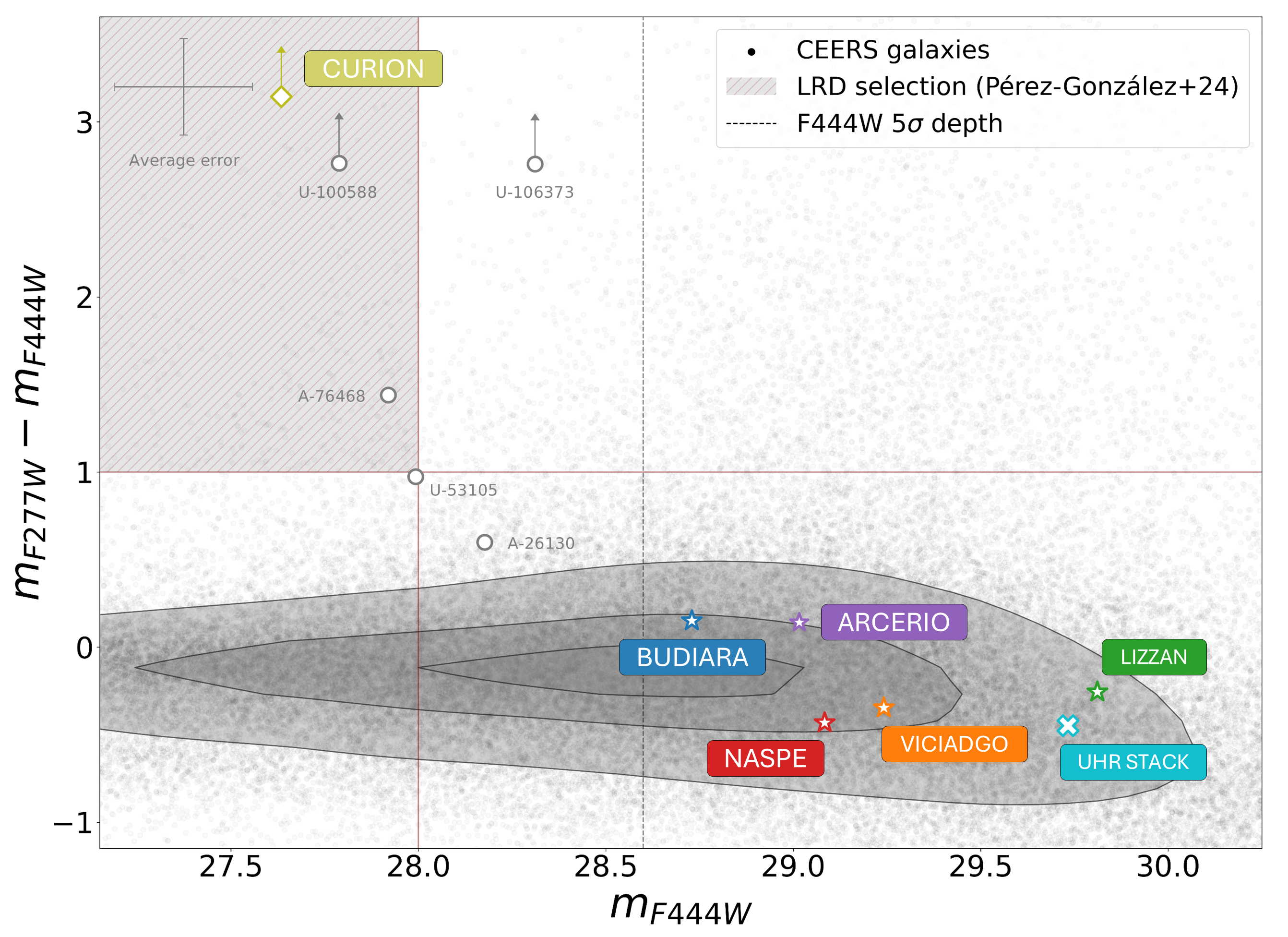}    
    \caption{[F277W - F444W] color-magnitude diagram for our F200W-dropouts. The F200W-dropout sample's objects are shown as colored stars (UHR candidates), pea green diamonds (CURION, a strong line emitter), cyan crosses (stacked UHR galaxies) or gray circles (all other objects in the sample). For CURION, U-100588 and U-106373 we report 1$\sigma$ upper limits. Black dots in the background represent the colors of CEERS galaxies, with gray-shaded contours enclosing 50\% (inner), 80\% (middle), and 95\% (outer) of the total CEERS source density. The F444W 5$\sigma$ depth is marked as a black dashed line, while the brown shaded area represents the color-magnitude space occupied by LRDs in \cite{2024ApJ...968....4P} (albeit such selection requires also F150W-F200W<0.5). Gray error bars represent the average errors for objects with colors that are not upper limits.}
    \label{fig:colormagplot1}
\end{figure*}

\begin{figure*}
    \centering
    \includegraphics[width=0.75\textwidth]{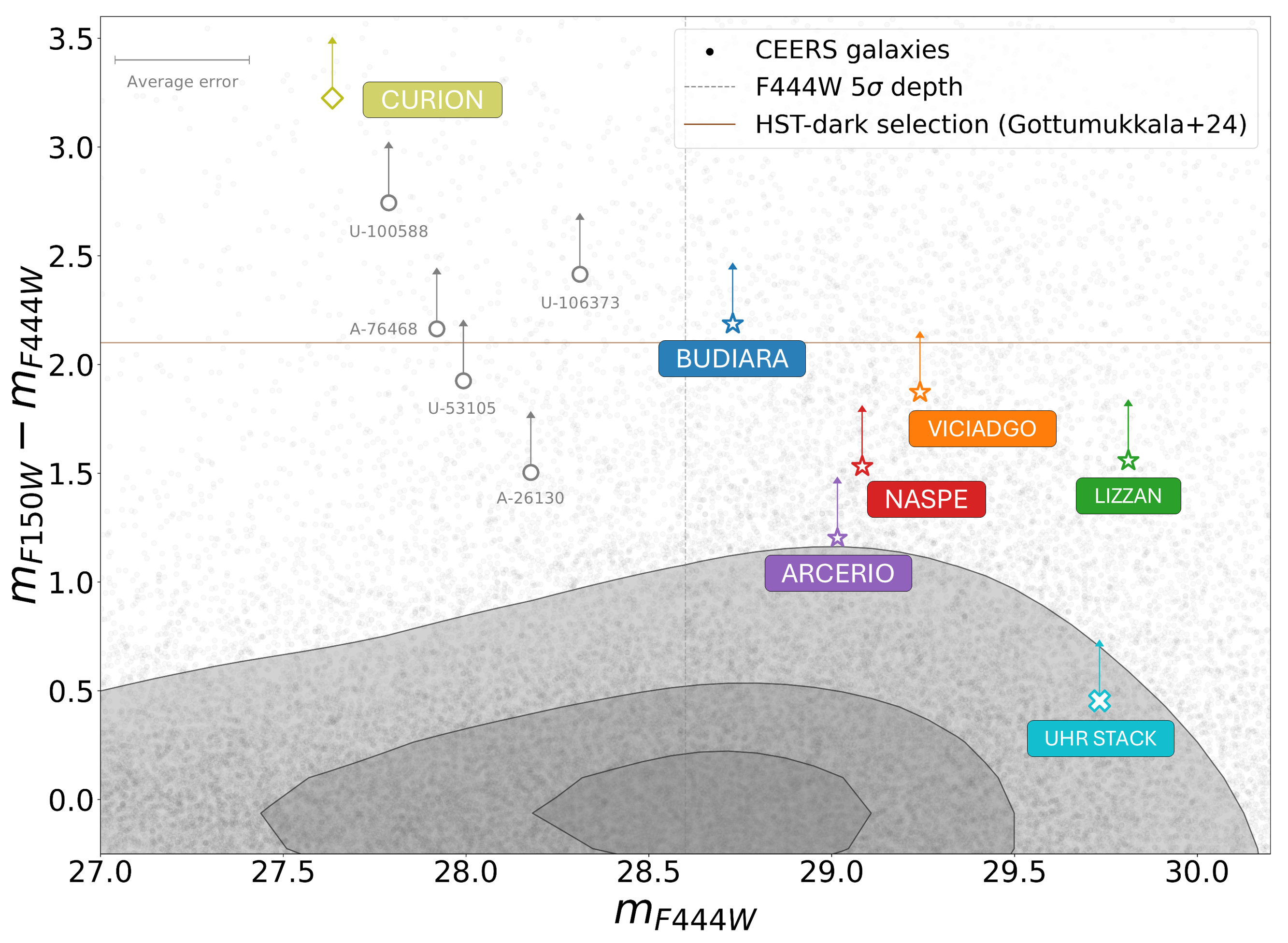}
    \caption{[F150W - F444W] color-magnitude diagram for our F200W-dropouts. The F200W-dropout representation scheme is the same adopted in Figure~\ref{fig:colormagplot1}, as well as the F444W 5$\sigma$ depth threshold and x-axis average errors. In addition, the HST-dark galaxy selection technique from \cite{2024MNRAS.530..966G} is represented by a brown continuous line.}
    \label{fig:colormagplot2}
\end{figure*}

\begin{figure*}
    \centering
    \includegraphics[width=0.75\textwidth]{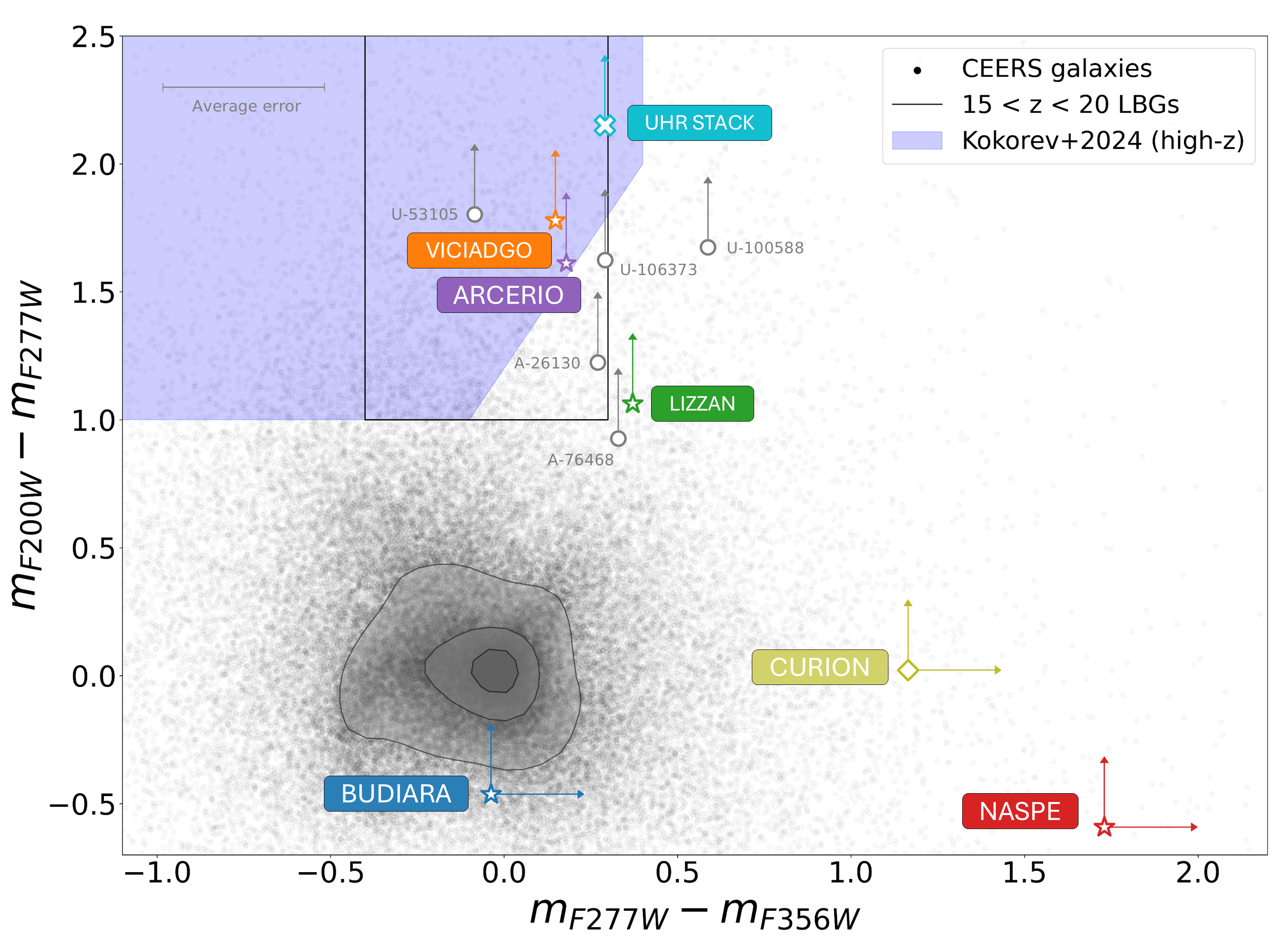}
    \caption{[F277W - F356W] versus [F200W - F277W] color-color diagram for our sources. The F200W-dropout representation scheme is the same adopted in Figure~\ref{fig:colormagplot1} and Figure~\ref{fig:colormagplot2}, as well as the x-axis average errors. The black rectangle highlights the UHR LBG selection for $15 < z < 20$ galaxies by \cite{Castellano2025}, whereas the blue shaded area corresponds to the $15 < z < 20$ LBG selection adopted in \cite{2024arXiv241113640K}.}
    \label{fig:colormagplot3}
\end{figure*}

This section summarizes the key color-magnitude/color-color properties of our F200W-dropout sample, subsumed in Figure~\ref{fig:colormagplot1}, Figure~\ref{fig:colormagplot2} and Figure~\ref{fig:colormagplot3}. Our sources are rather heterogeneously distributed in the [F277W - F444W] diagram (Figure~\ref{fig:colormagplot1}), with five of them lying within the 95\% CEERS source density contours, and only three of them with [F277W - F444W] > 2.5. Four sources (ARCERIO, NASPE, VICIADGO and LIZZAN) fall within the average errors below the 5 $\sigma$ depth limit in the F444W band, making their selection particularly uncertain. However, we will still attempt to characterize their physical properties to enhance our preliminary results with future, deeper observations targeting this field. Notably, four of our objects (CURION, U-53105, A-76468 and U-100588) are within 1$\sigma$ of the color-magnitude space utilized in \cite{2024ApJ...968....4P} to select LRDs in the SMILES/MIRI survey (i.e., [F277W - F444W] > 1 and F444W < 28, plus the additional requirement of F150W-F200W<0.5). Additionally, a fifth, object (U-106373) is at a 2$\sigma$-distance from the LRDs color-magnitude selection space. We will assess the possibility of these objects being LRDs in Section~\ref{5|sec:sedfitting} during our SED fitting runs.

The [F150W - F444W] color-magnitude diagram (Figure~\ref{fig:colormagplot2}) expresses rather well the extreme properties of our sources. In this color-magnitude space, all our sources lay outside the contour encompassing the 95\% of CEERS sources density adopting 1$\sigma$ upper limits, suggesting that they are outliers in terms of [F150W - F444W] colors with respect to the bulk of other CEERS sources (note that our objects are all undetected in the F150W band due to the adopted selection technique). Moreover, approximately half of our sources (CURION, BUDIARA, U-100588, U-106373, A-76468) exhibit colors falling above the selection threshold for HST-dark galaxies adopted in \cite{2024MNRAS.530..966G} (i.e., F150W - F444W > 2.1), further emphasizing the remarkable color properties of our objects.

Finally, a [F277W - F356W] versus [F200W - F277W] color-color diagram (Figure~\ref{fig:colormagplot3}) proves useful for evaluating which of our objects exhibit colors consistent with those expected for $15 < z < 20$ Lyman-break galaxies. Typically, $9 < z < 15$ Lyman-break galaxies are identified as red [F115W - F444W] objects with flat SED continua redward of the break (see, e.g., \citealt{2022ApJ...938L..15C}). For comparing our sources to the expected colors of $15 < z < 20$ Lyman-break galaxies, we employ the selection adopted in \cite{Castellano2025} to select such objects. This color-magnitude selection space is illustrated in Figure~\ref{fig:colormagplot3} as a black rectangle. We also reported as a blue shaded area the color-magnitude space corresponding to the selection adopted in \cite{2024arXiv241113640K} to select $15 < z < 20$ galaxies. Such area corresponds to requesting [F200W - F277W] > 1.0, [F200W - F277W] > 1.2 + 2.0[F277W - F356W], [F277W - F356W] < 0.5. Remarkably, five of our objects (VICIADGO, ARCERIO, U-53105, U-106373 and A-26130) are consistent within 1$\sigma$ (i.e., considering the average errors) with the selection of \cite{2024arXiv241113640K} and seven of them with the reddened selection exploited in \cite{2022ApJ...938L..15C}, indicating that their colors align with those of UHR galaxies in the $15 < z < 20$ range. Moreover, as reported in Figure~3 of \cite{2024MNRAS.529.1067H}, the F200W-F277W color of BDs in CEERS is clustered around 0 --- the color distribution of our objects would hence tend to exclude that they are BDs in the Milky Way. 

\section{SED-fitting procedure}
\label{5|sec:sedfitting}
To characterize the physical properties of our objects, we fitted their available photometry using different SED-fitting codes, and compared their results. First, we performed SED fits with \texttt{Bagpipes} (\citealt{2018MNRAS.480.4379C}), relying on the \cite{2003MNRAS.344.1000B} stellar population synthesis (SPS) model. For each object, we run separate SED-fitting runs exploiting all the available parametric Star Formation Histories (SFHs) in \texttt{Bagpipes} --- an exponentially declining SFH (\citealt{2017MNRAS.465..672M, 2018ApJ...855...85W, 2018MNRAS.479...25M}), a delayed exponentially declining SFH (\citealt{2017A&A...608A..41C, 2019MNRAS.483.2621C}), a log-normal SFH (\citealt{2017ApJ...839...26D, 2018MNRAS.478.2291C}), and a double power-law SFH (\citealt{2017A&A...608A..41C, 2018MNRAS.480.4379C}). As discussed in \cite{2019ApJ...873...44C}, some of these models may be biased against very high-z solutions due to their inherent inability to reproduce rising SFHs (especially the exponential and the delayed exponential models which trace declining SFHs; see e.g. \citealt{2012ApJ...754...25R, 2022ApJ...929....1H}). However, we are interested in investigating the impact of different SFHs on the final results of our SED-fitting runs, and we will discuss this matter further in Section~\ref{6|sec:results}. Note, however, that the stellar mass depends marginally on the shape of the parametric form of the adopted SFH, and we expect the mass estimates between different SFHs to be broadly consistent (see, e.g., \citealt{2015ApJ...801...97S, 2024A&A...686A.128C}). We also performed our \texttt{Bagpipes} runs testing two different dust attenuation laws: a Small Magellanic Cloud (SMC) dust attenuation law (\citealt{1984A&A...132..389P, 1985A&A...149..330B}) and a standard Calzetti law (\citealt{2000ApJ...533..682C}) to compare the results. To enhance sensitivity to strong line emitter solutions, which may occupy a narrow region in the probability distribution of galaxies' physical parameters, we increase the number of \texttt{Bagpipes}'s live points (i.e., the walkers used in the program's Markov Chain Monte Carlo sampling exploited by \texttt{MultiNest} algorithm; \citealt{2014A&A...564A.125B, 2019OJAp....2E..10F}) from the default 400 to 2000 (A. Carnall, private communication), ensuring a robust exploration of this parameter space. Additionally, we include a nebular emission component in all \texttt{Bagpipes} model fits (modeled through the \texttt{CLOUDY} photoionization code; \citealt{2017RMxAA..53..385F}), allowing the ionization parameter to reach $\log \text{U} = -1$. A detailed list of the model fit parameters and their prior ranges is provided in Table~\ref{tab:priors}. To account for systematic errors, we add 5\% of the flux estimates in quadrature to the photometric uncertainties for each run.

Since \texttt{Bagpipes} does not natively account for an AGN component, we also fitted our objects with \texttt{CIGALE} \citep{2019A&A...622A.103B}, which instead allows to model an AGN contribution to the galaxies' overall emission. Since the aim of these \texttt{CIGALE} runs is to reveal potential lower-z AGN contaminants, we will not adopt a wide set of SFHs as done for our \texttt{Bagpipes} runs, and we limit ourselves to only explore the effect of a double exponential SFH. Furthermore, our \texttt{CIGALE} SED-fitting runs rely on the single stellar populations library of \cite{2003MNRAS.344.1000B}, including a nebular emission component with \texttt{CLOUDY} and a modified \cite{2000ApJ...539..718C} dust attenuation law. Finally, we include in the \texttt{CIGALE} model fit template AGN emission models by \cite{2006MNRAS.366..767F}, allowing AGN fractions, which correspond to the AGN-heated dust contribution to the total infrared luminosity (integrated over 1-1000 $\mu$m), to vary between 0\% and 75\%. A complete list of the free parameters of our \texttt{CIGALE} fit and the related value grids are available in Table~\ref{tab:cigalegrid}. For each run, we accounted for systematic errors by adding 5\% of the flux estimates in quadrature to the photometric uncertainties.

Finally, we carried out SED fitting using the Python implementation of \texttt{EAZY}\footnote{\url{https://eazy-py.readthedocs.io}} (\citealt{2008ApJ...686.1503B}), employing the following broad set of SED-fitting templates:

\begin{itemize}
    \item The \texttt{EAZY} template set from \cite{2008ApJ...686.1503B}, based on the input set of \cite{2006A&A...449..951G} following the \cite{2007AJ....133..734B} algorithm plus a template accounting for dusty galaxies;
    \item Default \texttt{fsps} templates based on flexible SPS models (specifically, we adopted the \texttt{tweak\textunderscore fsps
    \textunderscore QSF\textunderscore 12 \textunderscore v3.param} template);
    \item \texttt{sfhz} template set, which incorporates redshift-dependent SFHs, excluding those that begin earlier than the Universe's age at any given epoch, along with a realistic emission-line model at $z \sim 8$ by \cite{2023MNRAS.518L..45C} (specifically, we tested for every object the \texttt{agn\textunderscore blue\textunderscore sfhz\textunderscore 13.param}, \texttt{blue\textunderscore sfhz\textunderscore 13.param}, \texttt{carnall\textunderscore sfhz\textunderscore 13.param} and \texttt{corr\textunderscore sfhz\textunderscore 13.param} templates);
    \item High-Redshift Optimized Templates\footnote{\url{https://github.com/e-m-garcia/hot-templates}} (\texttt{hot}; \citealt{2023ApJ...951L..40S}), a custom set of two synthetic templates optimized to reproduce the expected astrophysics of galaxies at high redshifts. These templates assume hotter gas temperatures of 45K for redshifts $z = 8-12$ and 60K for $z > 12$ in star-forming regions, resulting in bottom-lighter or top-heavier IMFs.
\end{itemize}

Our \texttt{EAZY} fits were conducted between z=0.01 and z=25 with a step of 0.01.

\section{Results}\label{6|sec:results}

\begin{table*}[t]
    \centering
    \small
    \caption{UHR candidates sample's best-fit properties and integrated high-redshift probabilities from \texttt{Bagpipes} and fully open priors reported in Table~\ref{tab:priors}.}
    \begin{tabular}{ll|ccc|ccc|c}
        \hline\hline
        ID & Name & $z_\text{high}$ & $\log M_\text{high}$ & $A_{V,\text{high}}$ & $z_\text{low}$ & $\log M_\text{low}$ & $A_{V,\text{low}}$ & $\int P(z > 8)\,dz$ \\
        \hline
        U-31863 & BUDIARA  & $17.76^{+1.6}_{-1.5}$   & $8.12^{+0.5}_{-0.3}$  & $0.17^{+0.13}_{-0.10}$ & $1.01^{+0.32}_{-0.19}$ & $7.68^{+0.3}_{-0.3}$  & $4.60^{+1.0}_{-1.8}$ & 0.87 \\
        U-34120 & VICIADGO & $17.56^{+1.4}_{-1.3}$   & $7.91^{+0.5}_{-0.3}$  & $0.09^{+0.12}_{-0.07}$ & $0.91^{+0.26}_{-0.15}$ & $7.49^{+0.2}_{-0.3}$  & $4.47^{+1.1}_{-1.6}$ & 0.88 \\
        U-75985 & LIZZAN   & $17.80^{+1.48}_{-1.45}$ & $7.63^{+0.42}_{-0.28}$& $0.08^{+0.11}_{-0.06}$ & $4.31^{+0.34}_{-3.49}$ & $7.23^{+0.27}_{-0.24}$& $2.04^{+2.65}_{-0.67}$ & 0.62 \\
        U-80918 & NASPE    & $17.55^{+1.47}_{-1.30}$ & $8.00^{+0.43}_{-0.29}$& $0.07^{+0.10}_{-0.05}$ & $0.89^{+0.4}_{-0.15}$  & $7.45^{+0.42}_{-0.27}$& $4.00^{+1.26}_{-2.22}$ & 0.80 \\
        A-22691 & ARCERIO  & $17.32^{+2.05}_{-2.98}$ & $8.19^{+0.47}_{-0.42}$& $0.27^{+0.34}_{-0.18}$ & $1.28^{+1.54}_{-0.37}$ & $7.74^{+0.49}_{-0.37}$& $4.01^{+1.39}_{-2.07}$ & 0.98 \\
        \hline
    \end{tabular}
        \tablefoot{For each object, we consider all \texttt{Bagpipes} runs performed over the full redshift prior range ($0 \leq z \leq 25$) and for all combinations of star-formation histories and dust attenuation laws explored. Given the typically bimodal shape of the resulting $P(z)$ distributions, we identify the best-fit solutions at $z \leq 8$ (“low”) and $z > 8$ (“high”) as those yielding the lowest $\chi^2$ value within each redshift regime. The reported best-fit parameters (redshift $z$, stellar mass $\log M/\mathrm{M}_\odot$, and dust attenuation $A_V$) are those corresponding to these two solutions, while their uncertainties are derived as the 16th--84th percentile range around the best-fit values of each physical quantities. The integrated probability $\int P(z>8),dz$ is instead computed over the full prior range ($0 \leq z \leq 25$) and considering the combination of SFH and dust law that minimizes $\chi^2$ at $z>8$. The full list of physical parameters is reported in Table~\ref{tab:bestfitsingle}.\label{summary}}
\end{table*}

\begin{table*}[t]
    \centering
    \caption{Best-fit \texttt{Bagpipes} parameters for the remaining F200W-dropouts.}
    \small
    \begin{tabular}{ll|ccc}
        \hline\hline
        ID & Name & $z$ & $\log M/\text{M}_\odot$ & $A_V$ \\
        \hline
        U-53105   & —       & $2.60^{+2.70}_{-1.08}$   & $8.31^{+0.64}_{-0.53}$  & $2.94^{+1.78}_{-1.47}$ \\
        U-112842  & CURION  & $5.53^{+0.02}_{-0.04}$   & $7.21^{+0.11}_{-0.12}$  & $1.01^{+0.29}_{-0.28}$ \\
        A-26130   & —       & $2.04^{+2.77}_{-0.76}$   & $8.78^{+0.62}_{-0.46}$  & $3.25^{+1.78}_{-1.60}$ \\
        A-76468   & —       & $3.63^{+4.98}_{-1.06}$   & $8.94^{+0.67}_{-0.53}$  & $3.73^{+1.42}_{-1.83}$ \\
        \hline
    \end{tabular}
    \tablefoot{Best-fit redshifts, stellar masses, and dust attenuations inferred from \texttt{Bagpipes} SED fits for the four additional F200W-dropouts in our sample. All fits adopt the best-fitting SFH type reported in Table~\ref{tab:bestfitother}.}
    \label{tab:bestfitother_compact}
\end{table*}

\subsection{SED-fitting results}

Our \texttt{Bagpipes} analysis highlights five sources in our F200W-dropout sample with a significant P(z) volume at ultra-high redshifts ($z > 15$; see Table~\ref{summary}): BUDIARA (U-31863), VICIADGO (U-34120), LIZZAN (U-75985), NASPE (U-80918) and ARCERIO (A-22691). For every SFH and dust extinction model assumed, these objects always showcase bi-modal P(z)s, featuring one HELM-like solution at $z < 8$ (i.e., a dusty, low-mass galaxy solution) and a UHR solution with negligible dust, meaning that the available JWST/NIRCam, HST/ACS and HST/WFC3 photometry alone cannot describe exact results of their nature. Due to the bi-modal nature of their P(z)s, we did not simply adopt the best-fit values output by \texttt{Bagpipes} as the definitive properties of our objects. This is because these values would reflect the physical characteristics of a median solution between the low- and high-redshift peaks of the bimodal P(z) distributions. Instead, we divided each object’s P(z) into two separate components: one for $z \leq 8$ and another for $z > 8$. For each of these redshift regimes, we determined the best-fit solutions maximizing their likelihood. This method allows us to rigorously account for both possible interpretations of each object’s nature. Consequently, we report two distinct sets of best-fit physical parameters for each source: one representing the low-redshift, $z \leq$ 8 solution and the other the high-redshift, $z > 8$ solution. The likelihood maximization for the $z\leq$ 8 and the $z > 8$ solutions draws from the different assumed SFHs, which means that the solutions at $z \leq$ 8 and $z > 8$ for the same object could be associated with two different SFHs. We report our fiducial \texttt{Bagpipes} best fit SEDs and normalized $z \leq$ 8 and $z > 8$ P(z)s obtained following this procedure in Figure~\ref{fig:singlefits}, with the inferred physical parameters for both solutions being listed in Appendix~\ref{uhrbestfitresults}.

\begin{figure*}[h!]
    \centering
    \includegraphics[width=0.86\textwidth]{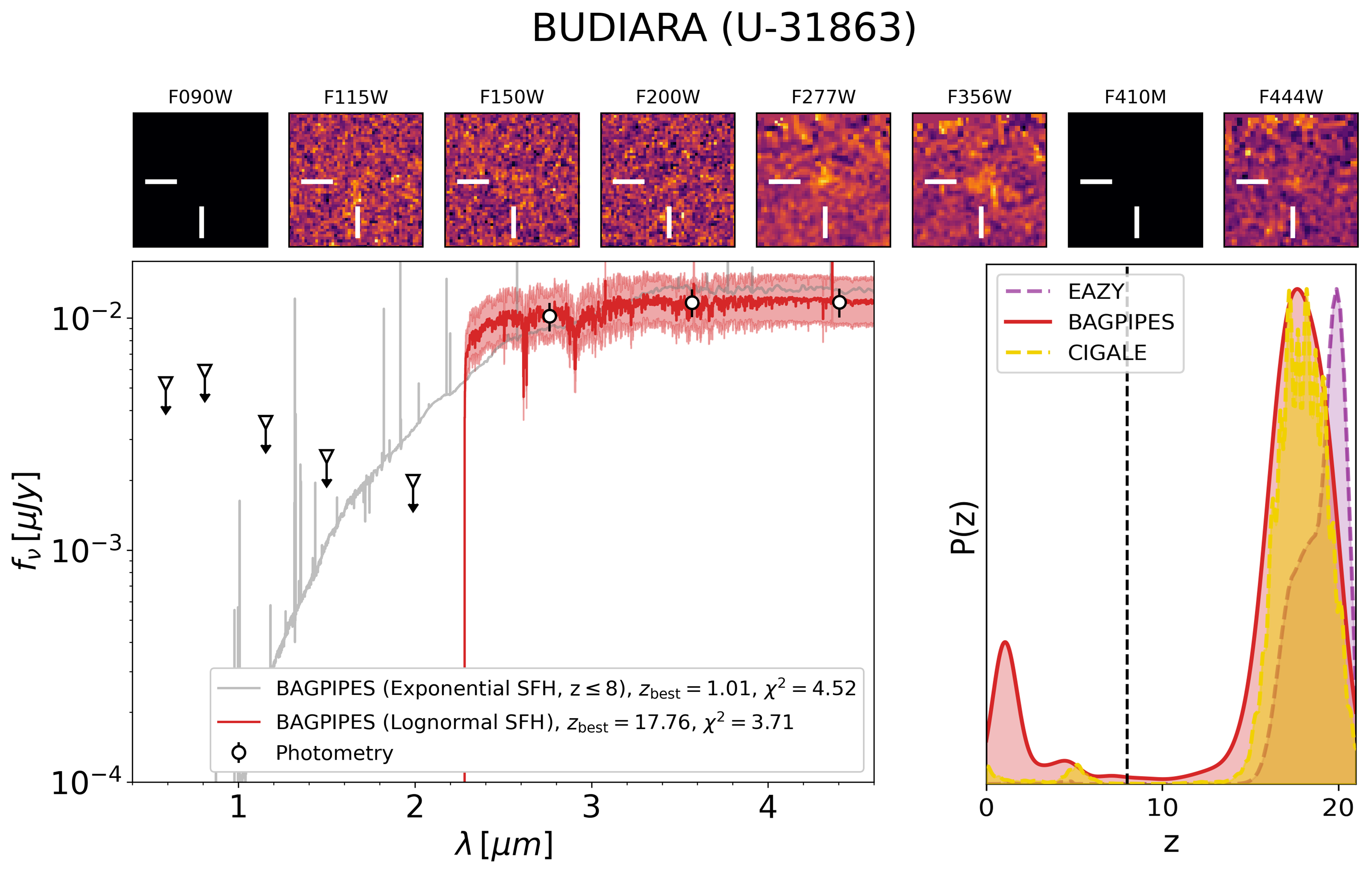}
    \includegraphics[width=0.86\textwidth]{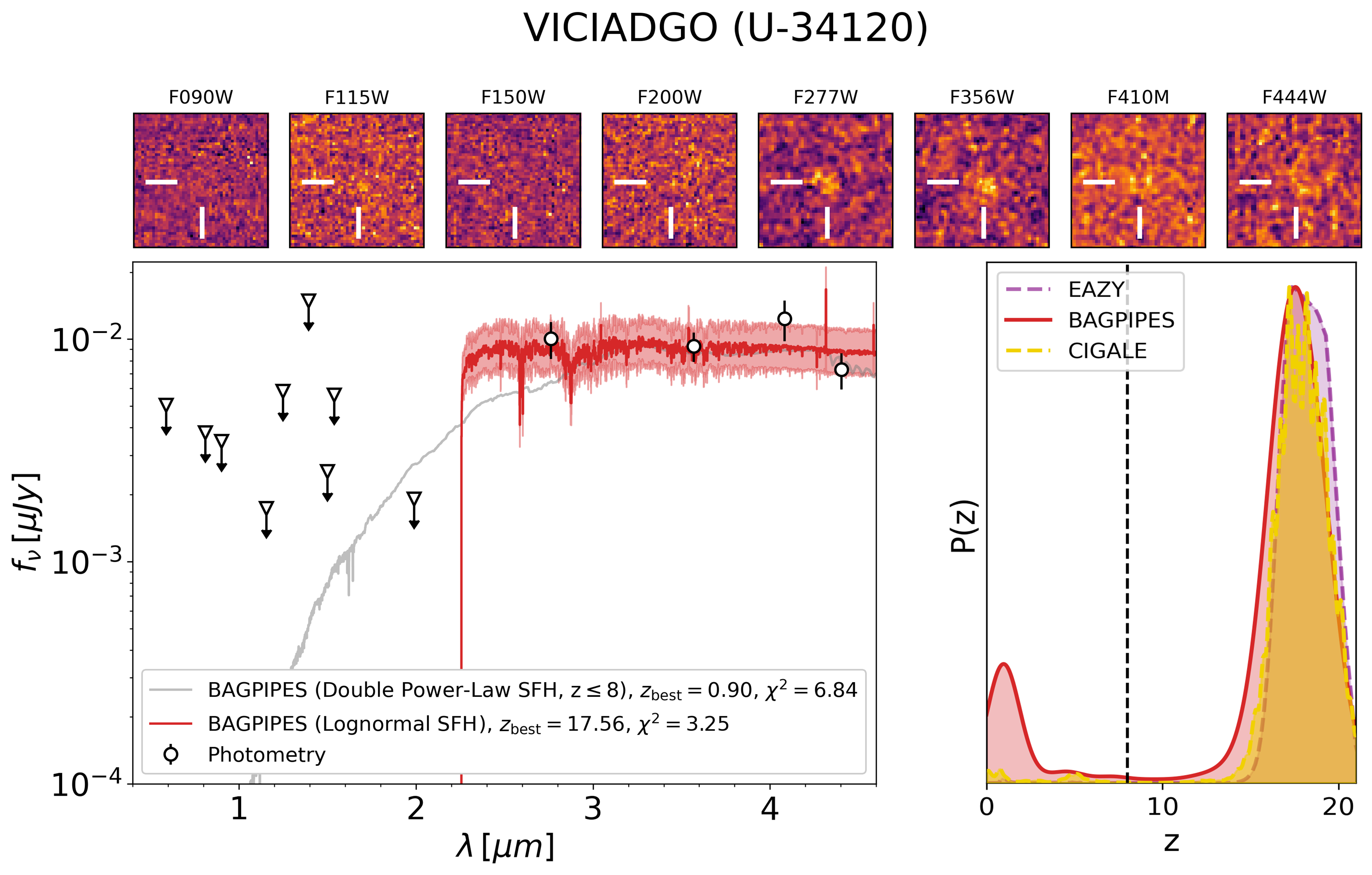}

    \caption{Best-fit SEDs and redshift probability distributions $P(z)$ obtained with the procedure described in Section~\ref{5|sec:sedfitting} for the UHR galaxy candidates and their median stack. The top section of each panel shows $1.5'' \times 1.5''$ NIRCam cutouts of the source. The left inset displays the observed photometry (black circles) with $1\sigma$ upper-limit uncertainties (black triangles). The \texttt{Bagpipes} best-fit SED at $z>8$ is shown in red, with its 0.1 dex confidence interval (red shaded area), while the best-fit SED at $z\leq8$ --- obtained independently over the same full prior range --- is shown in gray. The right inset reports the normalized redshift probability distributions from different SED-fitting codes: the $P(z)$ of the \texttt{Bagpipes} model that minimizes $\chi^2$ at $z>8$, shown in red over the full prior range ($0 \leq z \leq 25$); the \texttt{CIGALE} posterior (yellow dashed line); and the mean \texttt{EAZY} posterior averaged over all templates described in Section~\ref{5|sec:sedfitting} (purple). The vertical dashed black line marks the $z=8$ threshold.}
    \label{fig:singlefits}
\end{figure*}

\begin{figure*}[h!]
    \ContinuedFloat
    \centering
    \includegraphics[width=\textwidth]{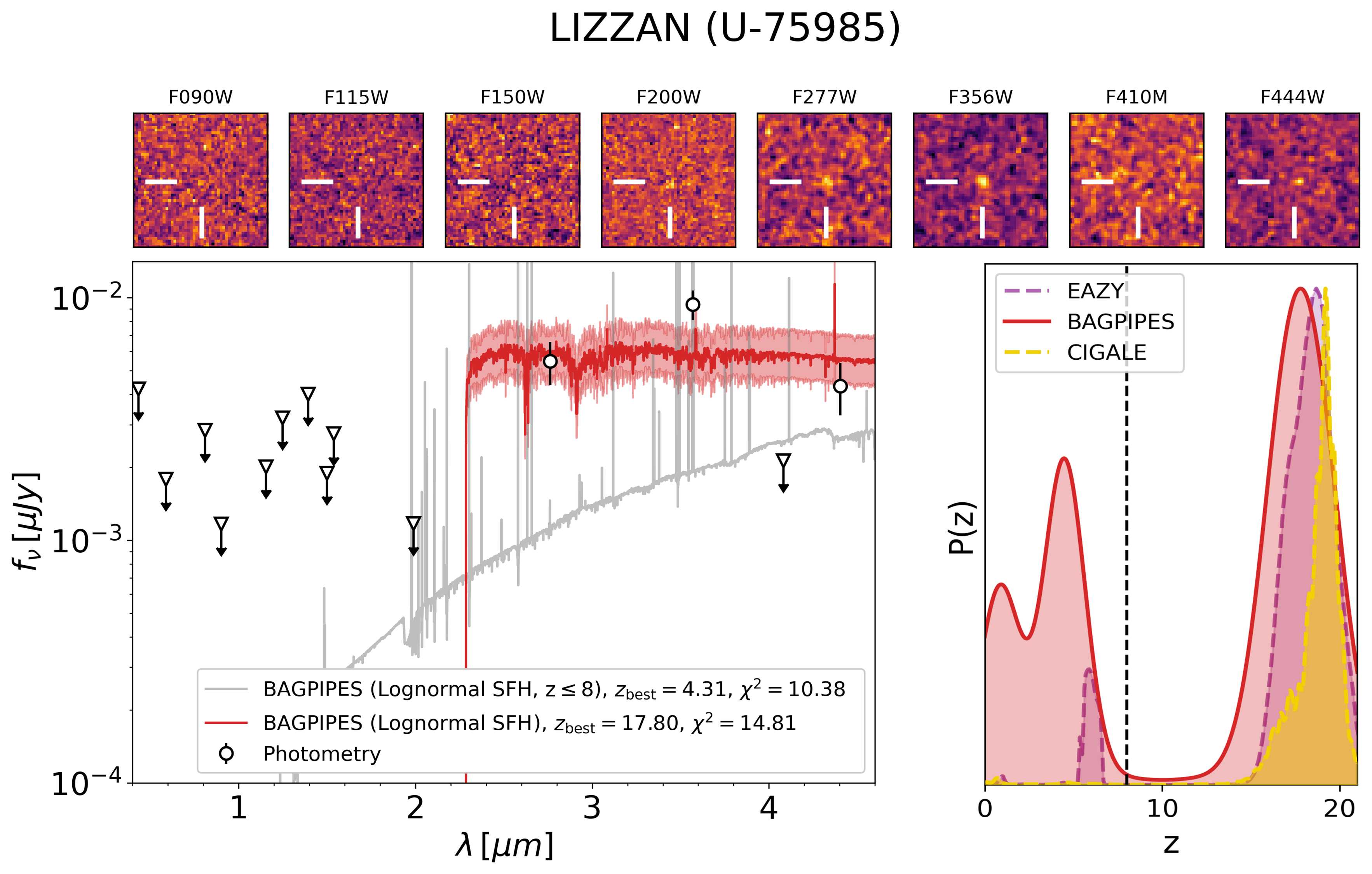}
    \includegraphics[width=\textwidth]{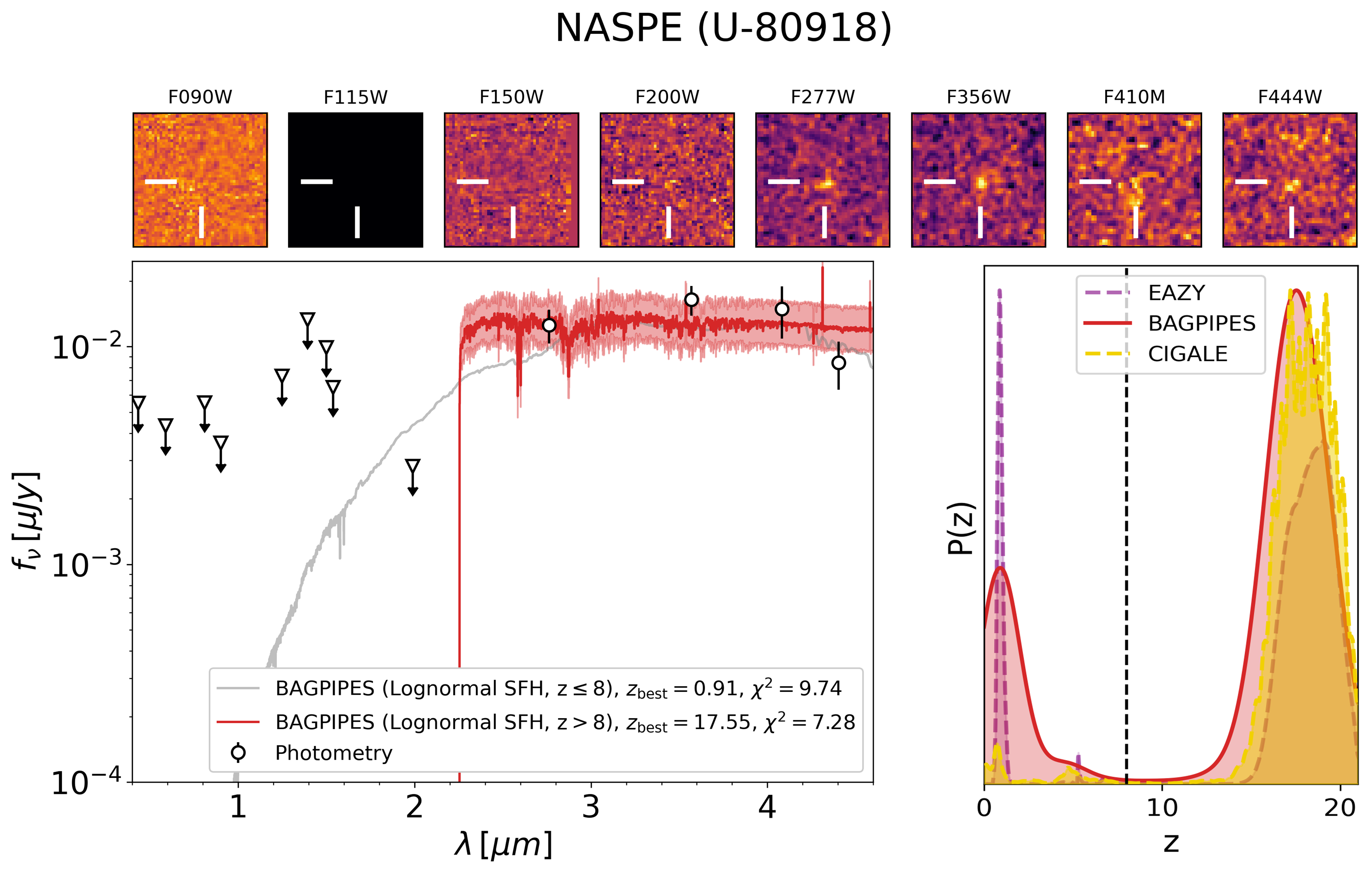}
    \caption{Continued.}
\end{figure*}

\begin{figure*}[h!]
    \ContinuedFloat
    \centering
    \includegraphics[width=0.9\textwidth]{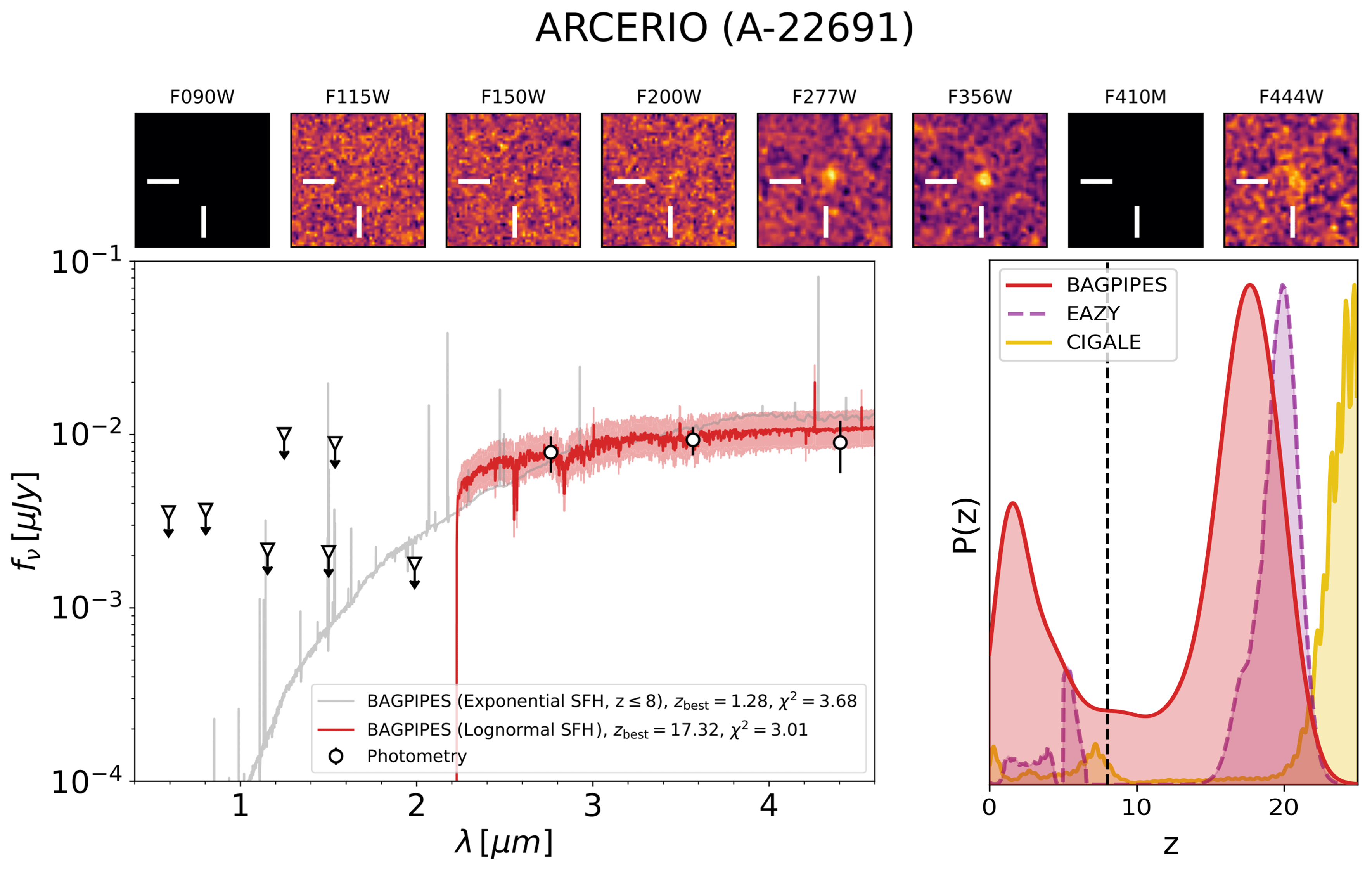}
    \caption{Continued.}
\end{figure*}

\begin{figure*}[h!]
    \centering
    \includegraphics[width=0.9\textwidth]{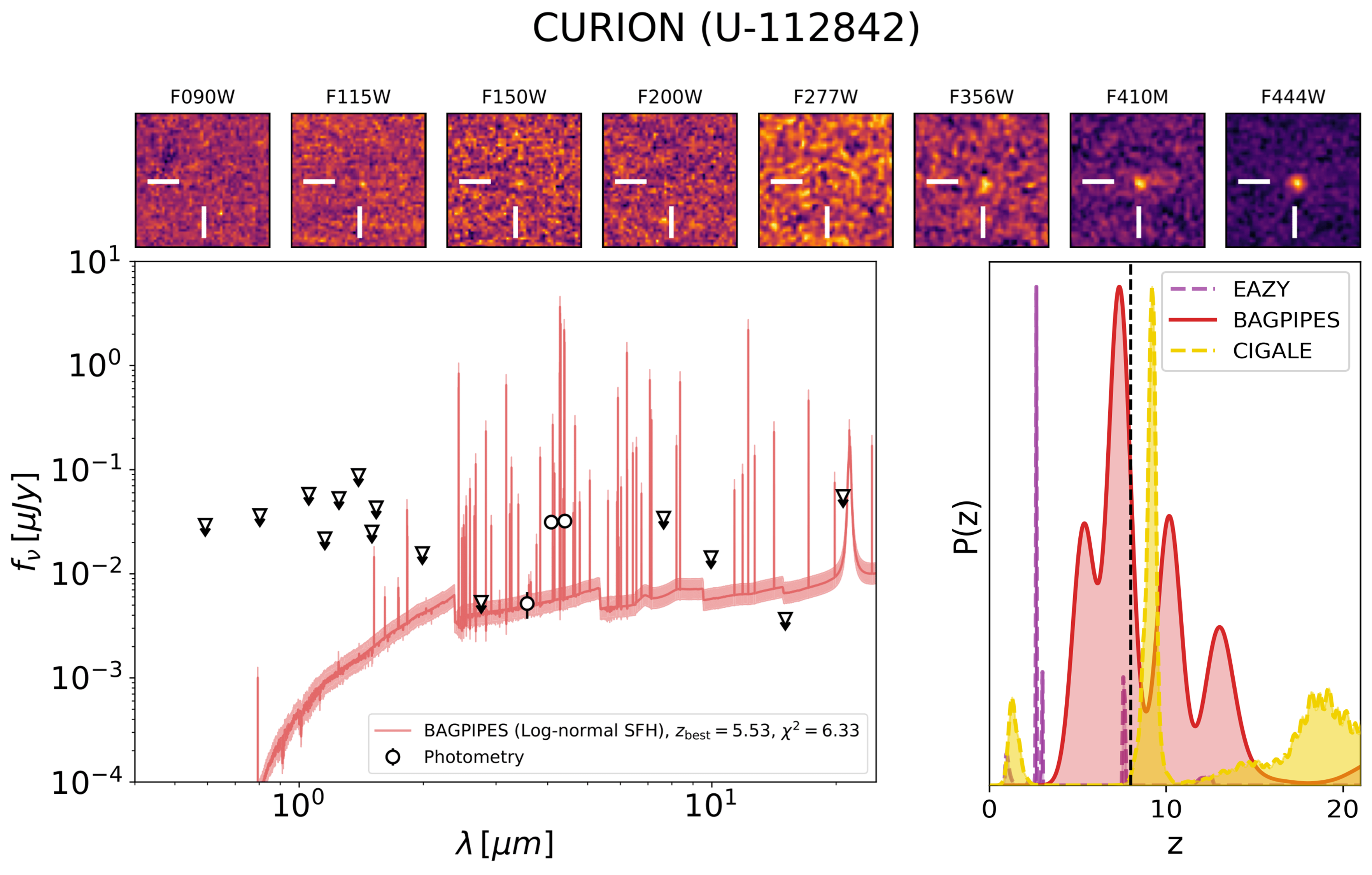}
    \caption{Best-fit SEDs, P(z)s and NIRCam cutouts for CURION. The plot's color scheme is the same followed in Figure~\ref{fig:singlefits}.}
    \label{fig:curion}
\end{figure*}

Another highlight from our analysis is CURION (U-112842). For this particular object we were able to include in our analysis data from the Cycle 2 Proposal MEGA Mass Assembly at Cosmic Noon: MIRI EGS Galaxy and AGN Survey (\cite{2023jwst.prop.3794K}; P.I. A. Kirkpatrick, obtained via private communication). These MIRI observations target the EGS field, covering ERS NIRCam observations in the F770W, F1000W, F1500W, and F2100W bands. The best-fit SED produced by our \texttt{Bagpipes} runs including MIRI data favors a log-normal SFH, with a best-fit redshift of $z \sim$ 5.4, an intermediate dust content ($\text{A}_\text{v} \sim 0.87$), a low mass ($\log \text{M} / \text{M}_\odot \sim 7$) and extreme line emission, with $\log \text{U} \sim -3.9$, approaching the lower limits of the priors assigned to this parameter in the fits. All the best-fit parameters inferred by these SED-fitting runs are listed in Appendix~\ref{appendix:otherdropouts}. As shown in Figure~\ref{fig:mainsequence}, CURION is consistent within 1$\sigma$ with the main sequence of star-forming galaxies. Fits yielded by \texttt{EAZY} and \texttt{CIGALE} do not hint at the presence of an AGN, favoring dusty solutions at $z < 10$, yielding, however, best-fit chi-squared values up to 7 or 8 times the ones returned by \texttt{Bagpipes} in the $z \sim$ 5.5 strong line emitter scenario. Finally, since CURION is unresolved in both the F356W and F444W bands (see Appendix~\ref{sizes}), we explored the possibility of it being a sub-stellar Milky Way object such as a BD. To do so, we fitted its photometry with L and T dwarf models from \cite{2006ApJ...640.1063B}, spanning different metallicities ($-0.5\leq \text{[Fe/H]}\leq 0.5)$, effective temperatures (700 K $\leq \text{T} \leq$ 2300 K) and surface gravity accelerations ($10^{4.5} \text{cm} \ \text{s}^{-2} \leq \text{g} \leq 10^{5.5} \text{cm} \ \text{s}^{-2}$). Our BD template fits return a best-fit chi-squared value of $\chi^2_{\rm BD}\sim20.96$, up to 6 times higher than the \texttt{Bagpipes} one.

\texttt{Bagpipes} fits for the source U-53105 yield a primary low-redshift peak at $z\sim$2.6 (adopting an exponential SFH, which yields the best chi-squared value for this object), with a low-probability P(z) up to $z\sim$15, while a compatible main redshift solution is obtained by varying the adopted SFH. A log-normal SFH maintains the primary peak at $z < 3$, while raising the P(z) volume predicted at $z > 5$, with two strong secondary peaks at $z\sim$9 and $z\sim$15. However, \texttt{EAZY} indicates a $z < 5$ solution for this object, even if some templates assign a strong probability beyond $z > 20$, with the \texttt{blue\textunderscore sfhz}, \texttt{carnall \textunderscore sfhz}, \texttt{corr \textunderscore sfhz}, \texttt{hot \textunderscore 45k} and \texttt{hot \textunderscore 60k} all yielding $\int_{z>15} \mathrm{d}z\, P(z) \geq 0.95$. A similar trend is followed by \texttt{CIGALE}'s P(z), showcasing a marked bi-modal distribution, with principal P(z) peaks falling at $z<8$ and $z>15$, not indicating the presence of an AGN. However, we note that the high solution volume at $z > 20$ predicted by \texttt{EAZY} and \texttt{CIGALE} is not reproduced by our \texttt{Bagpipes} runs even when adopting a log-normal SFH. Furthermore, U-53105 appears resolved in all NIRCam long-wavelength bands (see Appendix~\ref{sizes}), ruling out the possibility of it being a BD.

Source A-26130's \texttt{Bagpipes} fits adopting delayed, exponential, and double power-law SFHs suggest a best-fit $\log \text{M}/\text{M}_\odot \sim$ 8.78, with fairly high dust-obscuration ($\text{A}_\text{v} \sim$ 3), located at $z\sim$2. While these fits also reveal a high-redshift tail, the solution volume for such scenarios remains minimal. In contrast, a log-normal SFH fit implies a slightly higher mass $\log \text{M}/\text{M}_\odot \sim$8.7, retaining a primary peak at $z=2$ with the addition of a secondary $15 < z < 20$ peak. The \texttt{EAZY} P(z) further supports the low-redshift solution but also assigns a primary peak at $z\sim$20 for the majority of the adopted templates, backed up by \texttt{CIGALE}, which does not reveal a significant contribution by AGN emission. The source is resolved in the F277W band (see Appendix~\ref{sizes}), and we can exclude it being a BD.

The analysis of source A-76468 with \texttt{Bagpipes} indicates a primary peak at $z = 3$, with a stellar mass of $\log \text{M}/\text{M}\odot \sim 9$ and a dust attenuation of $\text{A}\text{v} \sim 4$. These results are consistent across all tested SFHs. When a log-normal SFH is adopted, the fits also reveal a more extended high-redshift tail, along with secondary peaks at $z = 8$ and $z = 15$, reproduced by our \texttt{CIGALE} runs. The \texttt{EAZY} $P(z)$ complements these findings by displaying a narrow primary peak at $z\sim$ 7 and a secondary peak at $z = 3$. This source is resolved in both the F277W and F356W bands, excluding the possibility of it being a BD. We report the best-fit \texttt{Bagpipes} SEDs of U-53105, A-26130 and A-76468 in Figure~\ref{fig:otherf200wdropouts}.

Unfortunately, source U-106373 yields a rather unconstrained P(z)s, and we refrain from reporting its best-fit physical parameters. \texttt{Bagpipes} fits of its photometry, using delayed, exponential, and double power-law SFHs, suggest that this is a relatively massive (9 < $\log \text{M}/\text{M}_\odot$ < 10) and dusty ($\text{A}_\text{v} \sim 4$) source at redshift $z\sim$5, with a secondary $10 < z < 15$ P(z) peak. Adopting a log-normal SFH introduces a tertiary P(z) peak at $z > 20$. However, the $\text{A}_\text{v}$ posteriors for all SFHs, while showing a weak probability peak, remain generally unconstrained. Additionally, both \texttt{EAZY} and \texttt{CIGALE} yield rather scattered P(z)s --- more data are needed to further constrain the nature of this source.

Finally, source U-100588 is detected in only two NIRCam bands, making it a F356W-dropout. This object occupies the edge of the UHR $15 < z < 20$ candidate color-magnitude selection space (see Figure~\ref{fig:colormagplot3}), and was characterized in a dedicated work \citep{2025arXiv250901664G} with the help of a $\sim0.8$h-exposure NIRSpec spectrum made available by the CAPERS collaboration.

\subsection{UV luminosity function at z$\sim$17}

\begin{figure}[!h]
    \centering
    \includegraphics[width=0.49\textwidth]{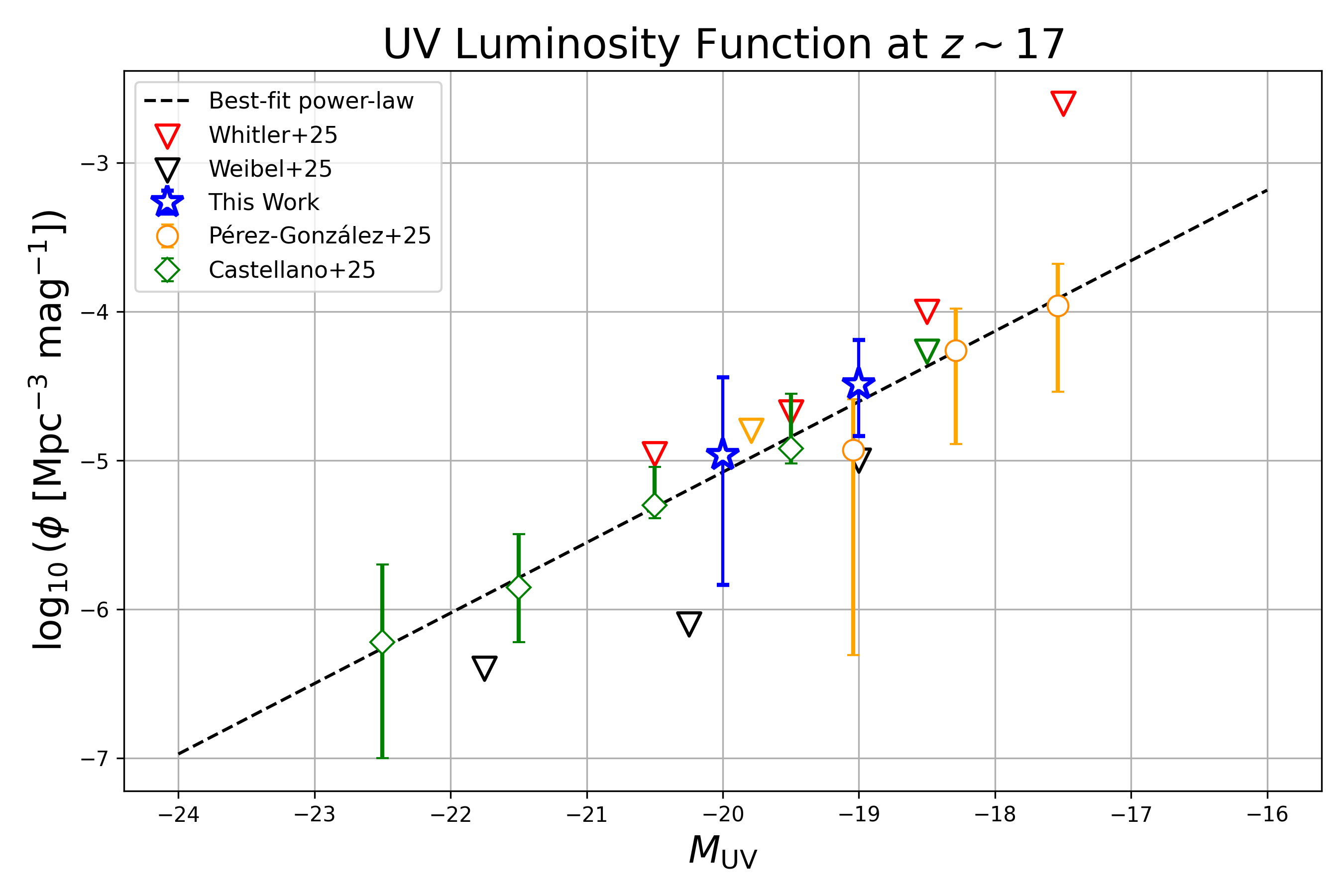}
    \caption{UV luminosity function estimate from our work (blue stars), alongside estimates from \cite{Castellano2025} (green diamonds), \cite{2025arXiv250315594P} (yellow circles), \cite{2025arXiv250706292W} (black triangles, representing upper limits) and \cite{2025arXiv250100984W} (red triangles, again representing upper limits). The best-fit power-law is represented as a black dashed line.}
    \label{fig:uvlum}
\end{figure}

In this section, we compute the UV luminosity function of our UHR candidates under the assumption that their true redshift corresponds to the high-redshift solution ($z > 8$). Given that the best-fit photo-$z$ values in this scenario are mutually consistent within their uncertainties, we assume that all sources lie within the same redshift bin. We then divide our sample into two absolute UV magnitude bins: $-20.5 < M_\mathrm{UV} < -19.5$ and $-19.5 < M_\mathrm{UV} < -18.5$. Given that LIZZAN has a $M_\mathrm{UV}\sim-18.491$ (i.e., falling outside the faintest bin) as well as having the lowest P(z) volume at $z>8$ (see Table~\ref{summary}), we remove it from our computations, therefore using the remaining four sources. We then compute the luminosity function using the standard $1/V_\mathrm{max}$ method \citep{1980ApJ...235..694A}, adopting the statistical formalism described in \cite{1986ApJ...303..336G} (see Table~\ref{tab:uvlf_uhr}). The resulting luminosity function is shown in Figure~\ref{fig:uvlum}, alongside points from other works that have attempted to extend the UVLF measurements to similar redshifts \citep{Castellano2025, 2025arXiv250315594P, 2025arXiv250706292W, 2025arXiv250100984W}. Overall, we find good agreement with values obtained by comparable studies in CEERS and other deep extragalactic fields.

\begin{table}[!h]
\centering
\caption{UV luminosity function measurements for the UHR sample assuming all galaxies lie at $z>8$. Here, $M_{\mathrm{UV}}$ represents the reference magnitude of each bin, $N_{\text{gal}}$ the number of galaxies in the bin, and $\phi$ the corresponding UV luminosity function value with its upper and lower bounds ($\sigma_{\phi,i}$). All values of $\phi$ and the related uncertainties are in units of $\mathrm{Mpc}^{-3}\,\mathrm{mag}^{-1}$.}
\small
\begin{tabular}{ccccc}
\hline\hline
$M_{\mathrm{UV}}$ & $N_{\text{gal}}$ & $\phi$ & $\sigma_{\phi,\text{up}}$ & $\sigma_{\phi,\text{low}}$ \\
\hline
$-20.00$ & 1 & $1.08 \times 10^{-5}$ & $2.52 \times 10^{-5}$ & $9.39 \times 10^{-6}$ \\
$-19.00$ & 3 & $3.25 \times 10^{-5}$ & $3.18 \times 10^{-5}$ & $1.80 \times 10^{-5}$ \\
\hline
\end{tabular}
\label{tab:uvlf_uhr}
\end{table}

Due to the narrow range in absolute UV magnitudes probed by our sample, and the relatively large associated uncertainties, a double power-law fit is not well constrained. For this reason, we adopt a single power-law model to describe the data (i.e., $\log \phi = \alpha M_{\text{UV}} + \beta$, a trending similar to the one followed by the faint end of the UV luminosity function). However, this approach comes with a caveat: caution should be exercised when integrating the resulting luminosity function over a much broader magnitude range than directly observed. In fact, the single-slope approximation may not reliably capture the true shape of the luminosity function at both the bright and faint ends. That said, the observed data are well fit by a single power law characterized by a slope $\alpha\sim0.4$, roughly corresponding to a double power-law faint-end slope as $\alpha = -\left( \alpha_{\text{DBL}}/{0.4} + 1 \right)$, meaning that  $\alpha_{\text{DBL}} \sim -2.18$ (i.e., roughly the same faint-end slope retrieved in \citealt{2024MNRAS.533.3222D} at $z>9$). The best-fit power-law intercept is $\beta\sim4.39$. Assuming $M_* = -19.88$ (i.e., the $M_*$ value reported by \citealt{2024MNRAS.533.3222D} at $z = 9.7$), we estimate the equivalent normalization $\phi_*$ of a DPL as $\log(\phi_*) = \alpha M_* + \beta = -4.95\,\mathrm{Mpc}^{-3}\,\mathrm{mag}^{-1}$, between the estimates normalizations by \citealt{2024MNRAS.533.3222D} ($-4.74\,\mathrm{Mpc}^{-3}\,\mathrm{mag}^{-1}$) and \citealt{2025arXiv250706292W} (< $-5.05\,\mathrm{Mpc}^{-3}\,\mathrm{mag}^{-1}$) at $z=17$.

\section{Discussion}\label{discussion}

In this section we discuss our results, weighing the possible solutions for each candidate examined through our SED fitting procedure.

\subsection{Five galaxies far, far away?}
\label{uhrproperties}
The $z > 8$ solution for all of our five UHR candidates consistently favors a log-normal SFH. This is not surprising, since log-normal SFHs may reproduce both rising and declining star formations, while the other models may be less flexible in representing the SFH shapes found in the early Universe. All the $z > 8$ best-fits redshifts of our sources are included in the $17 < z < 18$ range, and are accompanied by negligible dust, with the highest $\text{A}_\text{v}$ best-fit value being the one of ARCERIO ($\text{A}_\text{v} \sim 0.27$), while in general other candidates are less dusty ($\text{A}_\text{v} < 0.2$). The predicted best-fit metallicities are rather unconstrained, showing errors $>$ 60\% of the measurements, indicating that further data are needed in order to obtain reliable estimates in this sense. The best-fit values for the ionization parameter $\log\text{U}$ are compatible with the expected SED's flat continuum at these redshifts, with no strong line emission. We note that the best-fit galaxy age parameters defining the SFH shape are quite unconstrained and degenerate with each other --- this is a consequence of the limited number of available detections in our dataset, and we preferred not to assume any particular recipe to constrain these values and bias our results. However, our results clearly favor SFHs that started within the first Gyr of the Universe's life. Finally, we checked the consistency of the $z>8$ best-fit stellar mass of each object with $\Lambda$CDM in terms of baryons-to-star conversion efficiency in Section~\ref{sec:lcdmconsistency}.

The $z\leq$ 8 solutions of our sample of UHR candidates tend to favor HELM-like solutions, similar to the ones obtained for objects discussed in Section~\ref{dustydwarfs}. The inferred typical best-fit masses amount to $\log \text{M}/\text{M}_\odot$ < 8, high dust content ($\text{A}_\text{v}$ > 4) and $z < 2$. The main difference between our sample and the HELMs of \citetalias{2023A&A...676A..76B} is that the majority of our objects in the $z \leq$ 8 solution are dusty enough to all be within $1.5\sigma$ from the upper limit we set for the dust attenuation index ($\text{A}_\text{v}$ = 6) --- our objects could therefore be a particularly dusty class of HELMs, which is a rather intriguing possibility. We thus performed a comparison between the $z\leq$ 8 solution of our UHR candidates and the CEERS HELM galaxies sample of Bisigello et al., (2025; \textit{in prep.}); encompassing 4141 sources) in terms of stellar masses versus dust attenuation (see Figure~\ref{fig:helmcomp}). The $\text{A}_\text{v}$ measurements' centroids for all our UHR candidates in their $z \leq 8$ solutions fall on the edge or beyond the contour enclosing 95\% of the CEERS HELM galaxies. In the $z \leq 8$ scenario, our objects could be probing an extreme dust extinction regime if compared to other HELM galaxies given their stellar mass, which could enhance our understanding of how cosmic dust is produced in such low-mass systems. This regime is similar to the one probed by the $z\sim4.0$ solution of the NIRCam-dark ``Cerberus'' galaxy \citep{2024ApJ...969L..10P}. If the $z \leq 8$ nature of our UHR candidates was to be confirmed, their intrinsic faintness would not be due to their extreme distance, but it would rather be attributable to their exceptional dust content. However, we stress that the lack of data in MIR and (sub-)mm wavelengths or spectroscopic observations makes the dust content characterization of these sources particularly uncertain --- we will therefore conduct further studies of their dust properties as soon other data become available.

\begin{figure*}
    \centering
    \includegraphics[width=0.7\textwidth]{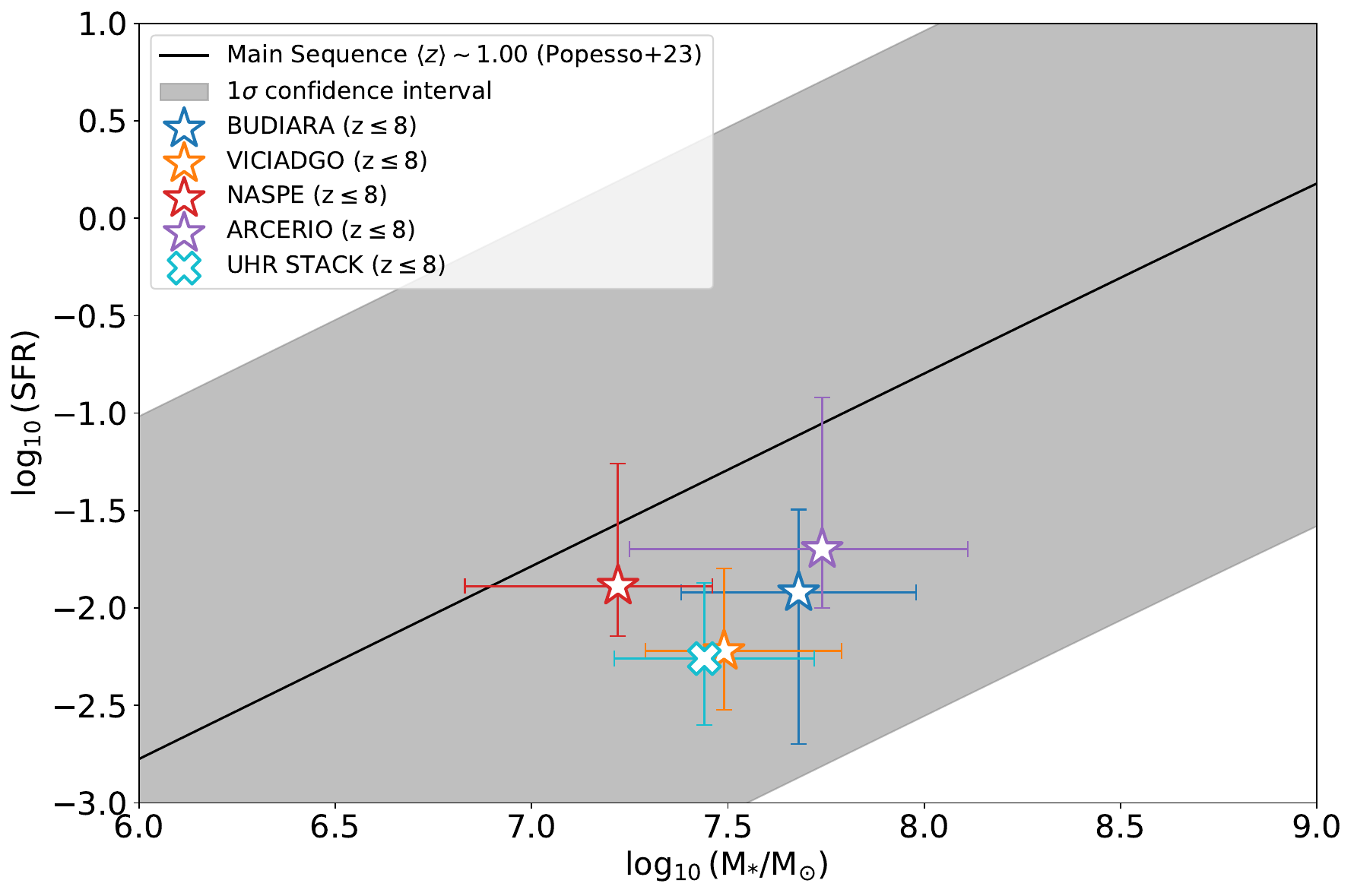}
    \includegraphics[width=0.7\textwidth]{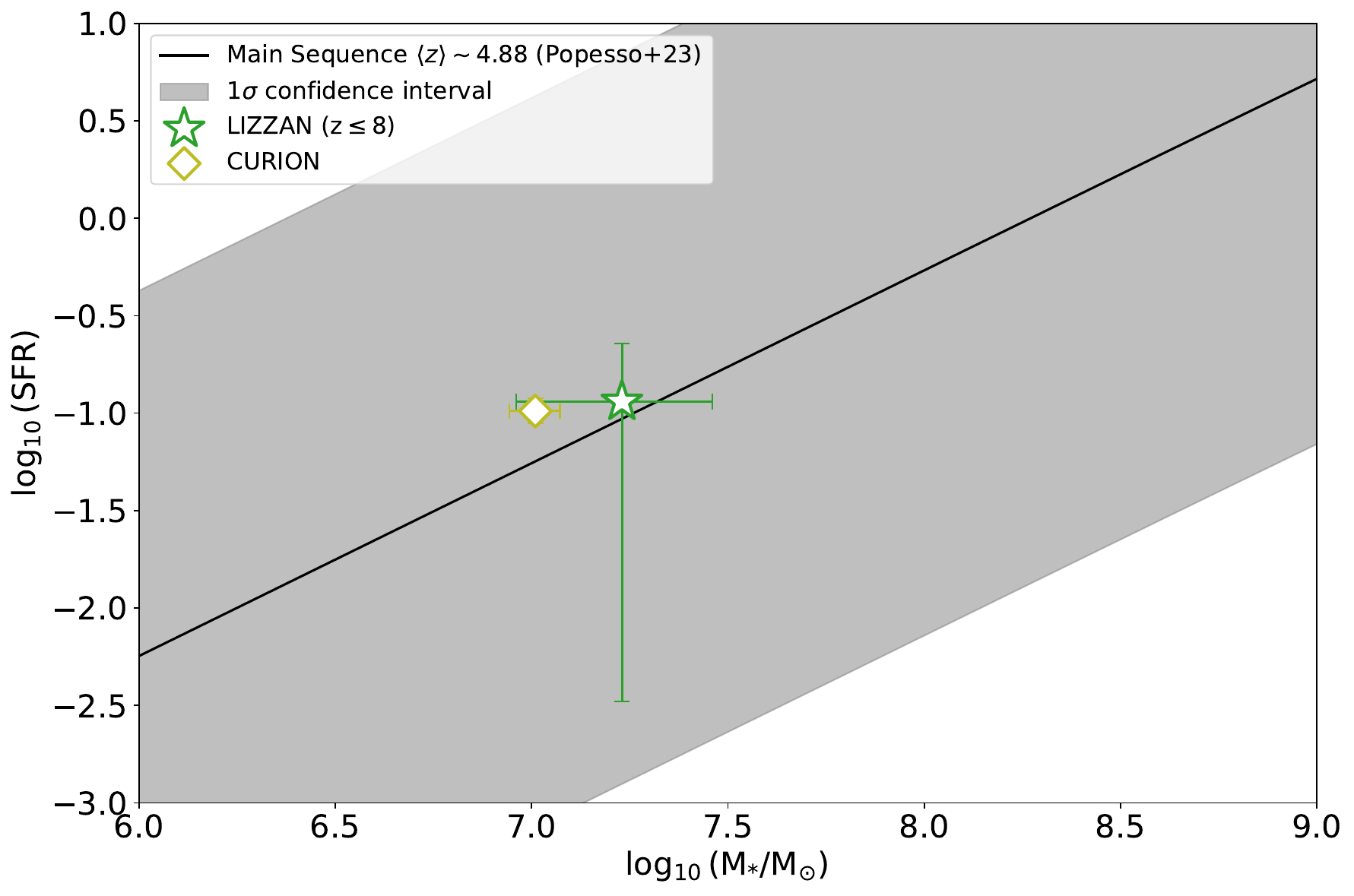}
    \caption{Comparison between galaxies in our sample and the main sequence of star-forming galaxies. The upper panel displays four of our UHR candidates assuming their best-fit $z\leq$8 solutions (represented as colored stars) alongside their median stack (represented as a cyan cross). Since the best-fit $z<8$ solution for LIZZAN is at a higher redshift ($\text{z} \sim 4.31$) with respect to its companion UHR candidates, we checked its consistency with the main sequence in a separate plot (bottom panel) alongside CURION (portrayed as a pea green diamond), falling in a similar redshift window. The main sequence is modeled after the parametrization of \cite{2023MNRAS.519.1526P} and is represented in each panel as a black line computed at the average redshift between the displayed objects ($\langle \text{z}\rangle \sim 1.00$ and $\langle \text{z}\rangle \sim 4.89$ respectively), alongside a 1$\sigma$ confidence interval (gray shaded area).}
    \label{fig:mainsequence}
\end{figure*}

The only exception to this is constituted by LIZZAN, whose $z\leq$ 8 solution is located at a higher redshift ($z \sim$ 4.31, albeit with a rather large lower error bound) and has a smaller dust attenuation ($\text{A}_\text{v} \sim$ 2.04) with respect to the other UHR candidates. LIZZAN's best-fit $z \leq$ 8 solution's SED indeed shows prominent emission lines, making LIZZAN a potential strong line emitter contaminant, and suggesting that our set-up is capable of correctly accounting for strong line emitter solutions for our galaxy sample. The middle and bottom panels of Figure~\ref{fig:mainsequence} highlight that in the $z<8$ solutions our UHR candidates are all consistent within 1$\sigma$ with the main sequence of star-forming galaxies.

We repeated our \texttt{Bagpipes} SED-fitting runs adopting a Calzetti dust attenuation law to assess the impact of different extinction assumptions on the inferred properties for our UHR candidates sample. We found that altering the dust attenuation law does not significantly change our results, yielding consistent physical parameters for our objects. The agreement between the two dust attenuation laws suggests that, in both the $z\leq 8$ and $z> 8$ scenarios, our UHR candidates lack a significant 2100 $\AA$ bump. Such bump is linked to the presence of carbon lattice grains in the interstellar medium (ISM), and is the key difference between the SMC and Calzetti dust models \citep{2023ApJ...948...55H}.  Indeed, some parameters inferred from our SED-fit analysis are intrinsically poorly constrained due to the limited number of photometric data points, leading to large uncertainties. Both the Calzetti and SMC attenuation laws yield similarly bimodal solutions for the UHR candidates studied here, though they differ slightly in how the solution space in the P(z) plane is distributed between HELM and UHR probabilities. Notably, the Calzetti law appears to favor UHR solutions in certain objects, even when using SFHs such as delayed or exponentially delayed models, further emphasizing the intriguing nature of our UHR candidates. Since the physical properties derived using the Calzetti law align with those obtained using the SMC law, we adopt the SMC law as the fiducial attenuation model for the remainder of our analysis. We plan to revisit the impact of different dust attenuation laws on the inferred properties of our galaxies as additional data at longer wavelengths become available, which should better highlight subtle distinctions between the two.

Our \texttt{EAZY} runs exploiting the wide set of templates listed in Section~\ref{5|sec:sedfitting} tend to favor, on average, the $z > 8$ solutions of our UHR galaxy candidates' samples in terms of the inferred P(z)s. Both BUDIARA and VICIADGO feature an integral probability $\int_{z>15} \mathrm{d}z\, P(z) > 0.99$ for all the templates used in the analysis, indicating a strong preference for $z > 15$ best-fit solutions. LIZZAN shows an integral probability $\int_{z>15} \mathrm{d}z\, P(z) > 0.99$ for the majority of the \texttt{EAZY} templates utilized in our runs, with the exception of \texttt{hot-45k} templates ($\int_{z>15} \mathrm{d}z\, P(z) > 0.68$) and \texttt{hot-60k} ($\int_{z>15} \mathrm{d}z\, P(z) > 0.79$). These latter two templates seem to allocate part of LIZZAN's P(z) volume to a secondary solution at $z < 10$, somewhat compatible with the secondary, lower-$z$ strong line emitter solution predicted by \texttt{Bagpipes} for this source. However, both the \texttt{hot-45k} and \texttt{hot-60k} still predict that the bulk of the P(z) volume for LIZZAN is located at $z > 15$, thus favoring the UHR nature of this source. \texttt{EAZY} templates predict a $\int_{z>15} \mathrm{d}z\, P(z) > 0.99$ for NASPE, with the exception of the \texttt{fsps} templates, yielding $\int_{z>15} \mathrm{d}z\, P(z) \sim 10^{-3}$. The P(z) predicted for NASPE by this particular template reproduces the one yielded by \texttt{Bagpipes} at $z < 8$, backing the HELM galaxy scenario for this source and assigning a negligible solution volume at $z > 8$. This makes NASPE the only source in our five UHR galaxies sample to have a strongly bimodal \texttt{EAZY} P(z). Finally, our \texttt{EAZY} runs predict for ARCERIO $0.78 < \int_{z>15} \mathrm{d}z\, P(z) < 0.95$, depending on the template used. This source thus shows a mild P(z) bimodality, with the bulk of the \texttt{EAZY} P(z)s always lying beyond $z = 15$, regardless of the template used.

Our \texttt{CIGALE} runs systematically back up the ultra high-z nature of our sources, with P(z) peaks aligning to the ones predicting by other SED-fitting codes (with the exception of ARCERIO, predicting an even higher best-fit redshift of $\sim 21.2$). Furthermore, our \texttt{CIGALE} runs did not reveal any significant AGN component associated with our UHR sample, yielding best-fit predictions for both the AGN fraction always $\leq0.4$ and, in may cases, compatible with zero.

Finally, we remark that all our UHR candidates are resolved in at least one NIRCam band (see Appendix~\ref{sizes}), which is enough to exclude the possibility of them possibly being Milky Way sub-stellar objects such as BDs.

\subsection{Stacking the UHR galaxy candidates sample with \texttt{CosMix}}\label{uhrstack}

\begin{figure*}
    \centering
    \includegraphics[width=\textwidth]{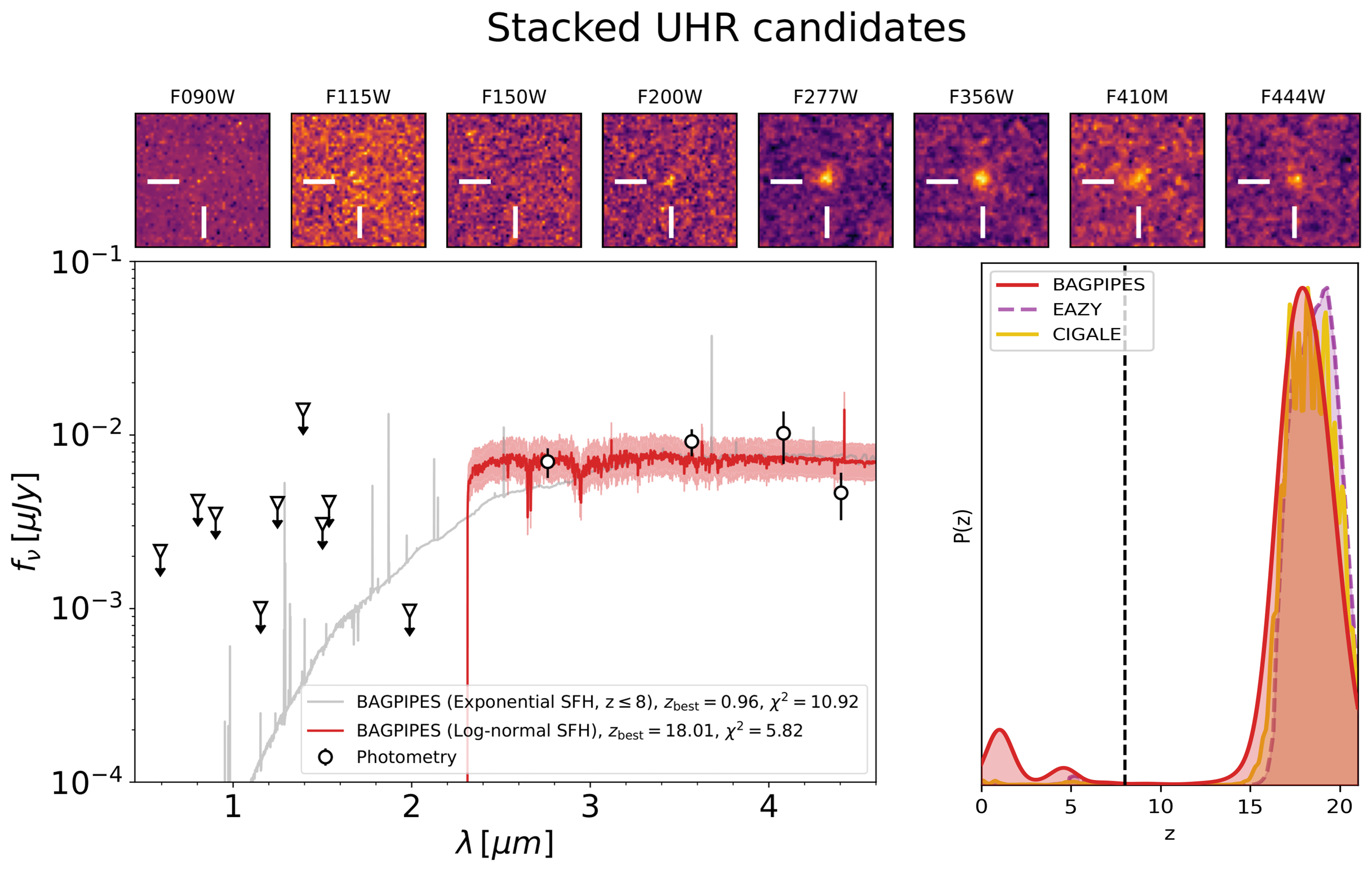}
    \caption{Best-fit SEDs, P(z)s and NIRCam cutouts for the median stacked UHR candidates sample, obtained exploiting \texttt{CosMix}. The plot's color scheme is the same followed in Figure~\ref{fig:singlefits}.}
    \label{fig:stacksed}
\end{figure*}

Despite the inherent degeneracy introduced by the bimodality of their P(z) distributions, the objects in our UHR galaxy sample exhibit remarkably consistent physical properties, allowing us to treat them as a homogeneous group for further analysis. To improve the S/N ratio of our five UHR candidates and to mitigate the degeneracy between the $z \leq$ 8 and $z > 8$ solutions, we chose to stack the sample. This approach allows us to analyze their combined median photometric properties, providing a more robust framework for understanding their nature.

To stack the sources in our UHR candidate sample, we used \texttt{CosMix} (Gandolfi et al., 2026; \textit{in prep.}), a new Python-based stacking tool designed to combine the multi-wavelength signal of faint sources with homogeneous physical properties and enhance their characterization. \texttt{CosMix} will soon be available as a free web application, running on a dedicated machine hosting a multi-wavelength set of data spanning different fields available for stacking. \texttt{CosMix} extracts image cutouts from input catalogs (ID, RA, DEC), combines them, and outputs downloadable \texttt{FITS} files of the desired size. It also computes stacked photometry with customizable settings, producing an SED and a \texttt{FITS} file with errors and upper limits, ready for fitting with any preferred program. We report the salient passages of our setup for stacking in Appendix~\ref{appendix:stack}. After obtaining the median stacked photometry through \texttt{CosMix}, we applied the same SED-fitting procedures described in Section~\ref{5|sec:sedfitting} for fitting single objects. The median UHR candidates stack shows a S/N < 3 in all bands blueward of F277W. Some residuals are visible in F200W due to the narrow contrast scale of the cutout, with a S/N $\sim$ 1.7, not enough to be associated with a detection. Furthermore, the F115W stack has a S/N $\sim$ 2.8 --- this value, which is not high enough to be associated with a sound detection, is likely attributable to noise accumulation during the stacking process, which could not be mitigated due to the limited number of objects in our sample. As visible in Figure~\ref{fig:colormagplot3}, the color-magnitude properties of the stacked UHR candidates are broadly in line with the ones of the single objects in the sample, being consistent with both the UHR galaxies' selections reported in the plot. 

Our \texttt{Bagpipes} stacked photometry fits highlight that the stacking procedure was not able to completely remove the P(z) bimodality of our UHR candidates, which is expected due to the narrow wavelength range of the available data. However, we now find that even delayed and delayed exponential SFHs admit a non-negligible solution volume at $z > 8$ (as visible by looking at the gray P(z) in Figure~\ref{fig:stacksed}), marking an important difference with respect to single-object fits. A strong improvement is also visible by looking at the median P(z) yielded by our \texttt{EAZY} fits, which now shows a completely negligible solution volume at $z < 10$, with $\int_{z>15} \mathrm{d}z\, P(z) \sim 0.99$ for all the templates used in the analysis. Our \texttt{CIGALE} fits yield consistent best-fit physical properties with respect to the other codes. These results further corroborate the intriguing nature of our UHR candidates and their potential as promising targets for future spectroscopic studies. To infer the physical properties of our stacked object, we split its \texttt{Bagpipes} P(z)s in two as we did with our single sources. Finally, the physical parameters inferred in this way (listed in Table~\ref{tab:bestfitstacked_properties}) are consistent with the $z \leq$ 8 and $z > 8$ solutions of our single UHR candidates. As we did with single objects, we assessed the stack's best-fit stellar mass compatibility with $\Lambda$CDM in terms of baryons-to-star conversion efficiency in Section~\ref{sec:lcdmconsistency}. We find that the stacked UHR galaxy sample is consistent with the main sequence of star-forming galaxies in the $z\leq$8 solution (see top panel of Figure~\ref{fig:mainsequence}).

\subsection{Consistency with $\Lambda$CDM}\label{sec:lcdmconsistency}

\begin{figure*}[h]
    \centering
    \includegraphics[width=0.9\textwidth]{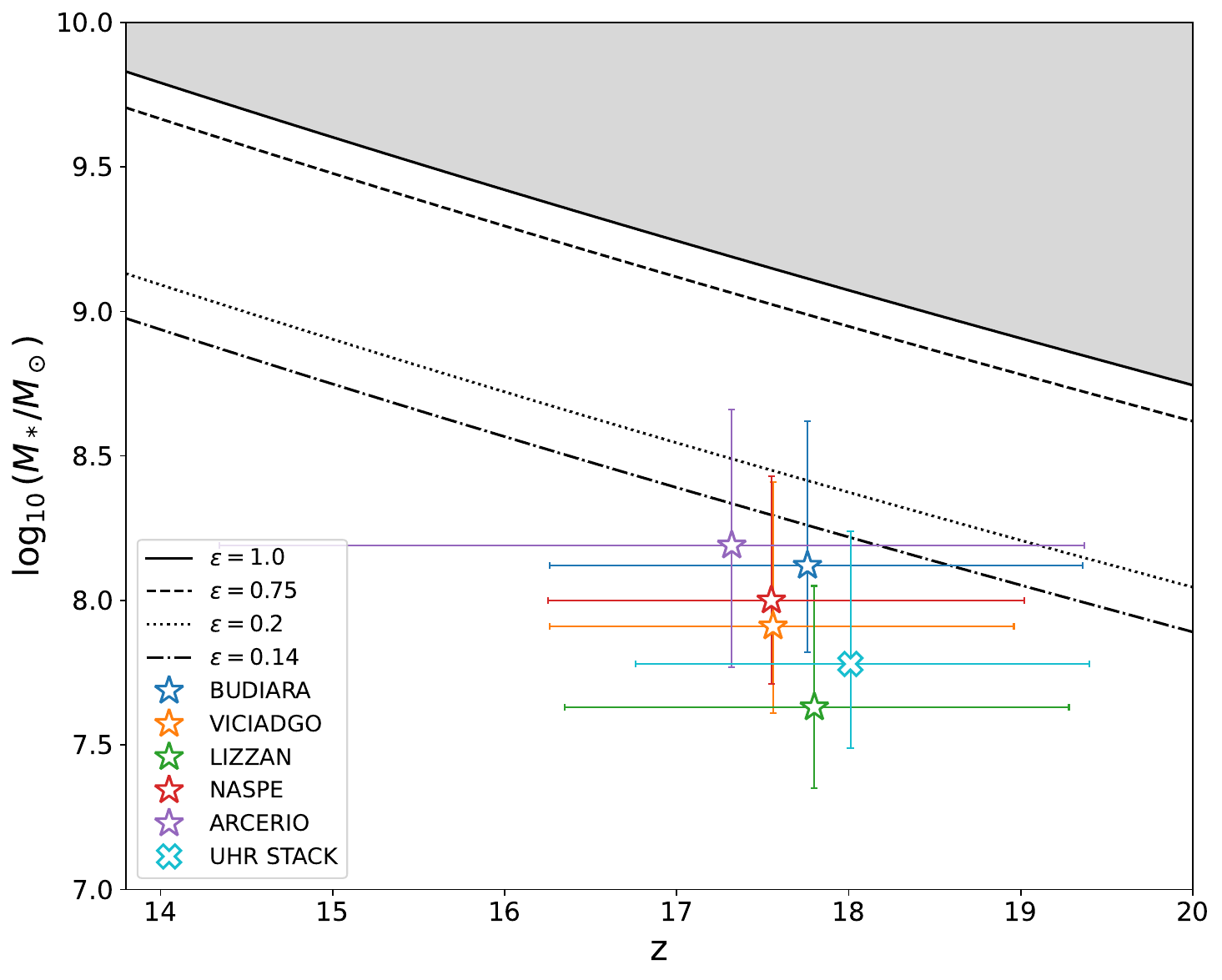}
    \caption{Best-fit stellar masses of our UHR candidates (and their median stack) in their $z > 8$ solutions versus best-fit redshift. Each candidate is represented as a colored star, with the median stack being depicted as a cyan cross. The dash-dotted, dotted, dashed, and continuous black lines represent different values of the baryons-to-star conversion efficiency (0.14, 0.2, 0.75 and 1 respectively), while the gray shaded area corresponds to nonphysical efficiency values > 100\%.}
    \label{fig:lcdmconsistency}
\end{figure*}

To corroborate whether the $z > 8$ solutions of our five UHR candidates could be realistic within the standard $\Lambda$CDM framework, we checked the consistency of their best-fit stellar masses and baryon-to-stars conversion efficiencies with the $\Lambda$CDM model. The underlying idea behind this test is simple: the hierarchical formation of cosmic structures predicted by $\Lambda$CDM sets a limit on the maximum dark matter halo mass expected at a given redshift. The mass of dark matter halos ($\text{M}_\text{halo}$) can be in turn related to the stellar mass ($\text{M}_*$) of their host galaxies as $\text{M}_* = \epsilon f_b \text{M}_\text{halo}$, where $\epsilon$ is the efficiency of baryon conversion into stars, and $f_b = 0.158$ \citep{2020A&A...641A...6P} is the cosmological baryon fraction. Assuming a certain dark matter halo mass function, one can thus calculate a theoretical stellar mass limit at each redshift.

In Figure~\ref{fig:lcdmconsistency}, we show the best-fit $z>8$ UHR candidates' (and their median stack's) stellar mass estimates versus their best-fit redshift in the $z>8$ solution. The black lines in the plot represent theoretical stellar mass limits at each redshift assuming different baryons-to-star conversion efficiencies $\epsilon$, whereas the gray area in the plot represents unrealistic values for the baryon-to-star conversion efficiency $\epsilon > 1$ --- a galaxy falling in this area would indicate a tension with $\Lambda$CDM. We computed the different $\epsilon$ value curves assuming a \cite{Sheth:1999mn} dark matter halo mass function and integrating it to determine the mass above which one halo is expected within CEERS' survey volume, assuming sliding redshift bins of $\Delta z = 2$ (i.e., following the approach outlined in \citealt{2024arXiv241112005L}). During this process, we made use of the \texttt{Python COLOSSUS}\footnote{https://bdiemer.bitbucket.io/colossus/} package (\citealt{2018ApJS..239...35D}) to deal with cosmological calculations.

Our analysis shows that all the UHR candidates in their $z > 8$ solutions are well below the $\epsilon=1$ threshold within 1$\sigma$, indicating that they are consistent with the $\Lambda$CDM framework. BUDIARA and ARCERIO require slightly larger efficiencies with respect to other objects in the sample in order to match the inferred stellar mass, being consistent within 1$\sigma$ with the $\epsilon = 0.2$ curve, implying higher efficiency values with respect to the ones inferred for $z < 4$ galaxies ($\epsilon \sim 0.14$, see \citealt{2024AJ....168..113C}). Other well-characterized high-redshift galaxies such as PENNAR (\citealt{2023MNRAS.518L..19R}; $z_\text{photom}=12.1$), GHZ2 (\citealt{2024ApJ...972..143C}; $z_\text{spec} = 12.34$) or JADES-GS-z14-0 (\citealt{2024arXiv240518485C}, $z_\text{spec} = 14.32$) yield $\epsilon$ values around $\epsilon \sim 0.3$. These high star-formation  efficiencies would also be in line with recent COSMOS-Web results showing an increased value for $\epsilon$ in $z>5$ galaxies \citep{2024arXiv241008290S}. Other UHR candidates' $z>8$ solutions (plus the stack) showcase instead a milder conversion of baryons into stars with $\epsilon \lesssim 0.1$, similar to the efficiency attained at a lower redshift by JADES-GS-z14-1 (\citealt{2024arXiv240518485C}, $z_\text{spec} = 13.90$). All in all, this test indicates that the high-redshift solutions of our objects are plausible within the $\Lambda$CDM framework.

As a further check, we estimated the physical sizes of our UHR candidates in all long-wavelength NIRCam bands assuming their $z>8$ solutions (see Appendix~\ref{sizes}). We then compared the inferred sizes with the empirical effective radius vs redshift relation found in \cite{2024arXiv241214970W}, extrapolated beyond $z=17$ (see Figure~\ref{fig:uhr_sizes}). All our UHR candidates in their $z>8$ turned out to be consistent with the extrapolated \cite{2024arXiv241214970W} relation within $2\sigma$ (with the exception of LIZZAN in the F277W and F356W bands, which instead falls within $2\sigma$ from the aforementioned relation).

\subsection{``CURION'' - a strong line emitter at $z \sim$ 5}
\label{sec:curion}

With a best-fit \texttt{Bagpipes} redshift of $z \sim$ 5.5, CURION is consistent with being a strong line emitter galaxy contaminant, underscoring the important of multi-wavelength data in UHR galaxy searches. In fact, preliminary \texttt{EAZY} SED-fitting runs relying on JWST/NIRCam, HST/ACS and HST/WFC3-only photometry yielded a primary peak in CURION's P(z) at $z \sim$ 13, with a secondary solution at $z \sim$ 9. \texttt{Bagpipes} runs were able to recover these solutions assuming a log-normal SFH, while for more conservative SFHs the primary P(z) peak turned out to be at $z \sim$ 5.5. However, the $z \sim$ 5.5 solutions predicted by \texttt{Bagpipes} relying on JWST/NIRCam, HST/ACS and HST/WFC3 photometry alone always yielded extremely high dust attenuation index values, compatible with the adopted parameter's prior upper limit ($\text{A}_\text{v} \geq $ 6). Even if CURION appears undetected in the available MEGA MIRI data, the inclusion of photometric information between 7.7 and 21 $\mu$m in the SED-fitting procedure drastically changes this picture, helping to constrain CURION's dust content and determine its redshift more robustly. Finally BD fits yield to the conclusion that CURION is indeed a galaxy and likely not a Milky Way sub-stellar object.

\subsection{Three dusty dwarf galaxies}\label{dustydwarfs}
The best-fit physical properties derived for U-53105 make us conclude that U-53105 is in all likelihood a dusty galaxy at $z\sim$2.6. A-26130 indeed falls within the $15 < z < 20$ UHR selection space (see Figure~\ref{fig:colormagplot3}), yet its P(z) does not exhibit the distinctly bimodal distribution typically associated with other UHR candidates discussed in this work. This lack of a pronounced high-redshift solution might stem from the absence of a significant spectral break in the best-fit inferred SED, consistent with the faint emission observed in the F200W band for this source. The faint visibility of the source in the F200W band also suggests that the high-redshift solution found by \texttt{EAZY} and \texttt{CIGALE} is likely spurious. Preliminary evidence, therefore, suggests that A-26130 is likely a dusty galaxy at $z\sim$2 rather than a bona fide UHR source in the $15<z<20$ range. Finally, for A-76468 both \texttt{Bagippes}, \texttt{EAZY} and \texttt{CIGALE} do not support the interpretation of this object as a UHR galaxy --- the high stellar masses ($\log \text{M}/\text{M}_\odot > 9.5$) predicted by the high-redshift solutions lead us to favor the conclusion that this source is a dusty galaxy located at $z = 3$. All three sources exhibit physical properties similar to the ones of other HELM galaxies in CEERS, tracing their high-stellar mass end.

\begin{figure}
    \centering
    \includegraphics[width=\linewidth]{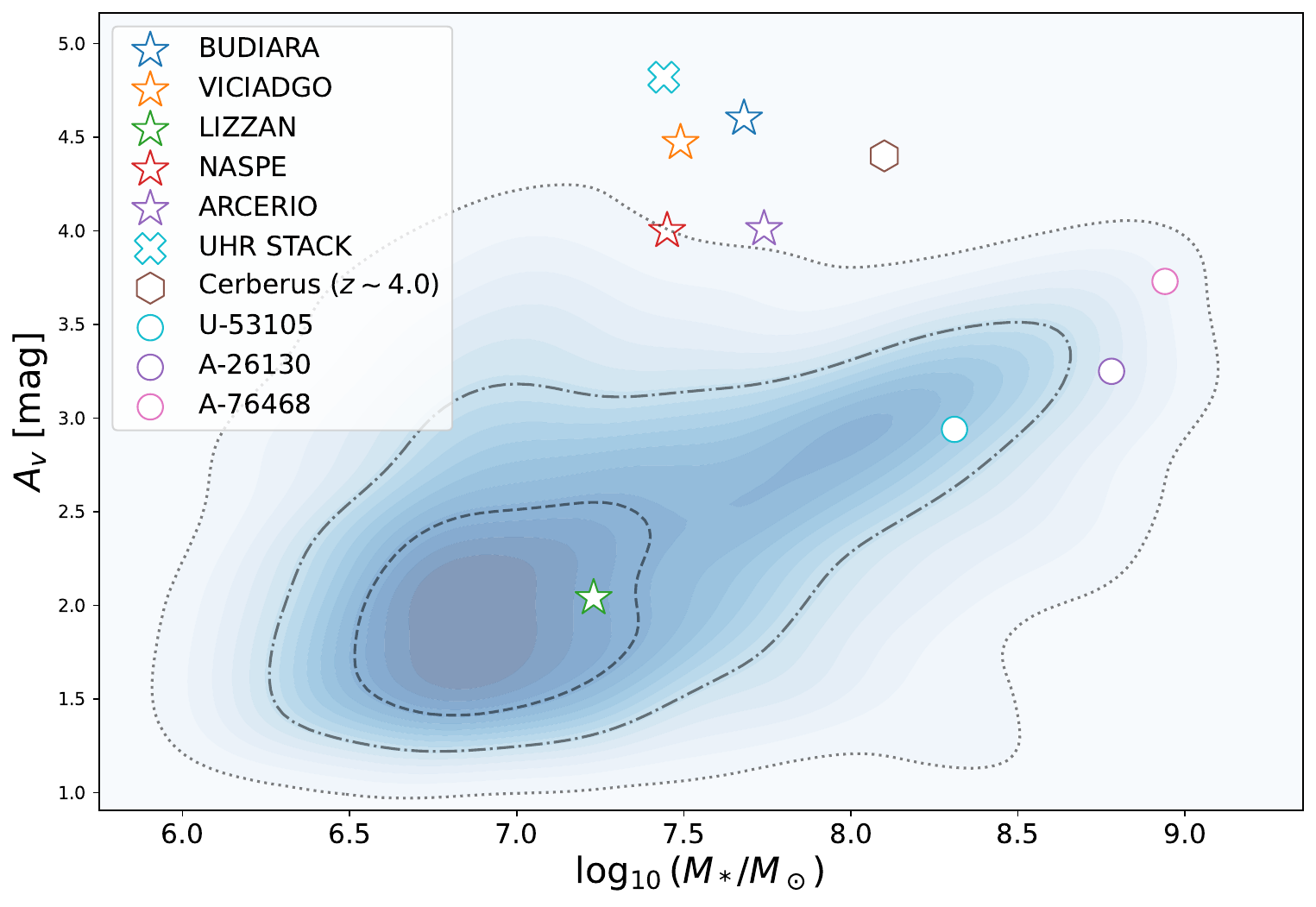}
    
    \caption{Comparison between the 4141 sources in the CEERS HELM galaxy sample by Bisigello et al., (2025, \textit{in prep.}) and the HELM solutions retrieved in this work in the form of a logarithmic stellar mass versus dust attenuation index plot. Blue contours represent isodensity lines in the distribution of the updated CEERS HELM galaxies sample by Bisigello et al., (2025, \textit{in prep.}). The dashed, dashdotted and dotted black lines are respectively enclosing 50\%, 80\% and 95\% of the total HELMs distribution. Colored star markers represent our UHR candidates adopting their $z \leq 8$ solutions, with their median stack being represented by a cyan cross. The three colored circles instead represent our strongest HELM candidates discussed in Section~\ref{dustydwarfs}. Finally, the brown hexagon represents the best-fit stellar mass and dust attenuation index for the NIRCam-dark ``Cerberus'' galaxy \citep{2024ApJ...969L..10P} in its $z\sim4$ solution (fitting with the \texttt{synthesizer-AGN} code and adopting a double-exponential SFH with a Calzetti dust attenuation law).}
    \label{fig:helmcomp}
\end{figure}

\begin{figure*}[!htbp]
    \centering
    \includegraphics[width=0.89\textwidth]{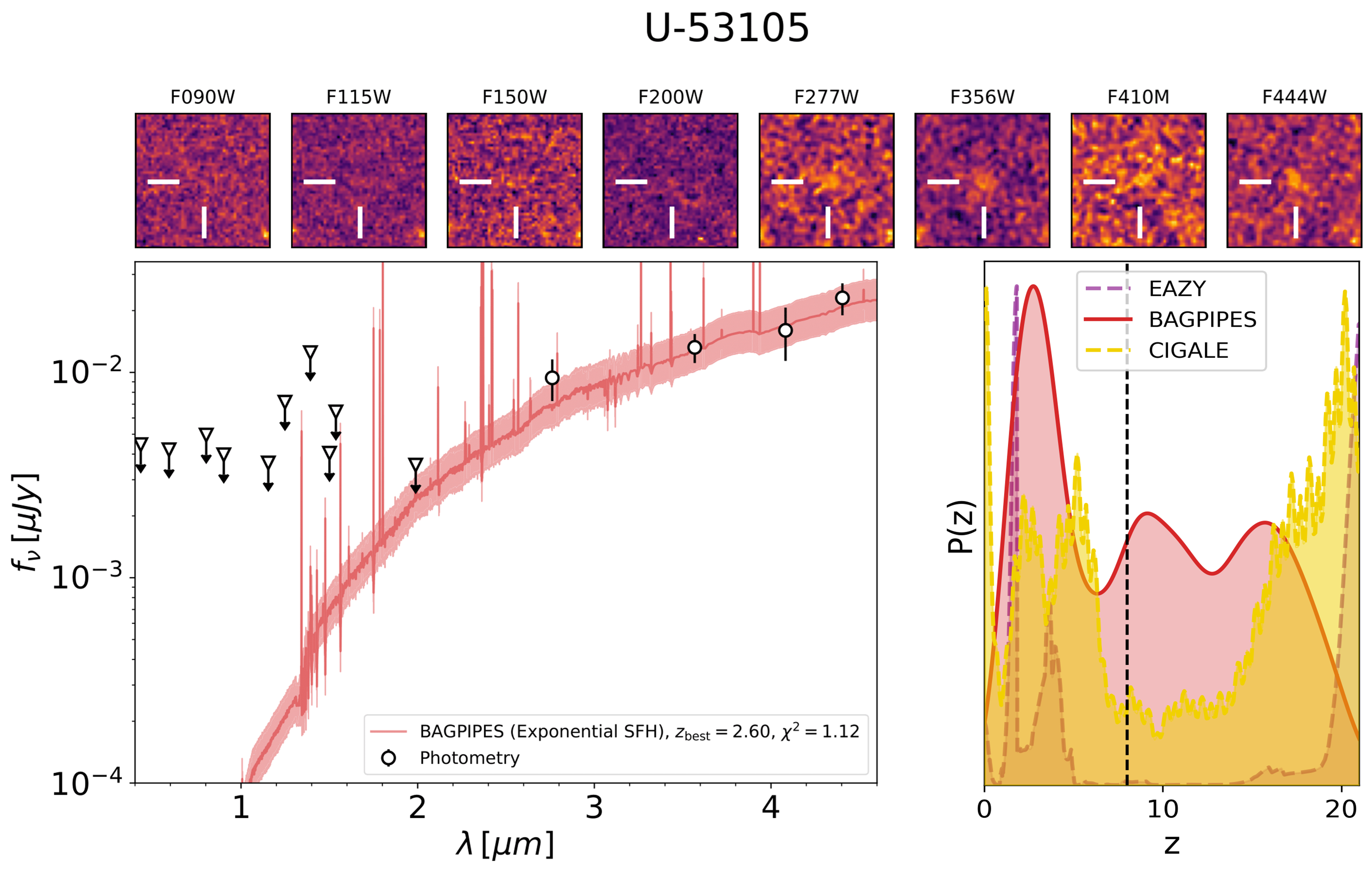}
    \includegraphics[width=0.89\textwidth]{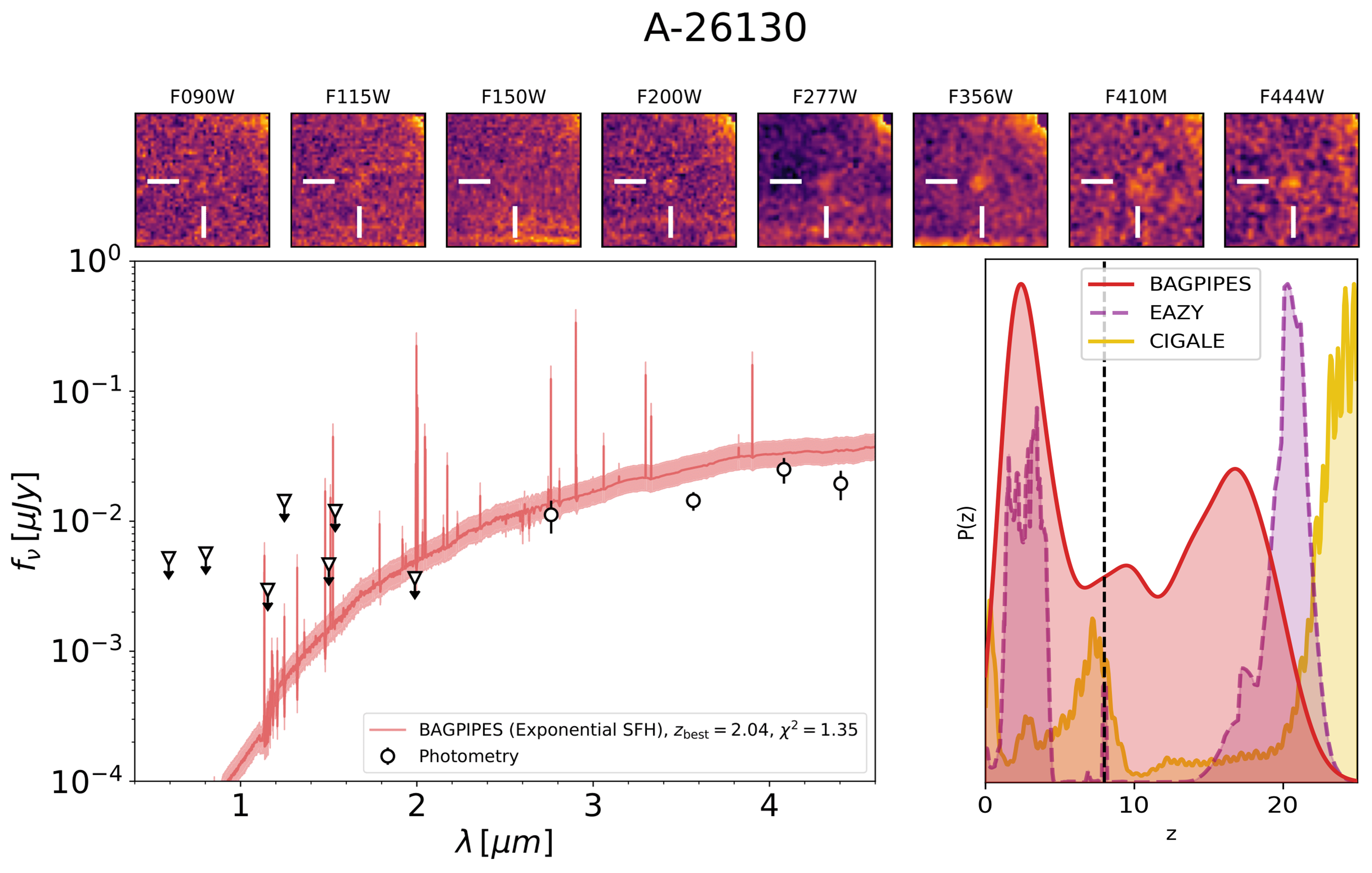}

    \caption{Best-fit SEDs and P(z)s for U-53105 (top), A-26130 (middle) and A-76468 (bottom). The top section of each plot shows 1.5'' $\times$ 1.5'' NIRCam cutouts for the source. The left inset displays the observed photometry (black circles) with $1\sigma$ upper-limit uncertainties (black triangles). The best-fit SED derived with \texttt{BAGPIPES} is shown as a red curve with a 0.1 dex confidence interval (red shaded area). The right inset reports the normalized redshift probability distributions from different SED-fitting codes: the $P(z)$ of \texttt{Bagpipes} in red; the \texttt{CIGALE} posterior in yellow; and the mean \texttt{EAZY} posterior averaged over all templates described in Section~\ref{5|sec:sedfitting} (purple). The vertical dashed black line marks the $z=8$ threshold.}
    \label{fig:otherf200wdropouts}
\end{figure*}

\begin{figure*}[!htbp]
    \ContinuedFloat
    \centering
    \includegraphics[width=0.89\textwidth]{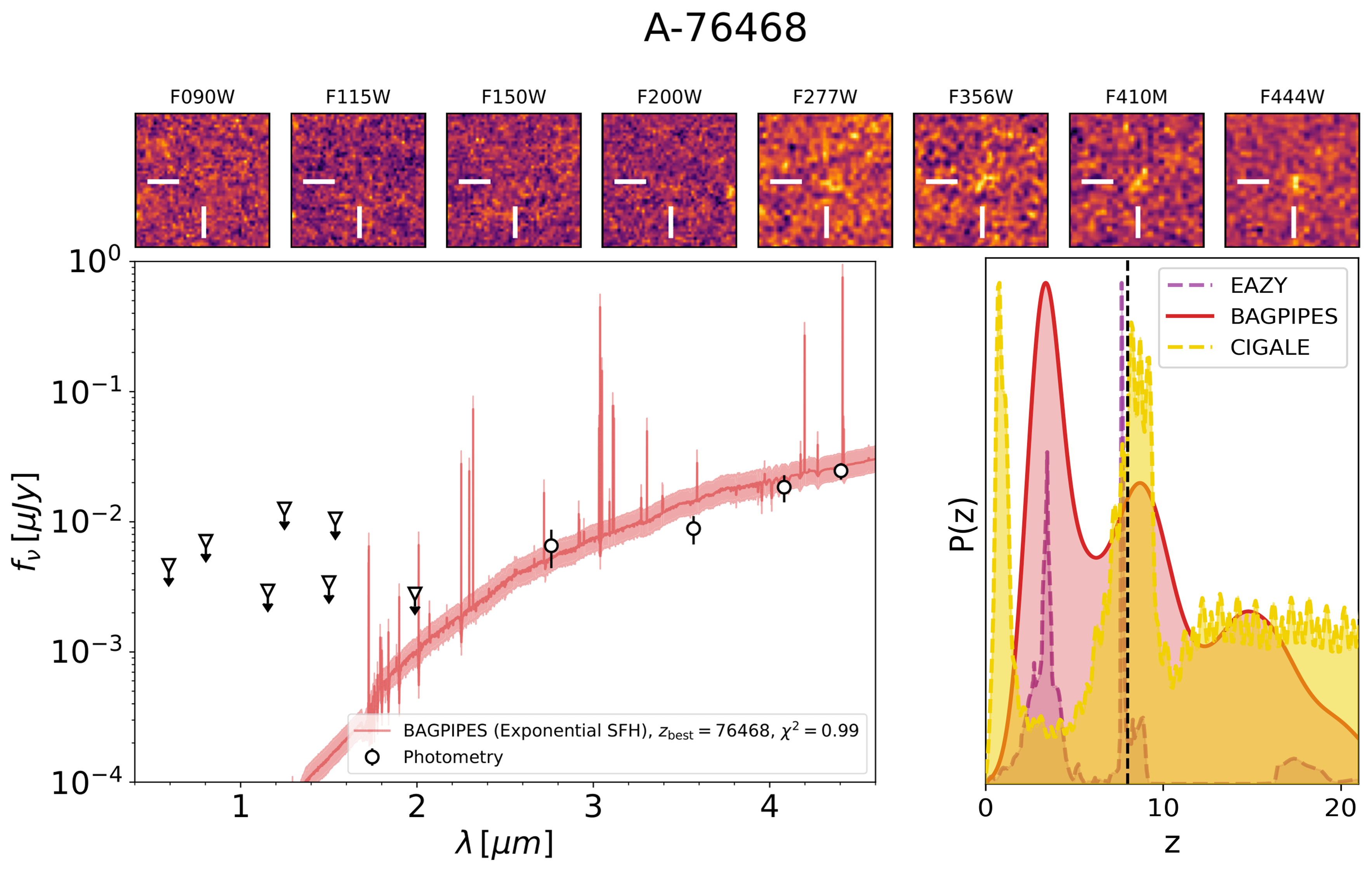}
    \caption{Continued.}
\end{figure*}

\begin{figure*}
    \centering
    \includegraphics[width=0.7125\textwidth]{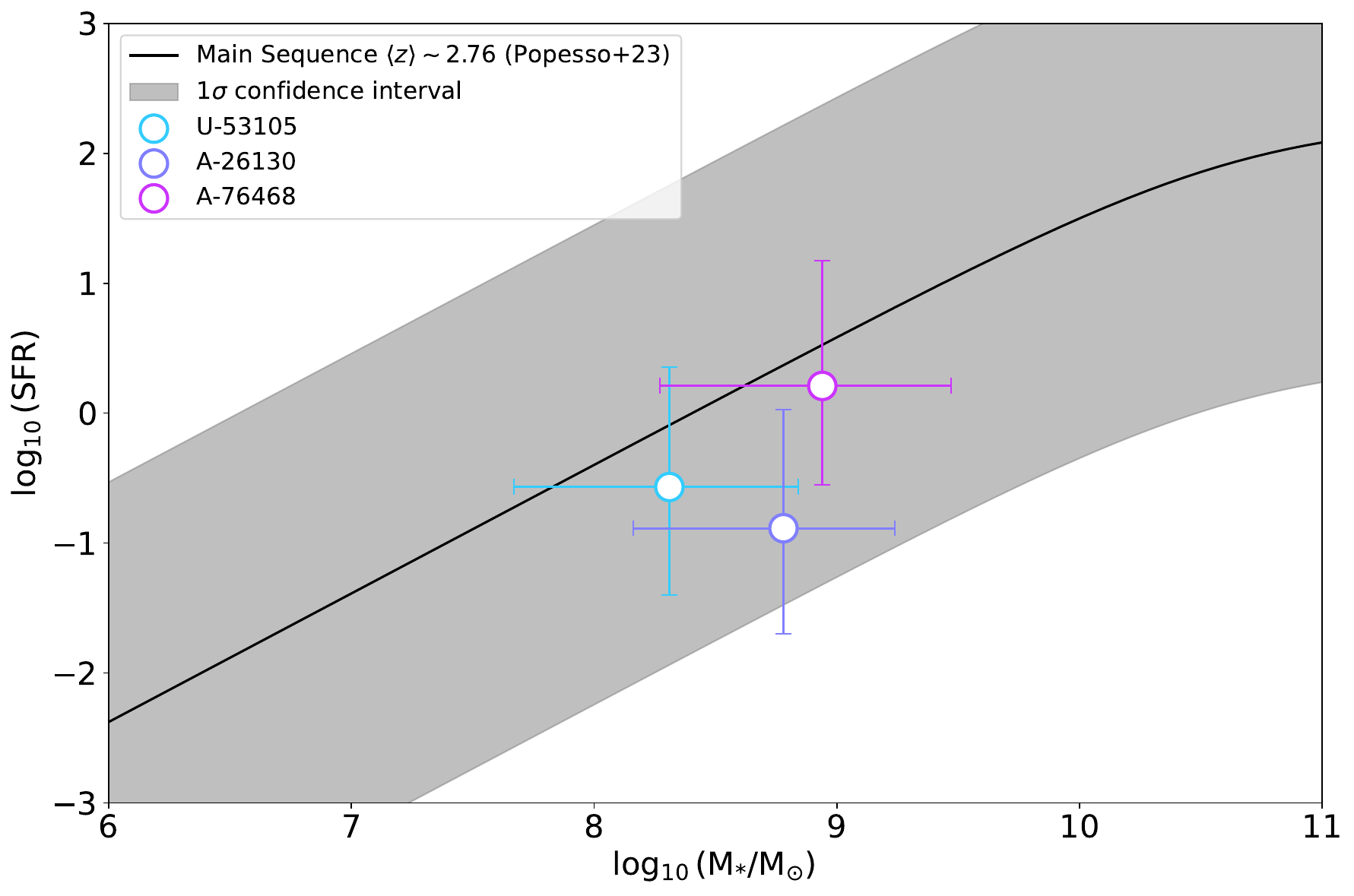}
    
    \caption{Comparison between U-53105, A-26130 and A-76468 to the main sequence of star-forming galaxies, represented as purple circles. The main sequence is modeled after the parametrization of \cite{2023MNRAS.519.1526P} and is represented in each panel as a black line computed at the average redshift between the displayed objects ($\langle z\rangle \sim 2.76$ alongside a 1$\sigma$ confidence interval (gray shaded area).}
    \label{fig:otherf200wms}
\end{figure*}

We show that U-53105, A-26130 and A-76468 are all compatible within 1$\sigma$ with the main sequence of star forming galaxies in Figure~\ref{fig:otherf200wms}. Furthermore, we detailed a comparison between the HELM galaxies discussed in this Section with the CEERS HELM galaxies sample of Bisigello et al., (2025; \textit{in prep.}) in terms of stellar masses versus dust attenuation (see Figure~\ref{fig:helmcomp}). Our three HELM galaxy candidates fall at more than 3$\sigma$ from the contour enclosing 50\% of the CEERS HELM galaxies in the stellar mass versus dust attenuation plane, indicating that they belong to the more massive end of the HELM galaxies distribution. In particular, A-26130 and A-76468 show higher best-fit stellar masses with respect to U-53105, and they lie within $1\sigma$ from the 95\% contour. Investigating this massive end of the HELM galaxies distribution could be crucial to characterize how stellar masses and dust attenuation correlate for these objects, highlighting potential differences with the expected mass versus dust attenuation relation for the other galaxies in the CEERS sample. However, data at longer wavelengths (such as in the MIR or (sub-)mm bands) or spectroscopic follow-ups would be needed to reduce the uncertainties affecting estimates of the dust content of the three HELM galaxies presented here.

\section{Conclusions}
\label{7|sec:conclusions}
Our analysis revealed a sample of faint, previously missed F200W-dropouts in the CEERS survey. These objects are interesting candidates for future observations, either aiming at soundly characterizing their SED in the MIR/(sub-mm) regime via the addition of new photometric information (thus extending their presently available wavelength coverage beyond the F444W band) , or via spectroscopic follow-ups. We investigated the physical properties of these sources mostly relying on the available NIRCam photometric data, complementing them with MIR/(sub-)mm data when available. Here, we recap the most significant results of our analysis:

\begin{itemize}
    \item We revealed three (U-53105, U-26130 and A-76468) $z<4$ HELM galaxies (i.e., dwarfs characterized by a puzzlingly high dust content) which occupy the high-mass end of the stellar mass distribution of HELM objects in CEERS. This regime can crucially help in characterizing how the stellar mass of these galaxies correlates with their dust content;
    \item We highlighted one (CURION, U-112842) strong line emitter at $z \sim 5$. Preliminary \texttt{EAZY} runs exploiting NIRCam-only photometry yielded a primary $z \sim 13$ solution, but the inclusion of MIRI upper limits from the MEGA collaboration revealed its true nature;
    \item We reported a sample of five faint objects with bimodal P(z)s --- our analysis suggests that these are either extremely dusty HELM galaxies or UHR objects between $15 < z < 20$, whose stellar masses are not in tension with $\Lambda$CDM cosmology. Either possibility is extremely intriguing --- yet, in order to be further thoroughly investigated, more longer-wavelength photometric data or detailed spectroscopic followups are needed;
    \item We outlined an analysis pipeline that combines different SED-fitting codes, templates, SFHs, dust extinction models and stacking which can be helpful in future UHR object searches, helping in discriminating between known lower-z contaminants and assessing the P(z) solution volume of each source.
\end{itemize}

We plan to extend our set of multi-wavelength photometric data describing the SEDs of the objects in the sample analyzed in this work. In particular, our object CURION highlights the discriminating power of MIR data in the SED-fitting procedure of these dropout objects, even when these data are just upper limits. Moreover, deeper observations in the Far-Infrared/(sub-)mm bands could help in constraining the dust content of our sources, both from current facilities such as NOEMA or future ones such as NASA's PRobe far-Infrared Mission for Astrophysics (PRIMA; see e.g., \citealt{2024A&A...689A.125B}). Extending the multi-wavelength coverage of these sources would be pivotal to test the impact of different and more detailed dust extinction models on their best-fit solutions, non-parametric SFHs and/or different SPS models. On the other hand, we also plan to refine our analysis including new SED-fitting codes such as \texttt{GalaPy} \citep{2024A&A...685A.161R} implementing a two-component dust absorption/re-radiation model which is unbiased against the physics of dust grains and integrated with high resolution stellar populations for the study of primordial galaxies.

Spectroscopic coverage of our sources would also be pivotal to soundly constrain their nature. A promising opportunity in this sense is constituted by the potential inclusion of our sources in the upcoming CAPERS program. A few of our objects (the five UHR candidates) are likely too faint to probe their continuum emission, yet it would still be possible to observe or rule out emission lines from solutions at $z < 10$, casting rather strong constraints on their true nature.

To date, only a very low number of HELM galaxies are spectroscopically confirmed \citep{2024arXiv241010954B, Castellano2025}. A larger sample of spectroscopically confirmed HELM galaxies would be important to better characterize dust production mechanisms in such low-mass galaxies, especially if, as hinted from our analysis, some of the targets have such an extreme dust content given their stellar mass with respect to other similar objects. Moreover, our analysis suggested that HELM galaxies can mimic the color-color properties of UHR, $15 < z < 20$ galaxies (see Figure~\ref{fig:colormagplot3}) --- an in-depth knowledge of their spectral properties will allow us to filter them out effectively from future UHR object searches, building a comprehensive set of templates.

On the other hand, a few of our objects (the ones discussed in Section~\ref{uhrproperties}) show a significant solution volume beyond $z>15$. Although our analysis is not sufficient to draw robust and definitive conclusions on the UHR nature of these sources, future follow-ups could enhance the UHR solution volumes for these galaxies, contributing to break the z = 15 galaxies barrier. Characterizing the statistics of galaxies beyond $z=15$ will have deep astrophysical implications, contributing to understanding how the first galaxies and their black holes formed and evolved. However, the discovery and characterization of UHR galaxies also has deep astroparticle implications. As shown by preliminary works using early release JWST data, the characterization of UHR sources falling near the rest-frame UV luminosity function's knee may help in constraining the particle properties of dark matter, to the point of succesfully discriminating between competing paradigms beyond the Cold Dark Matter framework \citep{Gandolfi:2022bcm}. 

Finally, the pipeline developed in this work to characterize the physical properties of faint, F200W-dropout objects in CEERS will be applied in other fields covered by deeper JWST observations. Finding other candidates with bimodal P(z)s would have the benefit of enlarging our stacked UHR candidates sample to further constrain the properties of these objects, while potentially outlining promising UHR candidates.

\begin{acknowledgements}
      G. G, G. R. and B. V. are supported  by the European Union --- NextGenerationEU RFF M4C2 1.1 PRIN 2022 project 2022ZSL4BL INSIGHT.
      M. G. acknowledges support from INAF under the following funding schemes: Large Grant 2022 (project "MeerKAT and LOFAR Team up: a Unique Radio Window on Galaxy/AGN co-Evolution") and Large GO 2024 (project "MeerKAT and Euclid Team up: Exploring the galaxy-halo connection at cosmic noon").
      We thank Adam Carnall for the stimulating discussions and the suggestions to refine our \texttt{Bagpipes} SED-fitting procedure.

      We thank D. Giacobazzi for suggesting the names of our UHR galaxy candidates sample.

      Finally, we thank the anonymous referee for the helpful suggestions and comments, which contributed in improving this manuscript.
\end{acknowledgements}

\bibliographystyle{aa}
\bibliography{main.bib}

\begin{appendix}

\section{Catalogs}\label{A0|catalogs}
Here, we describe the catalogs exploited in our analysis: the updated CEERS catalog and the ASTRODEEP-JWST catalog, hilighting key differences.

\subsection{The updated CEERS catalog}
The photometry process on which this catalog is based broadly follows what is described in \cite{2024ApJ...969L...2F}, and we summarize here the salient passages. This catalog, which relies on the recent CEERS DR 1.0, was built by PSF-matching ACS and NIRCam images blueward than F277W to the F277W image, while computing appropriate correction factors for images with larger PSFs. Photometry was then performed with Source Extractor (SExtractor; \citealt{1996A&AS..117..393B}; v2.25.0) in two image mode, using the co-added F277W+F356W native resolution image as the detection image. Co-addition was performed exploiting a weighted sum, using pipeline-generated error (ERR) map weights (i.e., error maps including Poisson noise) for the measurement image and 1/sqrt(WHT) weights (i.e., effective RMS maps not including Poisson noise) for the measurement one, so that Poisson noise, included in the ERR maps, does not affect detections. A first hot-mode run (i.e., a run optimized for the detection of faint objects; see, e.g., \citealt{2013ApJS..206...10G}) is performed setting \texttt{DETECT\textunderscore THRESH=1.2}, \texttt{DETECT\textunderscore MINAREA}=5, \texttt{DEBLEND\textunderscore NTHRESH}=32 and \texttt{DEBLEND\textunderscore MINCONT}=0.0001, with these values derived iteratively via visual inspection to minimize spurious sources while maximizing completeness. Then, a second cold-mode run (i.e., optimized to extract and deblend efficiently the brightest sources) is performed adopting \texttt{DETECT\textunderscore THRESH}=2.4, \texttt{DETECT\textunderscore MINAREA}=50, \texttt{DEBLEND\textunderscore NTHRESH}=32 and \texttt{DEBLEND\textunderscore MINCONT}=0.01. Both runs exploit a 4-pixel-wide top-hat convolution kernel with a kernel size of 5$\times$5 pixels. Fiducial fluxes are measured in small Kron apertures (Kron factor = 1.1 and minimum radius = 1.6, as in \citealt{2023ApJ...946L..13F}), applying two aperture corrections --- one based on the ratio of the default larger \texttt{MAG\textunderscore AUTO} Kron aperture to our small aperture and the second to compensate for the flux missed on larger scales, derived by source injection simulations. Flux uncertainties were estimated by randomly placing non-overlapping apertures, with diameters ranging from 0.1'' to 1.5'', in empty areas of the image, avoiding the segmentation map. The Normalized Median Absolute Deviation (NMAD) was then fitted exploiting a four-parameter function, repeating the fit for each aperture as a function of their area. The flux error for each object was determined using this fit based on the area of the object's aperture. To build the final catalog, cold mode objects were added to hot objects not included in the cold catalog. Moreover, a third run exploiting F444W observations as detection image was added to the two pre-existing catalogs. The final updated CEERS catalog consists of 62,752 hot run sources, 30,703 cold run sources and 11,326 F444W sources, for a total of 104,781 objects.

\subsection{The CEERS ASTRODEEP-JWST catalog}
The other part of our sample of objects relies on the CEERS ASTRODEEP-JWST catalog \citep{2024arXiv240900169M}. ASTRODEEP-JWST gathers data from eight JWST NIRCam observational programs (GLASS-JWST, UNCOVER, DDT2756, GO3990, PRIMER, JADES, NGDEEP and CEERS DR 0.5 and 0.6), exploiting carefully chosen detection parameters designed to maximize the detection of high-redshift faint extended sources. To build this catalog, the astrometries of JWST and HST images were re-matched down to the milli-arcsec level. The RMS and ERR maps were then adjusted by applying a correction factor, which was determined by comparing the flux dispersion of a set of simulated point-like sources (generated using \texttt{WebbPSF}) with their nominal errors. Science images were then smoothed with a Gaussian convolution filter with FWHM = 0.14'' (close to the F356W and F444W FWHMs) to optimize for lower surface brightness features. Due to this smoothing, the CEERS ASTRODEEP-JWST catalog may be more sensitive to extended sources, while the updated CEERS one is more sensitive to point sources. Photometry was then performed via SExtractor (v2.8.6) adopting a weighted stack of the convolved F356W and F444W images as detection image. The SExtractor run parameters closely resemble the ones used for the CANDELS campaign \citep{2011ApJS..197...36K, 2011ApJS..197...35G}, adopting \texttt{DETECT\textunderscore MINAREA}=5, \texttt{DEBLEND\textunderscore NTHRESH}=32 and \texttt{DEBLEND\textunderscore MINCONT}=0.0001 (see Table 3 of \citealt{2024arXiv240900169M} for further details). The final CEERS ASTRODEEP-JWST catalog amounts to 82,547 objects (flagged as detections with a S/N>2).

\section{SED-fitting priors}
We report here the priors and grids utilized during our SED-fitting runs with \texttt{Bagpipes} and \texttt{CIGALE}.

\begin{table*}
    \centering
    \caption{Uniform priors utilized for all our \texttt{Bagpipes} SED-fitting runs.}
    \begin{tabular}{p{6cm} p{2cm} p{9cm}} 
        \texttt{Bagpipes} fit parameters & Prior range & Description \\
        \cmidrule(lr){1-3}
        \textit{General} & \\
        $\log\text{M}_*/\text{M}_\odot$ & [1, 15] & Logarithmic stellar mass in solar mass units \\
        z & [0, 25] & Redshift \\
        $\text{A}_\text{v}$ & [0, 6] & Dust attenuation index (SMC attenuation law)\\
        $\log$U & [-4, -1] & Logarithmic ionization parameter \\
        Z & [0, 2.5] & Metallicity in solar units \\
        \cmidrule(lr){1-3}
        \textit{Delayed SFH} & \\
        $\text{Age}_\text{del}$ & [0.1, 14] & Time since the beginning of star formation in Gyr\\
        $\tau_\text{del}$ & [0.1, 14] & Time since the end of star formation in Gyr\\
        \cmidrule(lr){1-3}
        \textit{Exponential SFH} & \\
        $\text{Age}_\text{exp}$ & [0.1, 14] & Time since the beginning of star formation in Gyr\\
        $\tau_\text{exp}$ & [0.1, 14] & Timescale of star formation decrease in Gyr\\
        \cmidrule(lr){1-3}
        \textit{Log-normal SFH} & \\
        $\text{t}_\text{max}$ & [0.1, 15] & Age of the Universe at the star formation peak in Gyr\\
        FWHM & [0.1, 20] & Full width at half maximum star formation in Gyr\\
        \cmidrule(lr){1-3}
        \textit{Double powerlaw SFH} & \\
        $\alpha$ & [0.1, 1000] & Falling slope index\\
        $\beta$ & [0.1, 1000] & Rising slope index \\
        $\tau_\text{dbl}$ & [0.1, 14] & Age of the Universe at turnover in Gyr\\
        \cmidrule(lr){1-3}
        \label{tab:priors}
    \end{tabular}
\end{table*}

\begin{table*}
    \centering
    \caption{Grid utilized for the free parameters of all our \texttt{CIGALE} SED-fitting runs.}
    \begin{tabular}{p{9.5cm} p{3.5cm} p{4cm}} 
        \texttt{CIGALE} fit parameters & Grid values & Description \\
        \cmidrule(lr){1-3}
        \multicolumn{3}{l}{\textit{Double exponential SFH} [\texttt{sfh2exp} module]} \\
        $\tau_\text{main}$ & 1, 2, 6, 10, 13 & e-folding time of the main stellar population model in Gyr \\
        Age & 100, 500, 2000, 6000, 13000 & Age of the main stellar population in the galaxy in Myr \\
        \cmidrule(lr){1-3}
        \multicolumn{3}{l}{\textit{SSP component} [\texttt{bc03} module]} \\
        Z & 0.0001, 0.004, 0.008, 0.02, 0.05 & Metallicity\\
        \cmidrule(lr){1-3}
        \multicolumn{3}{l}{\textit{Nebular component} [\texttt{nebular} module]} \\
        $\log \text{U}$ & -4, -3, -2, -1 & Logarithmic ionization parameter \\
        \cmidrule(lr){1-3}
        \multicolumn{3}{l}{\textit{Dust attenuation component} [\texttt{dustatt\_modified\_CF00} module]} \\
        $A_\text{V, ISM}$ & 0.01, 0.1, 0.5, 1, 1.5, 2.5, 4, 5, 6 & V-band attenuation in the interstellar medium\\
        \cmidrule(lr){1-3}
        \multicolumn{3}{l}{\textit{AGN component} [\texttt{fritz2006} module]} \\
        $\beta$ & -1, -0.5, 0 & Dust density distribution parameter \\
        $\gamma$ & 0, 2, 4 & Dust density distribution parameter \\
        $\Psi$ & 0.001, 50.1, 80.1 & Angle between the equatorial axis and line of sight \\
        $\text{f}_\text{AGN}$ & 0., 0.1, 0.25, 0.5, 0.75 & AGN fraction \\
        \cmidrule(lr){1-3}
        \multicolumn{3}{l}{\textit{Redshifting component} [\texttt{redshifting} module]} \\
        z & [0 - 25] in steps of 0.2
        & Redshift of the source \\
        \cmidrule(lr){1-3}
        \label{tab:cigalegrid}
    \end{tabular}
    \tablefoot{In this table we limit ourselves to showing only the parameters we allowed to vary in our analysis. All the remaining parameters not included in the table are fixed and their value coincides with the default ones assigned by \texttt{CIGALE}, available in \cite{2019A&A...622A.103B}.}
\end{table*}

\begin{landscape}
\section{Photometry}\label{A1|photometry}
We report here the photometry of the F200W-dropouts found in our analysis in all the available HST (top table) and NIRCam (bottom table) bands.

\begin{table}[h]
    \centering
    \caption{Photometry of our F200W-dropouts.}
    \begin{tabular}{|c|c|c|c|c|c|c|c|c|}
        \hline
        ID & Names & $\text{f}_{\text{F435W}} \, \text{[nJy]}$ & $\text{f}_{\text{F606W}} \, \text{[nJy]}$ & $\text{f}_{\text{F814W}} \, \text{[nJy]}$ & $\text{f}_{\text{F105W}} \, \text{[nJy]}$ & $\text{f}_{\text{F125W}} \, \text{[nJy]}$ & $\text{f}_{\text{F140W}} \, \text{[nJy]}$ & $\text{f}_{\text{F160W}} \, \text{[nJy]}$ \\
        \hline
        U-31863  & BUDIARA & $-$ & $-3.60 \pm 5.20$ & $2.39 \pm 5.90$ & $-$ & $-$ & $-$ & $-$ \\
        U-34120  & VICIADGO & $-$ & $5.23 \pm 5.04$ & $-0.78 \pm 3.79$ & $-$ & $2.78 \pm 5.82$ & $-12.64 \pm 14.84$ & $-23.42 \pm 5.60$ \\
        U-53105  & - & $-0.16 \pm 4.45$ & $3.72 \pm 4.20$ & $0.89 \pm 4.95$ & $-$ & $-8.68 \pm 7.16$ & $23.44 \pm 12.48$ & $-14.71 \pm 6.42$ \\
        U-75985  & LIZZAN & $6.49 \pm 4.21$ & $2.83 \pm 1.79$ & $3.54 \pm 2.84$ & $-$ & $-6.22 \pm 3.19$ & $-0.41 \pm 4.01$ & $-0.58 \pm 2.75$ \\
        U-80918  & NASPE & $1.41 \pm 5.50$ & $0.82 \pm 4.33$ & $-3.99 \pm 5.51$ & $-$ & $12.79 \pm 7.30$ & $12.96 \pm 13.28$ & $-1.67 \pm 6.48$ \\
        U-100588 & - & $0.96 \pm 4.16$ & $-3.88 \pm 5.46$ & $0.20 \pm 5.79$ & $-$ & $11.57 \pm 7.28$ & $-3.61 \pm 13.06$ & $-6.38 \pm 6.68$ \\
        U-106373 & - & $-1.85 \pm 5.60$ & $0.50 \pm 3.81$ & $1.17 \pm 6.25$ & $-$ & $4.08 \pm 4.60$ & $-$ & $7.77 \pm 4.18$ \\
        U-112842 & CURION & $-$ & $-1.33 \pm 3.25$ & $-1.35 \pm 4.02$ & $-10.16 \pm 6.45$ & $6.15 \pm 5.87$ & $6.68 \pm 9.75$ & $-13.74 \pm 4.78$ \\
        A-22691 & ARCERIO & $-$ & $3.06 \pm 3.55$ & $4.13 \pm 3.66$ & $-$ & $5.74 \pm 10.02$ & $-$ & $5.39 \pm 8.89$\\
        A-26130 & - & $-$ & $0.00 \pm 5.17$ & $-6.18 \pm 5.62$& $-$ & $17.32 \pm 14.25$& $-$ & $12.02 \pm 11.93$\\
        A-76468 & - & $-$ & $-1.23 \pm 4.62$& $5.85 \pm 7.08$ & $-$ & $12.42 \pm 12.55$& $-$ & $-9.45 \pm 10.53$ \\
        \hline
    \end{tabular}
    \label{tab:photometry}
\end{table}

\begin{table}[h]
    \centering
    \begin{tabular}{|c|c|c|c|c|c|c|c|c|c|c|}
        \hline
        ID & Names & \( \text{f}_{\text{F090W}} \, \text{[nJy]} \) & \( \text{f}_{\text{F115W}} \, \text{[nJy]} \) & \( \text{f}_{\text{F150W}} \, \text{[nJy]} \) & \( \text{f}_{\text{F200W}} \, \text{[nJy]} \) & \( \text{f}_{\text{F277W}} \, \text{[nJy]} \) & \( \text{f}_{\text{F356W}} \, \text{[nJy]} \) & \( \text{f}_{\text{F410M}} \, \text{[nJy]} \) & \( \text{f}_{\text{F444W}} \, \text{[nJy]} \) \\
        \hline
        U-31863 & BUDIARA & $-$ & $5.64 \pm 3.55$ & $4.61 \pm 2.53$ & $3.18 \pm 1.98$ & $10.18 \pm 1.33$ & $11.66 \pm 1.49$ & $-$ & $11.70 \pm 1.56$ \\
        U-34120 & VICIADGO & $5.78 \pm 3.46$ & $3.14 \pm 1.73$ & $2.84 \pm 2.54$ & $1.57 \pm 1.91$ & $10.04 \pm 1.84$ & $9.28 \pm 1.32$ & $12.34 \pm 2.49$ & $7.30 \pm 1.30$ \\
        U-53105 & - & $3.31 \pm 3.97$ & $3.74 \pm 3.62$ & $7.66 \pm 4.04$ & $6.33 \pm 3.53$ & $9.41 \pm 2.10$ & $13.24 \pm 2.02$ & $16.02 \pm 4.58$ & $23.06 \pm 3.91$ \\
        U-75985 & LIZZAN & $1.15 \pm 1.17$ & $1.43 \pm 2.01$ & $-0.54 \pm 1.89$ & $1.64 \pm 1.17$ & $5.47 \pm 1.07$ & $9.40 \pm 1.23$ & $3.38 \pm 2.13$ & $4.32 \pm 1.03$ \\
        U-80918 & NASPE & $4.14 \pm 3.60$ & $-$ & $-10.81 \pm 9.89$ & $3.98 \pm 2.81$ & $12.56 \pm 2.13$ & $16.43 \pm 2.43$ & $14.88 \pm 3.94$ & $8.44 \pm 2.06$ \\
        U-100588 & - & $-2.86 \pm 4.14$ & $-1.75 \pm 3.02$ & $5.01 \pm 4.11$ & $0.58 \pm 3.34$ & $-1.12 \pm 2.18$ & $-0.93 \pm 2.11$ & $23.04 \pm 5.03$ & $27.83 \pm 2.22$ \\
        U-106373 & - & $2.83 \pm 1.88$ & $2.75 \pm 3.08$ & $3.08 \pm 2.83$ & $4.31 \pm 2.34$ & $1.03 \pm 1.35$ & $6.67 \pm 1.86$ & $18.34 \pm 2.64$ & $17.20 \pm 1.86$ \\
        U-112842 & CURION & $-$ & $0.83 \pm 2.41$ & $0.99 \pm 2.80$ & $-0.49 \pm 1.74$ & $-0.81 \pm 1.77$ & $5.18 \pm 1.46$ & $31.34 \pm 3.37$ & $32.08 \pm 1.64$ \\
        A-22691 & ARCERIO & $-$ & $1.02 \pm 2.16$ & $0.59 \pm 2.09$   & $2.65 \pm 1.79$ & $7.9 \pm 1.83$ & $9.32 \pm 1.66$ & $-$ & $8.99 \pm 2.97$\\
        A-26130 & - & $-$ & $4.05 \pm 2.95$ & $1.09 \pm 4.63$ & $0.98 \pm 3.63$   & $11.21 \pm 3.12$ & $14.38 \pm 2.23$& $25.04 \pm 5.45$ & $19.46 \pm 4.87$\\
        A-76468 & - & $-$ & $4.37 \pm 2.94$ & $3.75 \pm 3.41$ & $3.56 \pm 2.79$   & $6.55 \pm 2.12$ & $8.87 \pm 2.13$ & $18.45 \pm 4.25$ & $24.66 \pm 3.36$\\
        \hline
    \end{tabular}
    \tablefoot{We flagged with ``-'' all those observations that are not available in a certain band. Negative flux values are likely produced by local over-subtractions of the background, and they were treated in our analysis as upper limits.}
\end{table}
\end{landscape}

\begin{landscape}
\section{Physical properties of the UHR galaxy candidates sample}\label{uhrbestfitresults}
We report in this section the inferred physical properties yielded by our fiducial \texttt{Bagpipes} runs assuming a SMC dust attenuation law for both the $z > 8$ (top table) and $z \leq$ 8 (bottom table) solutions, alongside the inferred best-fit $\chi^2$ for both cases.

    \begin{table}[h]
        \centering
        \small
        \caption{\texttt{Bagpipes} fiducial best-fit parameters for our UHR candidates sample (z > 8 solutions).}
        \setlength{\tabcolsep}{6pt}
        \begin{tabular}{ccccccccccc}
        \hline \hline
        ID & Name & $\text{z}_\text{high}$ & $\log\text{M}_\text{high}/\text{M}_\odot$ & $\text{Z}_\text{high}/\text{Z}_\odot$ & $\text{A}_\text{v, high}$ & $\log \text{U}_\text{high}$ & $\text{SFR}_\text{high}$ [$\text{M}_\odot \ \text{yr}^{-1}$] & Best-fit SFH & SFH shape parameters [Gyr] & $\chi^2_{\text{best, high}}$\\
        \hline
        \vspace{2pt}
        U-31863 & BUDIARA & $17.76^{+1.6}_{-1.5}$ & $8.12^{+0.5}_{-0.3}$ & $1.24^{+0.9}_{-0.8}$ & $0.17^{+0.13}_{-0.10}$ & $-2.46^{+1.0}_{-1.1}$ & $1.31^{+1.8}_{-0.7}$ & log & $\text{FWHM} = 11.54^{+5.8}_{-6.3}$, $\text{t}_\text{max} = 7.67^{+5.0}_{-5.0}$ & 3.71 \\
        U-34120 & VICIADGO & $17.56^{+1.4}_{-1.3}$ & $7.91^{+0.5}_{-0.3}$ & $1.15^{+0.9}_{-0.8}$ & $0.09^{+0.12}_{-0.07}$ & $-2.37^{+0.9}_{-1.0}$ & $0.89^{+1.0}_{-0.5}$ & log & $\text{FWHM} = 11.97^{+5.4}_{-6.3}$, $\text{t}_\text{max} = 7.90^{+4.8}_{-5.0}$ & 3.25 \\
        U-75985 & LIZZAN & $17.80^{+1.48}_{-1.45}$ & $7.63^{+0.42}_{-0.28}$ & $1.07^{+0.95}_{-0.78}$ & $0.08^{+0.11}_{-0.06}$ & $-2.39^{+0.95}_{-1.07}$ & $0.42^{+0.64}_{-0.19}$ & log & $\text{FWHM} = 11.09^{+5.89}_{-6.06}$, $\text{t}_\text{max} = 8.47^{+4.41}_{-5.18}$ & 14.81 \\
        U-80918 & NASPE & $17.55^{+1.47}_{-1.30}$ & $8.00^{+0.43}_{-0.29}$ & $1.17^{+0.88}_{-0.82}$ & $0.07^{+0.10}_{-0.05}$ & $-2.35^{+0.93}_{-1.08}$ & $1.07^{+1.07}_{-0.61}$ & log & $\text{FWHM} = 11.39^{+5.69}_{-6.18}$, $\text{t}_\text{max} = 8.14^{+4.61}_{-4.98}$ & 7.28 \\
        A-22691 & ARCERIO & $17.32^{+2.05}_{-2.98}$ & $8.19^{+0.47}_{-0.42}$ & $1.27^{+0.83}_{-0.86}$ & $0.27^{+0.34}_{-0.18}$ & $-2.45^{+0.96}_{-1.01}$ & $1.50^{+2.33}_{-0.92}$ & log & $\text{FWHM} = 11.44^{+5.65}_{-6.23}$, $\text{t}_\text{max} = 7.70^{+4.66}_{-4.95}$ & 3.01 \\
        \hline
        \end{tabular}
        \label{tab:bestfitsingle}
    \end{table}

    \begin{table}[h]
        \centering
        \small
        \caption{\texttt{Bagpipes} fiducial best-fit parameters for our UHR candidates sample (z $\leq$ 8 solutions).}
        \setlength{\tabcolsep}{6pt}
        \begin{tabular}{ccccccccccc}
        \hline \hline
        ID & Name & $\text{z}_\text{low}$ & $\log\text{M}_\text{low}/\text{M}_\odot$ & $\text{Z}_\text{low}/\text{Z}_\odot$ & $\text{A}_\text{v, low}$ & $\log \text{U}_\text{low}$ & $\text{SFR}_\text{low}$ [$\text{M}_\odot \ \text{yr}^{-1}$] & Best-fit SFH & SFH shape parameters [Gyr] & $\chi^2_{\text{best, low}}$ \\
        \hline
        \vspace{2pt}
        U-31863 & BUDIARA & $1.01^{+0.32}_{-0.19}$ & $7.68^{+0.3}_{-0.3}$ & $1.28^{+0.8}_{-0.8}$ & $4.60^{+1.0}_{-1.8}$ & $-2.47^{+1.0}_{-1.0}$ & $0.012^{+0.02}_{-0.01}$ & exp & $\text{Age}_\text{exp} = 2.96^{+2.0}_{-1.9}$, $\tau_\text{exp} = 7.02^{+4.7}_{-4.9}$ & 4.52 \\
        U-34120 & VICIADGO & $0.91^{+0.26}_{-0.15}$ & $7.49^{+0.2}_{-0.3}$ & $1.40^{+0.8}_{-0.9}$ & $4.47^{+1.1}_{-1.6}$ & $-2.47^{+1.0}_{-1.0}$ & $0.006^{+0.01}_{-0.003}$ & exp & $\text{Age}_\text{exp} = 3.44^{+2.0}_{-2.2}$, $\tau_\text{exp} = 7.06^{+4.8}_{-4.9}$ & 6.84 \\
        U-75985 & LIZZAN & $4.31^{+0.34}_{-3.49}$ & $7.23^{+0.27}_{-0.24}$ & $0.77^{+1.1}_{-0.62}$ & $2.04^{+2.65}_{-0.67}$ & $-2.56^{+0.92}_{-0.96}$ & $0.11^{+0.11}_{-0.11}$ & log & $\text{FWHM} = 4.45^{+9.87}_{-2.56}$, $\text{t}_\text{max} = 9.96^{+3.34}_{-5.63}$ & 10.38 \\
        U-80918 & NASPE & $0.89^{+0.4}_{-0.15}$ & $7.45^{+0.42}_{-0.27}$ & $1.40^{+0.75}_{-0.88}$ & $4.00^{+1.26}_{-2.22}$ & $-2.56^{+1.08}_{-1.01}$ & $0.013^{+0.029}_{-0.007}$ & del & $\text{Age}_\text{del} = 3.33^{+2.10}_{-2.60}$, $\tau_\text{del} = 6.95^{+5.06}_{-5.07}$ & 5.06 \\
        A-22691 & ARCERIO & $1.28^{+1.54}_{-0.37}$ & $7.74^{+0.49}_{-0.37}$ & $1.29^{+0.83}_{-0.87}$ & $4.01^{+1.39}_{-2.07}$ & $-2.49^{+1.00}_{-1.05}$ & $0.02^{+0.10}_{-0.01}$ & exp & $\text{Age}_\text{exp} = 2.14^{+2.10}_{-1.45}$, $\tau_\text{exp} = 6.92^{+4.79}_{-4.70}$ & 3.68 \\
        \hline
        \end{tabular}
        \tablefoot{IDs starting with ``U-'' were selected in the updated CEERS catalog, whereas those starting with ``A-'' are from the CEERS ASTRODEEP-JWST catalog. In the best-fit SFH columns we used ``exp'' to indicate a simple exponential SFH, ``del'' for a delayed exponential SFH and ``log'' for a log-normal SFH.}
    \end{table}\label{bestfitsingle2}
\end{landscape}
\section{Stacked Cutouts}\label{appendix:stack}
\begin{figure*}[b]
    \centering
    \includegraphics[width=0.475\textwidth]{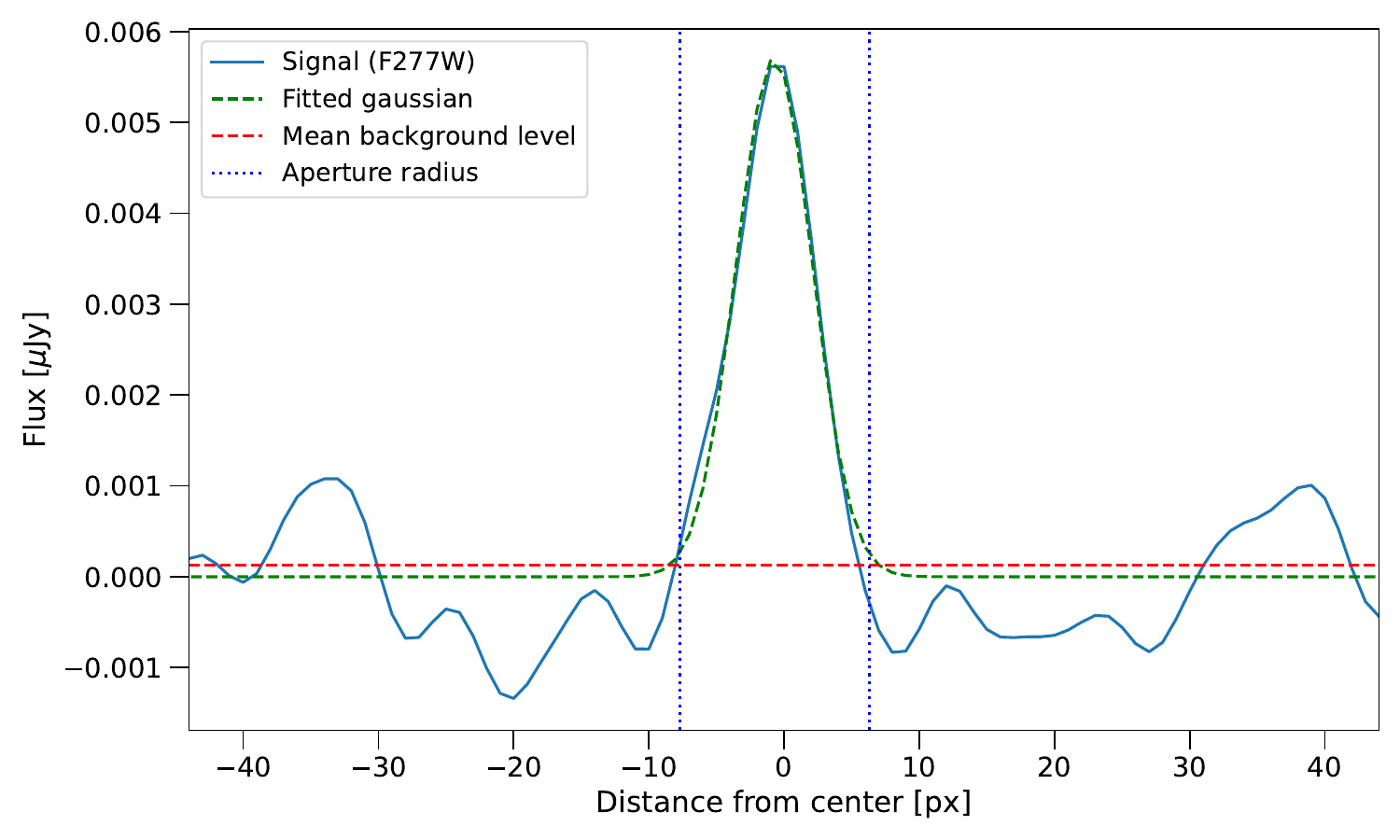}
    \includegraphics[width=0.475\textwidth]{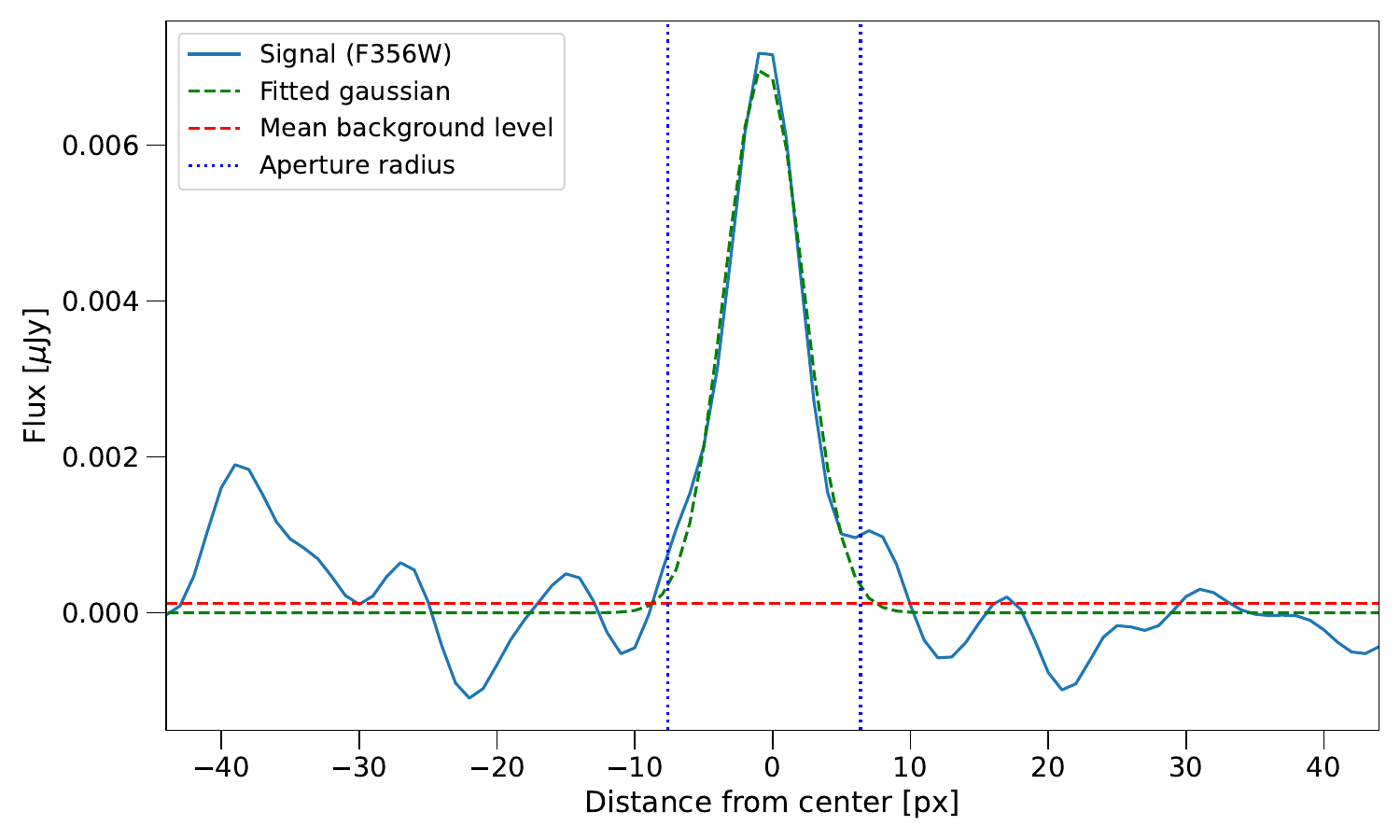}
    \includegraphics[width=0.475\textwidth]{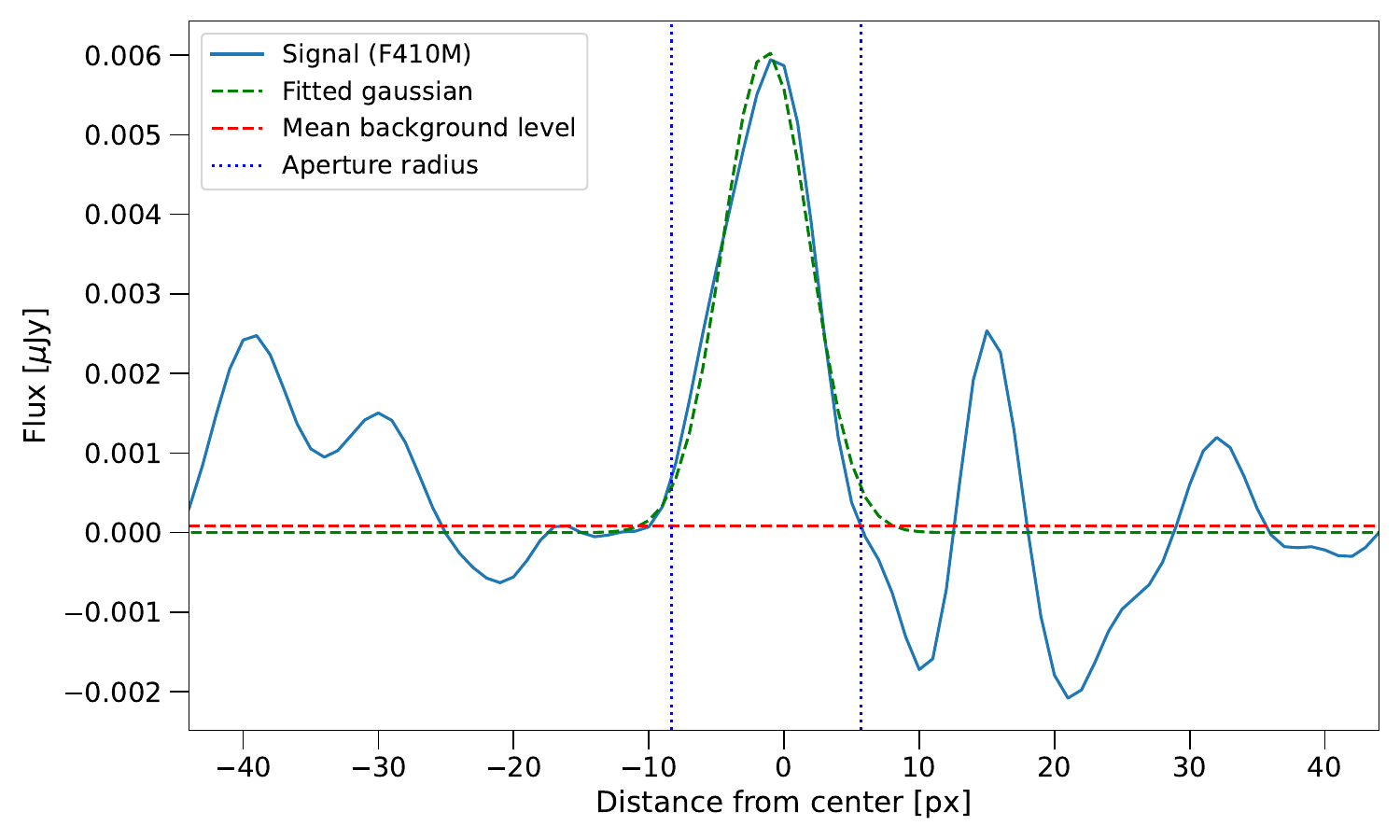}
    \includegraphics[width=0.475\textwidth]{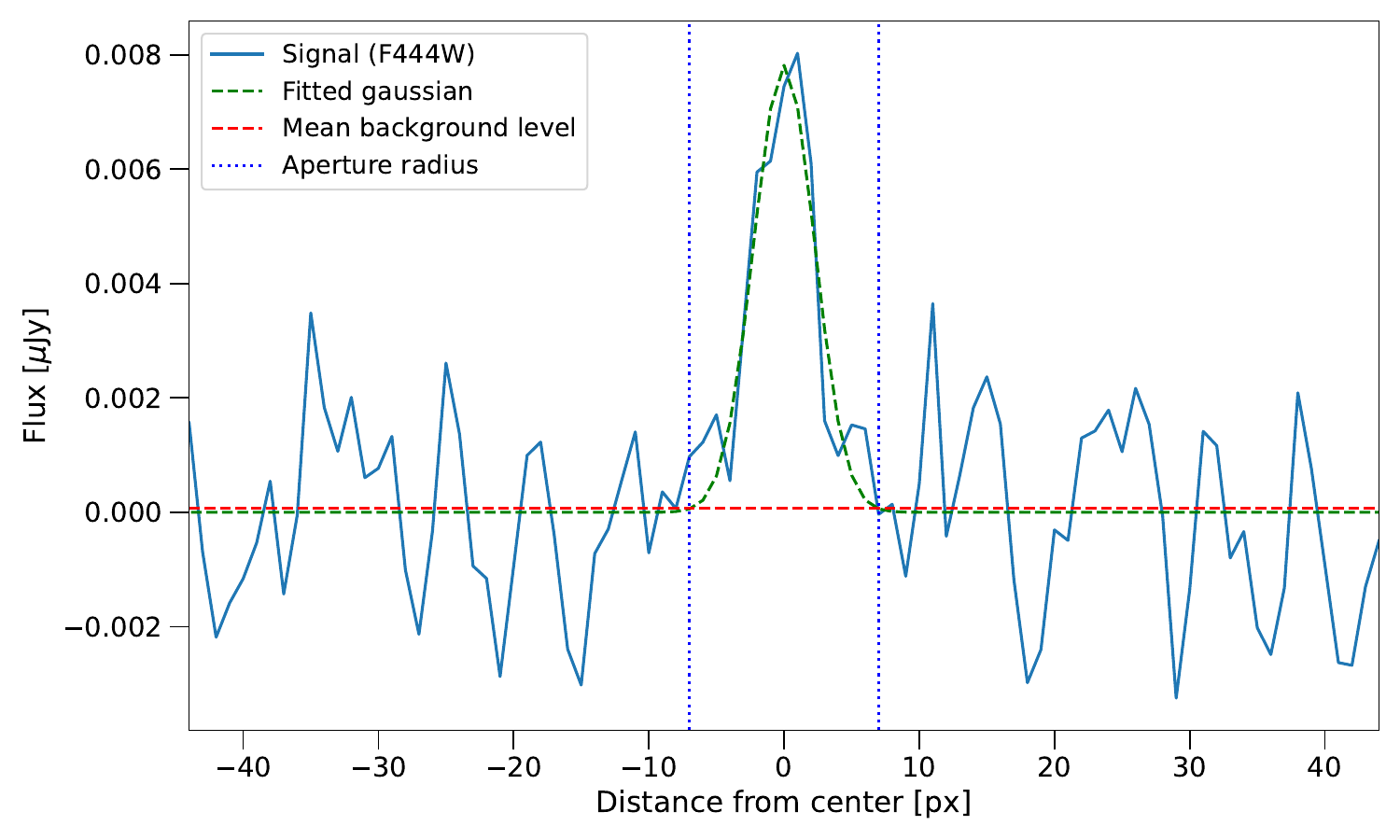}

    \caption{Signal in the convolved long-wavelength NIRCam cutouts is shown for our stacked UHR candidate sample, with insets corresponding to F277W (top left), F356W (top right), F410M (bottom left), and F444W (bottom right). The blue line represents the signal, in $\mu$Jy, for each pixel in the stacked image as a function of its distance from the center of the cutout (measured in pixels). A green dashed line depicts the Gaussian fit used to estimate the aperture photometry radius, while a red dashed line indicates the mean background level. The vertical blue dashed lines mark the inferred radius for fiducial aperture photometry.}
    \label{fig:aperture_photometry}
\end{figure*}

This appendix lists the salient passages we followed while stacking our UHR candidates sample. We used \texttt{CosMix} to create 2.7'' $\times$ 2.7'' (i.e., 90 $\times$ 90 pixels) median cutouts for our UHR candidates sample. We then degraded the PSF of such stacked cutouts to match the F444W's PSF --- this filter, the reddest available, has the largest PSF with a FWHM of 0.161''. All NIRCam filters from F090W to F444W and the three HST/ACS filters (F435W, F606W and F814W) have smaller FWHMs than F444W, meaning they will be PSF-matched (see Table~\ref{tab:depths}). However, the HST/WFC3 filters have larger FWHMs and thus do not require convolution. PSFs for each filter were obtained by identifying PSF stars in the CEERS images, following the procedure described by \cite{2022ApJ...928...52F}. We then used the \texttt{PYPHER}\footnote{\url{https://pypher.readthedocs.io}} Python routine (Boucaud et al. \citeyear{2016ascl.soft09022B}) to create convolution kernels and match the PSFs of all filters to F444W. The images were then convolved with their respective kernels to ensure consistency across all bands. After obtaining the PSF-matched stacked cutouts, we needed to select an appropriate photometry aperture size. We determined the aperture size by fitting a Gaussian function to the radial flux profiles for each band featuring a detection (see Figure~\ref{fig:aperture_photometry}). For each band, we used a circular region centered on the stacked source with a radius equal to one-third of the cutout size, ensuring the source was well-contained within this area. We then measured the average pixel value outside this circular region to estimate the average stacked background level. Using this background estimate, we interpolated each Gaussian profile to find the radius where the flux matched the background level, repeating the process for each band. We then averaged these aperture sizes across the different bands to obtain a fiducial aperture size passing this value to \texttt{CosMix} to compute the stacked photometry, finding a fiducial value of 0.21'' (i.e., 7 pixels). We then used \texttt{CosMix} to perform aperture photometry on the chosen aperture in each stacked band. Stacked photometry errors were estimated using a bootstrap technique, which involves random resampling with replacement from the measured fluxes to create multiple synthetic datasets. The median flux is calculated for each bootstrap sample, resulting in a distribution of medians. The overall median of these medians estimates the central tendency, while the standard deviation provides an approximation of the uncertainty in the measured flux. This method effectively captures variability and accommodates outliers or non-normal distributions in the data, yielding a robust error estimate. We summarize the so-obtained photometry in Table~\ref{tab:stacked_photometry}, while SED-fitting results are reported in Table~\ref{tab:bestfitstacked_properties} adopting a SMC dust attenuation law. Note that not all objects are covered in all bands. During stacking, we excluded from SED-fitting any stacked cutouts created from fewer than 2/3 of the total objects (e.g., the F105W band, which has limited coverage in CEERS), though we included the relative flux measurements in the stacked SED plot for completeness.

\begin{landscape}

\begin{table}[h]
    \centering
    \caption{Photometry of our median stacked UHR candidates sample}
    \begin{tabular}{|c|c|c|c|c|c|c|}
        \hline
        $\text{f}_{\text{F435W}} \, \text{[nJy]}$ & $\text{f}_{\text{F606W}} \, \text{[nJy]}$ & $\text{f}_{\text{F814W}} \, \text{[nJy]}$ & $\text{f}_{\text{F105W}} \, \text{[nJy]}$ & $\text{f}_{\text{F125W}} \, \text{[nJy]}$ & $\text{f}_{\text{F140W}} \, \text{[nJy]}$ & $\text{f}_{\text{F160W}} \, \text{[nJy]}$ \\
        \hline
        - & $0.7 \pm 2.1$ & $1.4 \pm 4.1$ & - & $2.3 \pm 4.0$ & $7.6 \pm 13.9$ & $-2.3 \pm 4.1$ \\
        \hline
    \end{tabular}
    \label{tab:stacked_photometry}

\end{table}

\begin{table}[h]
    \centering
    \begin{tabular}{|c|c|c|c|c|c|c|c|c|}
        \hline
        \( \text{f}_{\text{F090W}} \, \text{[nJy]} \) & \( \text{f}_{\text{F115W}} \, \text{[nJy]} \) & \( \text{f}_{\text{F150W}} \, \text{[nJy]} \) & \( \text{f}_{\text{F200W}} \, \text{[nJy]} \) & \( \text{f}_{\text{F277W}} \, \text{[nJy]} \) & \( \text{f}_{\text{F356W}} \, \text{[nJy]} \) & \( \text{f}_{\text{F410M}} \, \text{[nJy]} \) & \( \text{f}_{\text{F444W}} \, \text{[nJy]} \) \\
        \hline
        $3.9 \pm 3.5$ & $2.8 \pm 1.0$ & $1.6 \pm 3.0$ & $1.7 \pm 1.0$ & $7.0 \pm 1.3$ & $9.2 \pm 1.6$ & $10.2 \pm 3.4$ & $4.6 \pm 1.4$\\
        \hline
    \end{tabular}
\end{table}

\begin{table}[h]
    \centering
    \caption{UHR candidates median stack fiducial best-fit parameters yielded by \texttt{Bagpipes} (z>8 solution) assuming a SMC dust attenuation law.}
    \setlength{\tabcolsep}{6pt}
    \begin{tabular}{ccccccccc}
    \hline \hline
    $\text{z}_\text{high}$ & $\log\text{M}_\text{high}/\text{M}_\odot$ & $\text{Z}_\text{high}/\text{Z}_\odot$ & $\text{A}_\text{v, high}$ & $\log \text{U}_\text{high}$ & $\text{SFR}_\text{high}$ [$\text{M}_\odot \ \text{yr}^{-1}$] & Best-fit SFH & SFH shape parameters [Gyr] & $\chi^2_{\text{best, high}}$\\
    \hline
    \vspace{2pt}
    $18.01^{+1.39}_{-1.25}$ & $7.78^{+0.46}_{-0.29}$ & $1.15^{+0.89}_{-0.82}$ & $0.094^{+0.12}_{-0.07}$ & $-2.42^{+0.97}_{-1.03}$ & $0.59^{+0.97}_{-0.29}$ & log & $\text{t}_\text{max} = 11.56^{+5.57}_{-6.28}$, $\text{FWHM} = 8.20^{+4.57}_{-5.25}$ & 5.82\\
    \hline
    \end{tabular}
    \label{tab:bestfitstacked_properties}
\end{table}

\begin{table}[h]
    \centering
    \caption{UHR candidates median stack fiducial best-fit parameters yielded by \texttt{Bagpipes} (z$\leq$8 solution).}
    \setlength{\tabcolsep}{6pt}
    \begin{tabular}{ccccccccc}
    \hline \hline
    $\text{z}_\text{low}$ & $\log\text{M}_\text{low}/\text{M}_\odot$ & $\text{Z}_\text{low}/\text{Z}_\odot$ & $\text{A}_\text{v, low}$ & $\log \text{U}_\text{low}$ & $\text{SFR}_\text{low}$ [$\text{M}_\odot \ \text{yr}^{-1}$] & Best-fit SFH & SFH shape parameters [Gyr] & $\chi^2_{\text{best, low}}$\\
    \hline
    \vspace{2pt}
    $0.96^{+0.23}_{-0.14}$ & $7.44^{+0.23}_{-0.28}$ & $1.40^{+0.75}_{-0.88}$ & $4.82^{+0.83}_{-1.25}$ & $-2.50^{+0.99}_{-1.01}$ & $0.0055^{+0.008}_{-0.003}$ & exp & $\text{Age}_\text{exp} = 3.40^{+1.76}_{-1.99}$, $\tau_\text{exp} = 6.53^{+5.05}_{-4.60}$ & 10.92\\
    \hline
    \end{tabular}
\end{table}
\end{landscape}

\begin{landscape}
\section{Other F200W-dropout}\label{appendix:otherdropouts}
We list here the physical properties inferred for CURION (U-112842) and the three HELM galaxies (U-53105, A-26130 and A-76468) found in our analysis.

\begin{table}[h]
    \centering
    \small
    \caption{\texttt{Bagpipes} fiducial best-fit parameters for the other F200W-dropouts in our sample.}
    \setlength{\tabcolsep}{6pt}
    \begin{tabular}{cccccccccccc}
    \hline \hline
    ID & Name & $\text{z}$ & $\log\text{M}/\text{M}_\odot$ & $\text{Z}/\text{Z}_\odot$ & $\text{A}_\text{v}$ & $\log \text{U}$ & $\text{SFR}$ [$\text{M}_\odot \ \text{yr}^{-1}$] & Best-fit SFH & SFH shape parameters [Gyr] & $\chi^2_{\text{best}}$ & $\chi^2_{\text{BD}}$\\
    \hline
    \vspace{2pt}
    U-53105 & - & $2.60^{+2.70}_{-1.08}$ & $8.31^{+0.64}_{-0.53}$ & $1.24^{+0.80}_{-0.77}$ & $2.94^{+1.78}_{-1.47}$ & $-2.43^{+0.98}_{-1.09}$ & $0.27^{+2.00}_{-0.23}$ & exp & $\text{Age}_\text{exp}=1.08^{+1.62}_{-0.75}$, $\tau_\text{exp}=6.97^{+4.87}_{-4.73}$ & $1.12$ & $20.97$ \\
    U-112842 & CURION & $5.53^{+0.02}_{-0.04}$ & $7.21^{+0.11}_{-0.12}$ & $1.22^{+0.57}_{-0.28}$ & $1.01^{+0.29}_{-0.28}$ & $-3.83^{+0.18}_{-0.12}$ & $0.17^{+0.05}_{-0.04}$ & log & $\text{t}_\text{max}=8.57^{+4.46}_{-4.45}$, $\text{FWHM}=1.37^{+1.11}_{-0.91}$ & 6.33 & 20.96\\
    A-26130 & - & $2.04^{+2.77}_{-0.76}$ & $8.78^{+0.62}_{-0.46}$ & $1.18^{+0.90}_{-0.78}$ & $3.25^{+1.78}_{-1.60}$ & $-2.59^{+1.10}_{-0.93}$ & $0.13^{+0.94}_{-0.11}$ & exp & $\text{Age}_\text{exp}=1.45^{+1.65}_{-1.03}$, $\tau_\text{exp}=7.18^{+4.64}_{-4.58}$ & 1.35 & 16.77 \\
    A-76468 & - & $3.63^{+4.98}_{-1.06}$ & $8.94^{+0.67}_{-0.53}$ & $1.36^{+0.73}_{-0.89}$ & $3.73^{+1.42}_{-1.83}$ & $-2.53^{+1.02}_{-1.02}$ & $1.62^{+13.24}_{-1.34}$ & exp & $\text{Age}_\text{exp}=0.67^{+0.95}_{-0.42}$, $\tau_\text{exp}=7.00^{+4.53}_{-4.89}$ & $0.99$ & 12.00 \\
    \hline
    \end{tabular}
    \tablefoot{IDs starting with ``U-'' were selected in the updated CEERS catalog, whereas those starting with ``A-'' are from the CEERS ASTRODEEP-JWST catalog. In the best-fit SFH columns we used ``exp'' to indicate a simple exponential SFH, ``del'' for a delayed exponential SFH and ``log'' for a log-normal SFH.}
    \label{tab:bestfitother}
\end{table}
    
\end{landscape}

\section{Estimating the sizes of our galaxy candidates}\label{sizes}
In this section we report estimates of the sizes of our five UHR galaxy candidates (BUIDARA, VICIADGO, LIZZAN, NASPE, ARCERIO), our three HELM candidates (U-53105, A-26130 and A-76468) and CURION. We estimated their effective radii by fitting their two-dimensional flux distribution in each NIRCam long-wavelength broadband channel with a S\'ersic profile, given by

\begin{equation*}
    I(R) = I_e \exp \left\{ q_s \left[ \left( \frac{R}{R_e} \right)^{1/n_s} - 1 \right] \right\},
\end{equation*}

with $I_e$ being the intensity calculated at the effective radius $R_e$ (i.e., the radius enclosing half of the total light of the source). The overall shape of the profile is determined by the axis ratio $q_s$ and the S\'ersic index $n_s$. To perform the fit, we utilized the code \texttt{galight}\footnote{\texttt{https://galight.readthedocs.io/en/latest/index.html}} \citep{2020ApJ...888...37D}. This code provides estimations of the angular sizes of our objects in arcseconds in each band, as well as the best-fit values of the S\'ersic profile's axis ratio $q_s$ and index $n_s$, and estimates of the effective radii in kpc upon the assumption of a reference redshift. During our \texttt{galight} runs, we set the code's detection threshold to S/N$= 1$ in order to fit the shape of our sources despite their intrinsic faintness.

We report the inferred size values in Table~\ref{tab:F277Wuhrsizes} (F277W band), Table~\ref{tab:F356Wuhrsizes} (F356W band) and Table~\ref{tab:F444Wuhrsizes} (F444W band) for our UHR candidates adopting their $z>8$ best-fit redshifts. We found that all our UHR candidates are resolved in the F277W band (except for LIZZAN), meaning that these objects are unlikely BDs (which would appear unresolved in all NIRCam bands). The only resolved source in the F356W band is VICIADGO, while the only resolved source in the F444W band is ARCERIO. Our objects seem to have consistent physical sizes (within the measurement errors) across the different bands. Deeper observations would be needed to enhance this comparison, since the intrinsic faintness of our sources translates into large errors in the effective radii estimates. The best-fit S\'ersic index values we retrieved are generally $<1$, possibly indicating that our UHR candidates are galaxies with peculiar morphologies rather than compact spheroids (see, e.g., Figure~13 in \citealt{10.1093/mnras/stad3597}). However, we caveat that our measured S\'ersic indices could be also explained as a surface brightness dimming effect on a galaxy with an intrinsic sersic index of $\sim$ 1 (see, e.g., \citealt{2024arXiv241214970W}). 

\begin{figure*}
    \centering
    \includegraphics[width=0.4\textwidth]{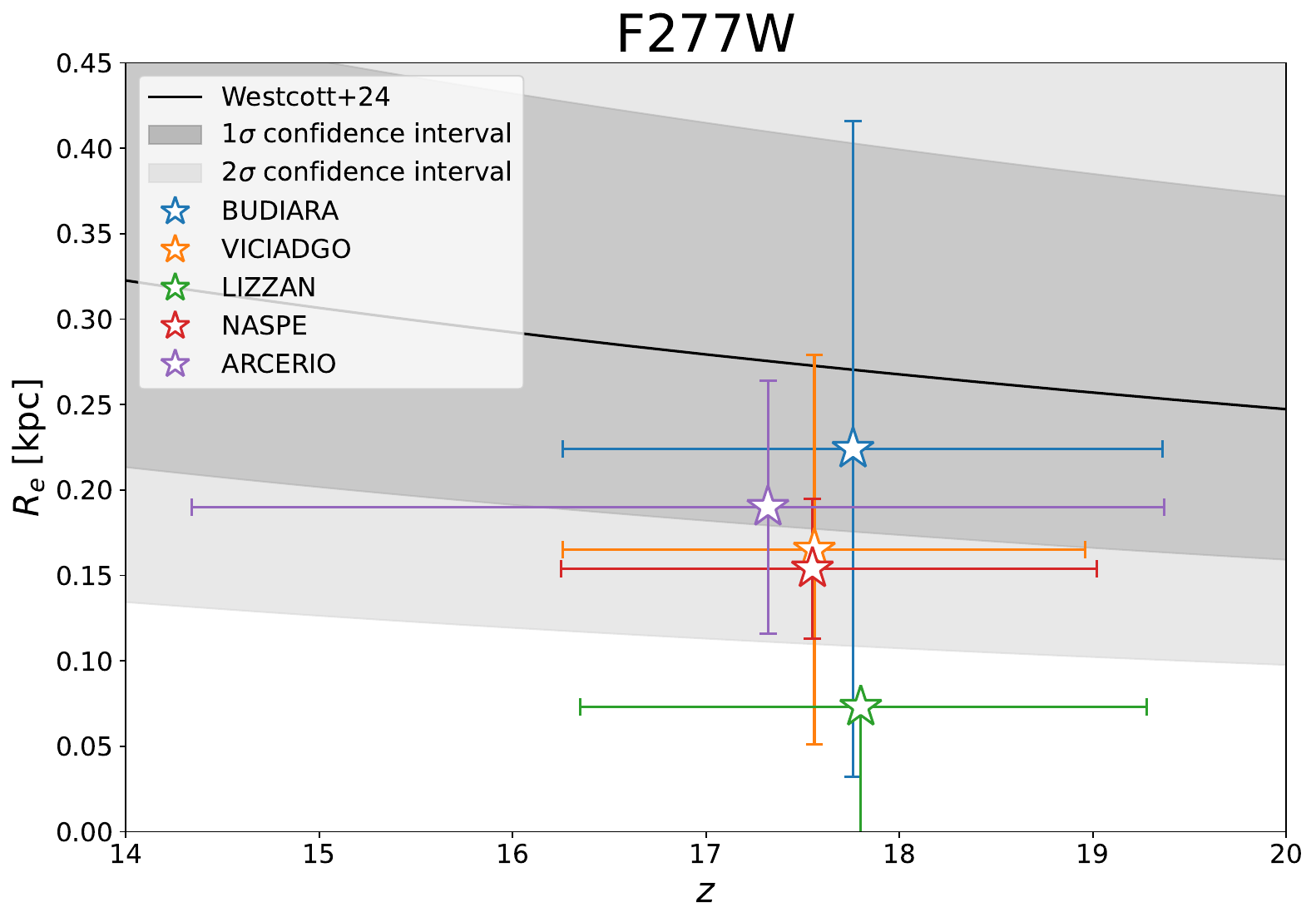}
    \includegraphics[width=0.4\textwidth]{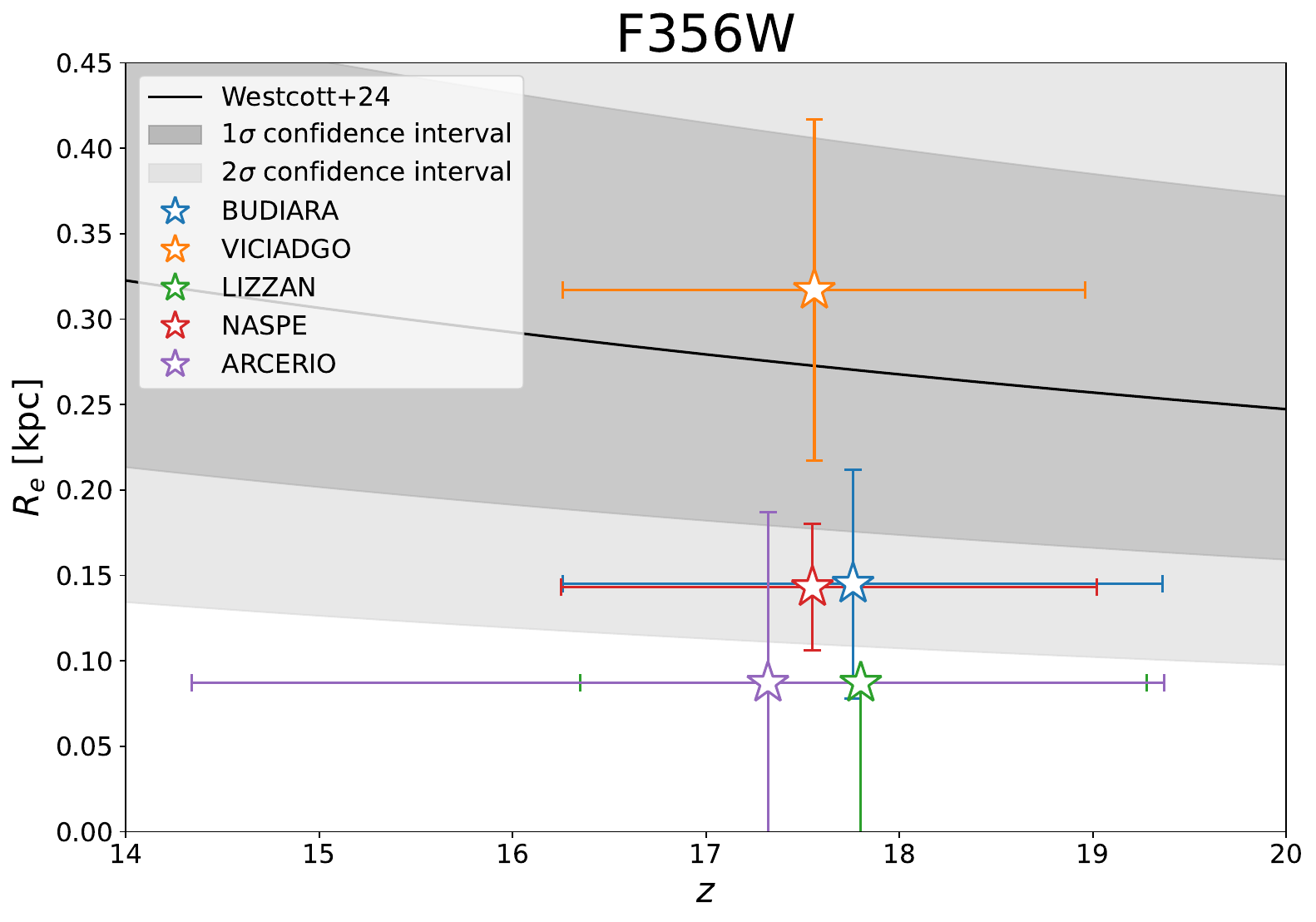}
    \includegraphics[width=0.4\textwidth]{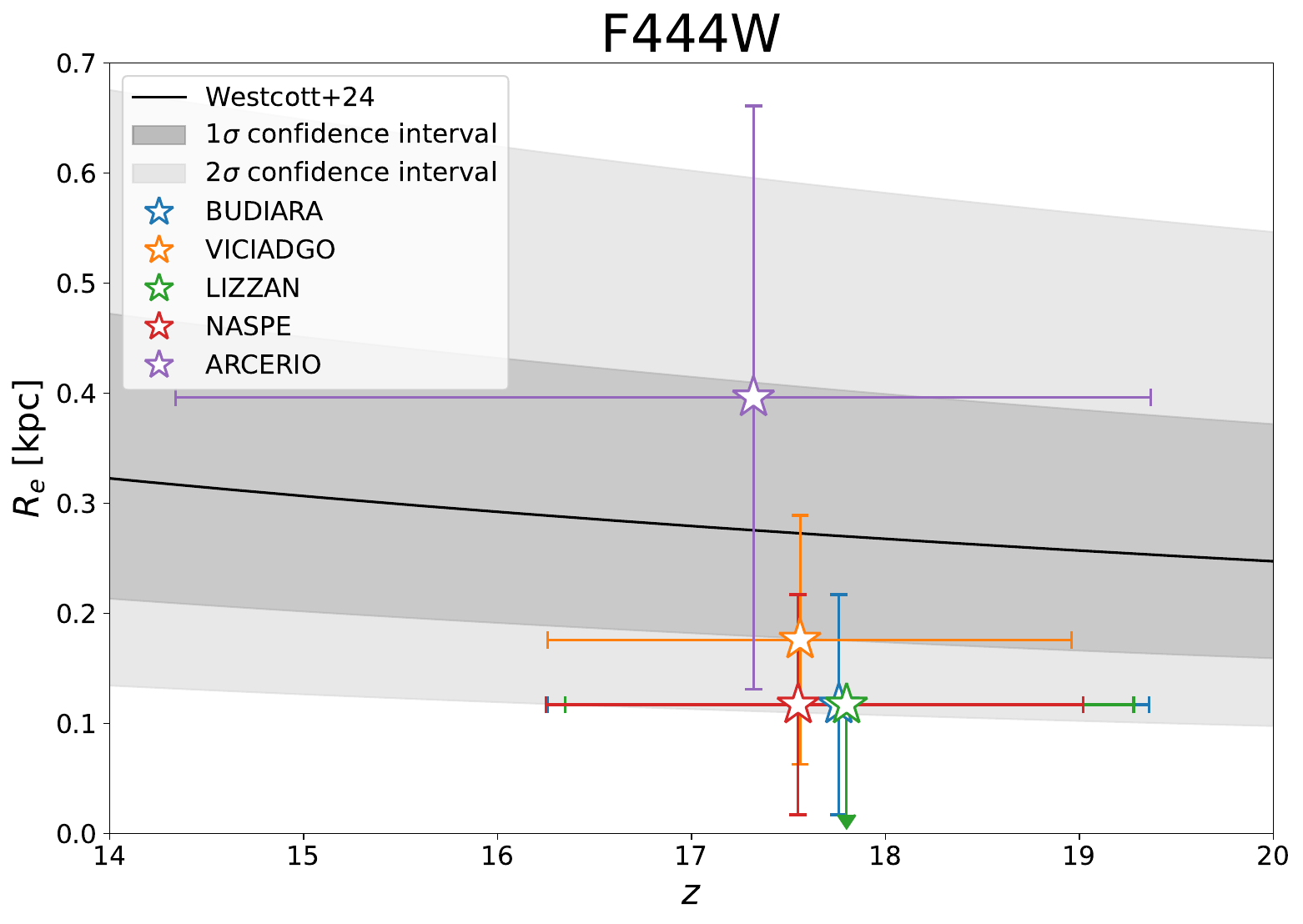}

    \caption{Effective radii (in kpc) vs redshift for our UHR candidates sample in the F277W, F356W and F444W bands respectively. We report their physical sizes estimated with \texttt{galight} vs their $z>8$ best-fit redshift solutions as colored stars. As a comparison, we report the empirical effective radius vs redshift scaling retrieved in \cite{2024arXiv241214970W} as a black, continuous line with $1\sigma$ (gray) and $2\sigma$ (light gray) confidence intervals.
    }
    \label{fig:uhr_sizes}
\end{figure*}

In the three panels of Figure~\ref{fig:uhr_sizes}, we checked the consistency of the effective radii of our UHR candidates in the F277W, F356W and F444W bands with the effective radius vs redshift relation found in \cite{2024arXiv241214970W}. Such relation was obtained by fitting the effective radii distribution of a sample of 521 galaxies between $6.5 < z < 12.5$ from different JWST-covered fields (CEERS, JADES GOOD-S, NGDEEP, SMACS0723, GLASS and PEARLS), and it reads as

\begin{equation*}
    R_e =  2.74 \pm 0.49 (1+z)^{-0.79\pm0.08}.
\end{equation*}

To allow the comparison with our UHR candidates, we extrapolated this relation to redshifts beyond $z \sim 17$ --- we stress that this redshift regime is not yet probed by spectroscopically confirmed candidates, and should be thus treated as indicative. We found that all of our UHR galaxy candidates in their $z>8$ solutions are consistent within $2\sigma$ with the relation by \cite{2024arXiv241214970W} extrapolated at ultra high-redshifts (with the only exception of LIZZAN, which is at a $3\sigma$ distance from the \citealt{2024arXiv241214970W} relation in both the F277W and F356W bands). Our analysis does not highlight any major inconsistency between our inferred best-fit UHR galaxy sizes and the extrapolated relation by \cite{2024arXiv241214970W}.

We repeated our size estimation with \texttt{galight} for our three HELM candidates and CURION, listing the obtained results in Table~\ref{tab:F277Wothersizes}, Table~\ref{tab:F356Wothersizes} and Table~\ref{tab:F444Wothersizes}. Our \texttt{galight} run was not able to converge while fitting the size of CURION in the F277W filter due to the extreme faintness of the object in this band; while all other sources are resolved in the F277W band. Instead, only U-53105 and A-76468 are resolved in the F356W band, whereas only U-53105 is resolved in the F444W band. Our angular size estimates seem to back up the hypothesis that these objects are not BDs, albeit we do not have this confirmation for CURION. Similarly to our UHR candidates, these objects showcase compatible physical sizes across the different bands considered.

\begin{figure*}
    \centering
    \includegraphics[width=0.4\textwidth]{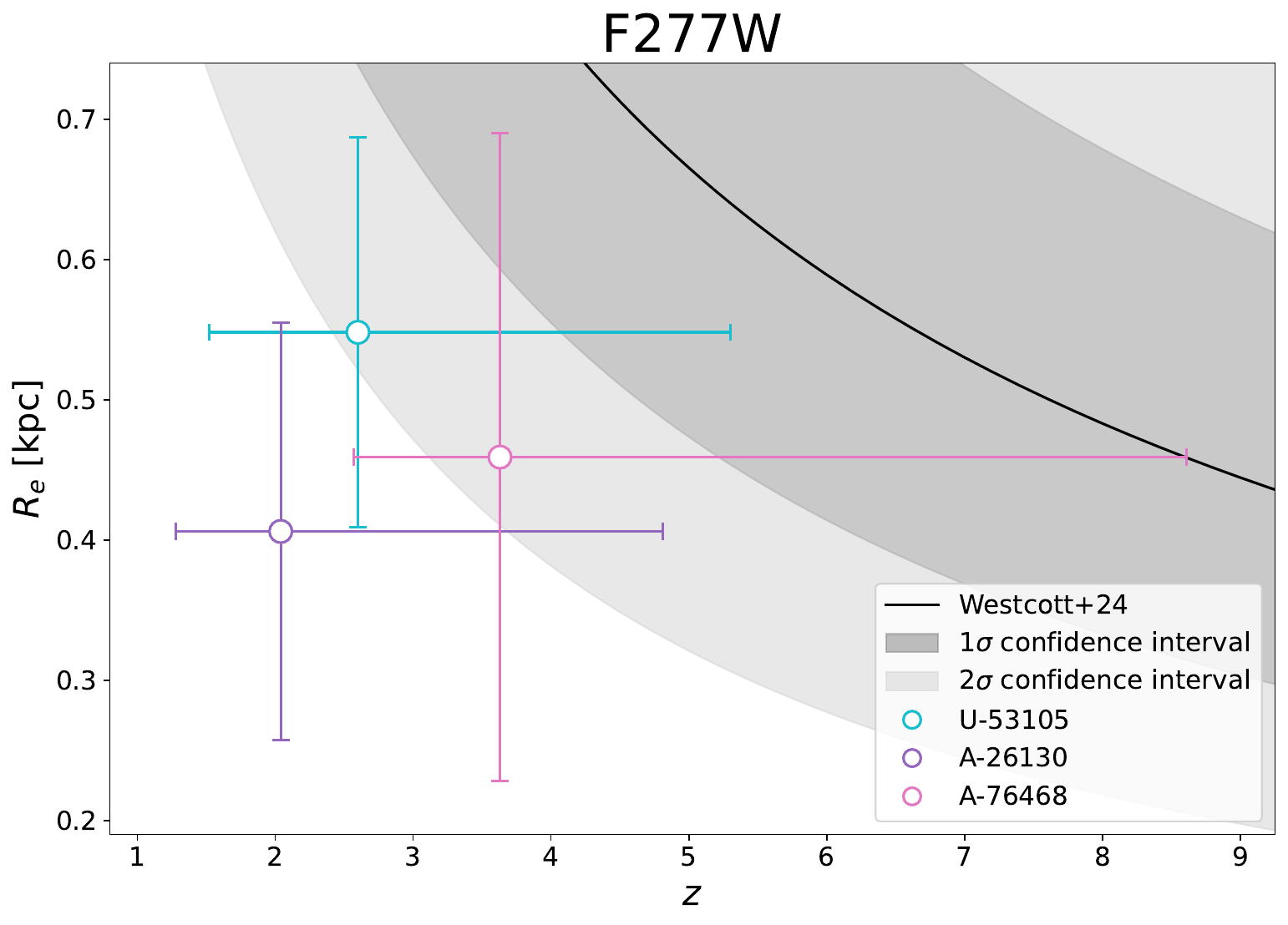}
    \includegraphics[width=0.4\textwidth]{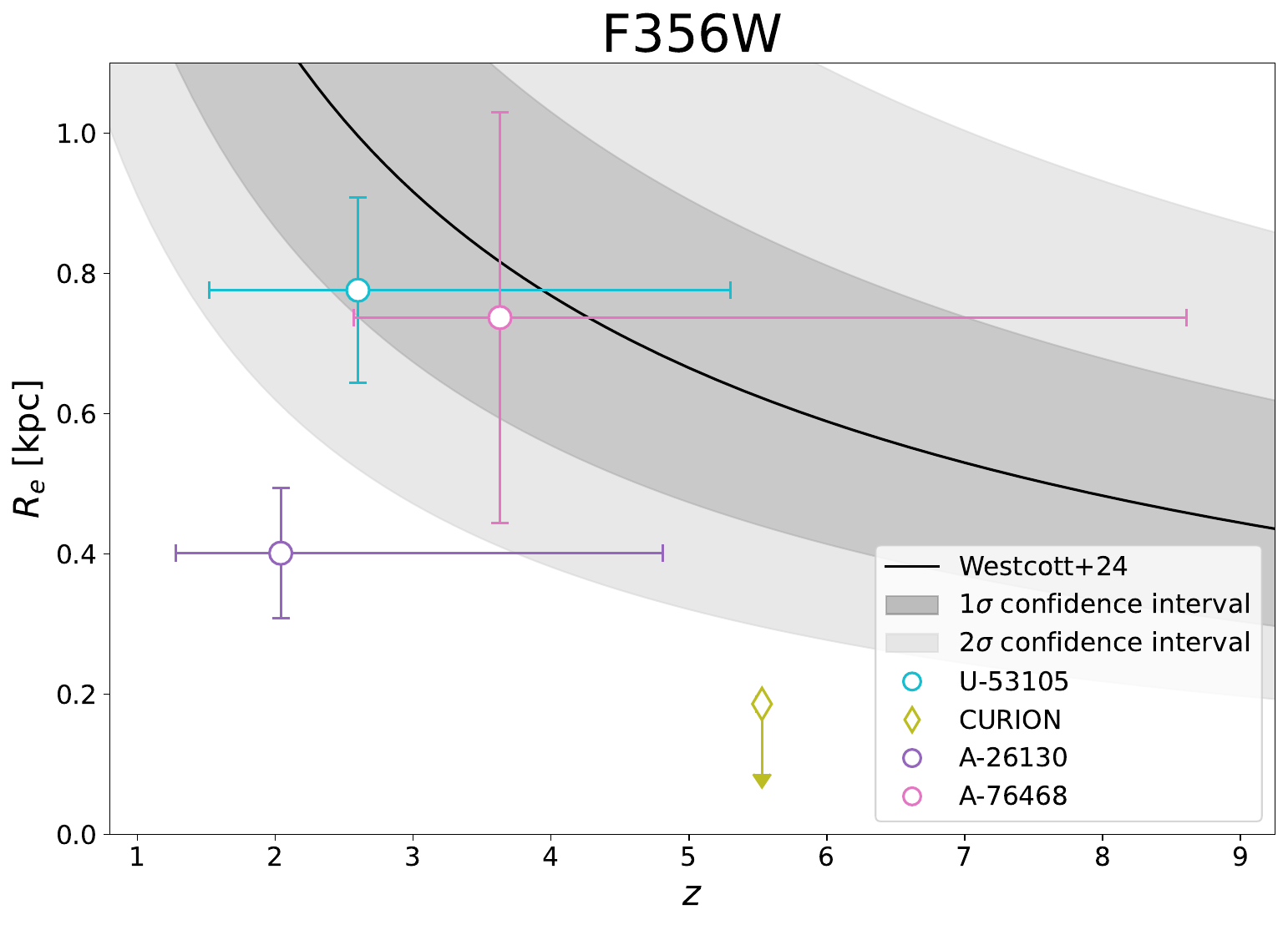}
    \includegraphics[width=0.4\textwidth]{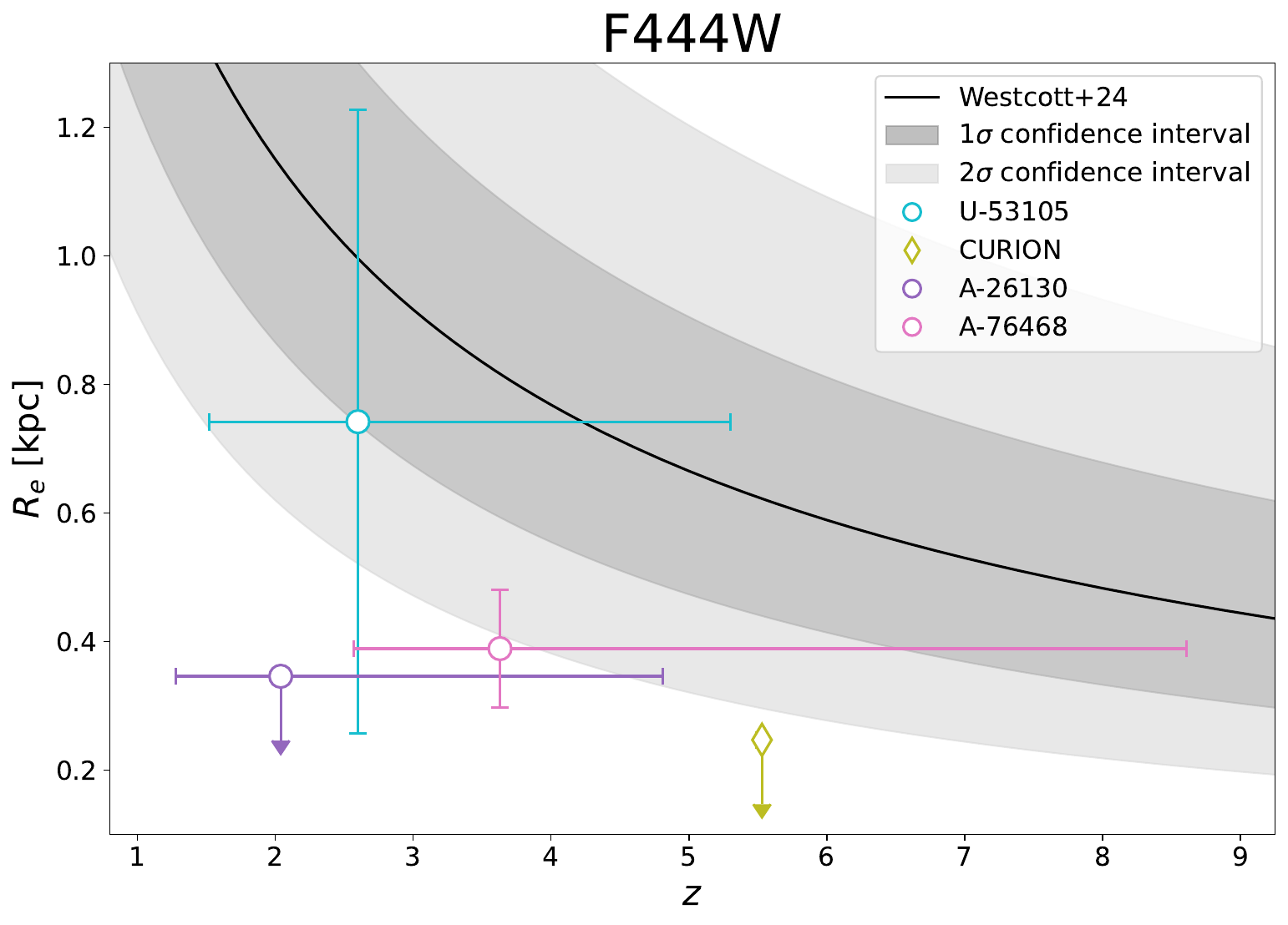}

    \caption{Same as Figure~\ref{fig:uhr_sizes} for our three HELM candidates and CURION.
    }
    \label{fig:other_sizes}
\end{figure*}

Finally, in the three panels of Figure~\ref{fig:other_sizes}, we perform a comparison between the best-fit sizes for our objects and the relation of \cite{2024arXiv241214970W} in an identical fashion to Figure~\ref{fig:uhr_sizes}. All of our HELM candidates are consistent within $2\sigma$ with the \cite{2024arXiv241214970W} relation, while CURION is consistent with it within $3\sigma$. As in the case of our UHR candidates, our analysis does not highlight any major inconsistency between our inferred best-fit galaxy sizes and the extrapolated relation by \cite{2024arXiv241214970W}.

\begin{landscape}

    \begin{table}[h]
        \centering
        \small
        \caption{F277W effective radii of our UHR galaxy candidates sample assuming their $z > 8$ solutions.}
        \setlength{\tabcolsep}{6pt}
        \begin{tabular}{ccccccc}
        \hline \hline
        ID & Name & $\text{z}_\text{high}$ & $\text{R}_\text{e}$ [arcsec] & $\text{R}_\text{e}$ [kpc] & $\text{n}_\text{s}$ & $\text{q}_\text{s}$\\
        \hline
        \vspace{2pt}
        U-31863 & BUDIARA & $17.76^{+1.6}_{-1.5}$ & 0.077 & $0.22 \pm 0.19$ & 0.329 & 0.319 \\
        U-34120 & VICIADGO & $17.56^{+1.4}_{-1.3}$ & 0.057 & $0.17 \pm 0.11$ & 0.467 & 0.319 \\
        U-75985 & LIZZAN & $17.80^{+1.48}_{-1.45}$ & 0.008 & $\leq 0.073$ & 3.502 & 0.319 \\
        U-80918 & NASPE & $17.55^{+1.47}_{-1.30}$ & 0.053 & $0.154 \pm 0.041$ & 0.703 & 0.319 \\
        A-22691 & ARCERIO & $17.32^{+2.05}_{-2.98}$ & 0.065 & $0.190 \pm 0.074$ & 0.552 & 0.292 \\
        \hline
        \end{tabular}
        \label{tab:F277Wuhrsizes}
    \end{table}

    \begin{table}[h]
        \centering
        \small
        \caption{F356W effective radii of our UHR galaxy candidates sample assuming their $z > 8$ solutions.}
        \setlength{\tabcolsep}{6pt}
        \begin{tabular}{ccccccc}
        \hline \hline
        ID & Name & $\text{z}_\text{high}$
 & $\text{R}_\text{e}$ [arcsec] & $\text{R}_\text{e}$ [kpc] & $\text{n}_\text{s}$ & $\text{q}_\text{s}$\\
        \hline
        \vspace{2pt}
        U-31863 & BUDIARA & $17.76^{+1.6}_{-1.5}$ & 0.050 & $0.145 \pm 0.067$ & 0.346 & 0.319 \\
        U-34120 & VICIADGO & $17.56^{+1.4}_{-1.3}$ & 0.109 & $0.3 \pm 0.1$ & 0.307 & 0.319 \\
        U-75985 & LIZZAN & $17.80^{+1.48}_{-1.45}$ & 0.028 & $\leq 0.087$ & 3.052 & 0.319\\
        U-80918 & NASPE & $17.55^{+1.47}_{-1.30}$ & 0.049 & $0.143 \pm 0.037$ & 0.558 & 0.319 \\
        A-22691 & ARCERIO & $17.32^{+2.05}_{-2.98}$ & 0.026 & $\leq 0.087$ & 5.2 & 0.371 \\
        \hline
        \end{tabular}
        \label{tab:F356Wuhrsizes}
    \end{table}

    \begin{table}[h]
        \centering
        \small
        \caption{F444W effective radii of our UHR galaxy candidates sample assuming their $z > 8$ solutions.}
        \setlength{\tabcolsep}{6pt}
        \begin{tabular}{ccccccc}
        \hline \hline
        ID & Name & $\text{z}_\text{high}$ & $\text{R}_\text{e}$ [arcsec] & $\text{R}_\text{e}$ [kpc] & $\text{n}_\text{s}$ & $\text{q}_\text{s}$\\
        \hline
        \vspace{2pt}
        U-31863 & BUDIARA & $17.76^{+1.6}_{-1.5}$ & 0.026 & $\leq 0.117$ & 0.361 & 0.319 \\
        U-34120 & VICIADGO & $17.56^{+1.4}_{-1.3}$ & 0.061 & $0.18 \pm 0.11$ & 0.318 & 0.319 \\
        U-75985 & LIZZAN & $17.80^{+1.48}_{-1.45}$ & 0.006 & $\leq 0.117$ & 0.75 & 0.319 \\
        U-80918 & NASPE & $17.55^{+1.47}_{-1.30}$ & 0.007 & $\leq 0.117$ & 0.856 & 0.319 \\
        A-22691 & ARCERIO & $17.32^{+2.05}_{-2.98}$ & 0.136 & $0.40 \pm 0.27$ & 0.37 & 0.256 \\
        \hline
        \end{tabular}
        \tablefoot{IDs starting with ``U-'' were selected in the updated CEERS catalog, whereas those starting with ``A-'' are from the CEERS ASTRODEEP-JWST catalog.}
        \label{tab:F444Wuhrsizes}
    \end{table}

\end{landscape}

\newpage

\begin{landscape}

    \begin{table}[h]
        \centering
        \small
        \caption{F277W effective radii of our three HELM galaxy candidates and CURION.}
        \setlength{\tabcolsep}{6pt}
        \begin{tabular}{ccccccc}
        \hline \hline
        ID & Name & $\text{z}_\text{high}$
 & $\text{R}_\text{e}$ [arcsec] & $\text{R}_\text{e}$ [kpc] & $\text{n}_\text{s}$ & $\text{q}_\text{s}$\\
        \hline
        \vspace{2pt}
        U-53105 & - & $2.60^{+2.70}_{-1.08}$ & 0.066 & $0.55 \pm 0.14$ & 0.48 & 0.319 \\
        U-112842 & CURION & $5.53^{+0.02}_{-0.04}$ & - & - & - & - \\
        A-26130 & - & $2.04^{+2.77}_{-0.76}$ & 0.047 & $0.41 \pm 0.15$ & 0.334 & 0.202 \\
        A-76468 & - & $3.63^{+4.98}_{-1.06}$ & 0.058 & $0.46 \pm 0.23$ & 0.73 & 0.319 \\
        \hline
        \end{tabular}
        \label{tab:F277Wothersizes}
    \end{table}

    \begin{table}[h]
        \centering
        \small
        \caption{F356W effective radii of our three HELM galaxy candidates and CURION.}
        \setlength{\tabcolsep}{6pt}
        \begin{tabular}{ccccccc}
        \hline \hline
        ID & Name & $\text{z}_\text{high}$
 & $\text{R}_\text{e}$ [arcsec] & $\text{R}_\text{e}$ [kpc] & $\text{n}_\text{s}$ & $\text{q}_\text{s}$\\
        \hline
        \vspace{2pt}
        U-53105 & - & $2.60^{+2.70}_{-1.08}$ & 0.093 & $0.78 \pm 0.13$ & 0.496 & 0.319 \\
        U-112842 & CURION & $5.53^{+0.02}_{-0.04}$ & 0.007 & $\leq 0.186$ & 0.31 & 0.589 \\
        A-26130 & - & $2.04^{+2.77}_{-0.76}$ & 0.046 & $0.401 \pm 0.093$ & 0.373 & 0.332 \\
        A-76468 & - & $3.63^{+4.98}_{-1.06}$ & 0.093 & $0.74 \pm 0.29$ & 0.455 & 0.319 \\
        \hline
        \end{tabular}
        \label{tab:F356Wothersizes}
    \end{table}

    \begin{table}[h]
        \centering
        \small
        \caption{F444W effective radii of our three HELM galaxy candidates and CURION.}
        \setlength{\tabcolsep}{6pt}
        \begin{tabular}{ccccccc}
        \hline \hline
        ID & Name & $\text{z}_\text{high}$
 & $\text{R}_\text{e}$ [arcsec] & $\text{R}_\text{e}$ [kpc] & $\text{n}_\text{s}$ & $\text{q}_\text{s}$\\
        \hline
        \vspace{2pt}
        U-53105 & - & $2.60^{+2.70}_{-1.08}$ & 0.089 & $0.74 \pm 0.49$ & 0.443 & 0.319 \\
        U-112842 & CURION & $5.53^{+0.02}_{-0.04}$ & 0.005 & $\leq 0.247$ & 1.479 & 0.425 \\
        A-26130 & - & $2.04^{+2.77}_{-0.76}$ & 0.020 & $\leq 0.346$ & 1.224 & 0.394 \\
        A-76468 & - & $3.63^{+4.98}_{-1.06}$ & 0.049 & $0.389 \pm 0.092$ & 0.382 & 0.319 \\
        \hline
        \end{tabular}
        \tablefoot{IDs starting with ``U-'' were selected in the updated CEERS catalog, whereas those starting with ``A-'' are from the CEERS ASTRODEEP-JWST catalog.}
        \label{tab:F444Wothersizes}
    \end{table}
    
\end{landscape}

\end{appendix}

\end{document}